\documentclass[preprint,12pt]{elsarticle}

\usepackage{amssymb}
\usepackage{epsfig}
\usepackage{color}
\usepackage{bm}
\usepackage{mathrsfs}

\biboptions{sort&compress}

\newcommand{\beq}[1]{\begin{equation}\label{#1}}
\newcommand{\eeq}{\end{equation}}
\newcommand{\bea}[1]{\begin{eqnarray} \label{#1}}
\newcommand{\eea}{\end{eqnarray}}
\newcommand{\ba}{\begin{array}}
\newcommand{\ea}{\end{array}}
\newcommand{\nn}{\nonumber}
\newcommand{\rf}[1]{(\ref{#1})}

\usepackage{epsfig}
\usepackage{color}

\newcommand{\Epho}{E_\gamma}
\newcommand{\Epi}{E_\pi}

\newcommand{\Fpho}{F_\gamma}

\def\be{\begin{equation}}
\def\ee{\end{equation}}

\def\bea{\begin{eqnarray}}
\def\eea{\end{eqnarray}}

\newcommand{\ga}{\gtrsim}

\newcommand{\postscript}[2]{\setlength{\epsfxsize}{#2\hsize}
   \centerline{\epsfbox{#1}}}

\newcommand{\eps}{\epsilon}

\newcommand{\rarr}{\rightarrow}

\newcommand{\nubar}{\bar\nu}
\newcommand{\nue}{\nu_{e}}
\newcommand{\numu}{\nu_{\mu}}

\newcommand{\numubar}{{\bar \nu}_\mu}

\newcommand{\epspi}{\epsilon_{\pi^\pm}}

\usepackage[usenames,dvipsnames]{xcolor}
\definecolor{rossoCP3}{cmyk}{0,.88,.77,.40}
\begin{document}
\begin{frontmatter}



\thispagestyle{empty}

\title{\begin{flushright}{\small \tt FERMILAB-PUB-13-541-A}\end{flushright}
\vspace{0.8cm} 
\color{rossoCP3} {\bf Cosmic Neutrino Pevatrons: A Brand New Pathway to Astronomy, 
Astrophysics, and Particle Physics} \color{black}}

\author[1]{Luis A.~Anchordoqui}
\author[2]{Vernon Barger}
\author[3]{Ilias Cholis}
\author[4]{Haim Goldberg}
\author[3,5]{Dan \nolinebreak Hooper}
\author[6,7]{Alexander Kusenko}
\author[8]{John G. Learned}
\author[8]{Danny \nolinebreak Marfatia}
\author[8]{Sandip Pakvasa}
\author[1,4]{Thomas \nolinebreak C. \nolinebreak Paul}
\author[9]{Thomas J. Weiler}

\address[1]{Department of Physics,\\ University of Wisconsin-Milwaukee, Milwaukee, WI 53201, USA}
\address[2]{Department of Physics,\\ University of Wisconsin, Madison, WI 53706, USA}
\address[3]{Center for Particle Astrophysics,\\ Fermi National Accelerator Laboratory, Batavia, IL 60510,USA}
\address[4]{Department of Physics,\\ Northeastern University, Boston, MA 02115, USA}
\address[5]{Department of Astronomy and Astrophysics,\\ Enrico Fermi
Institute, University of Chicago, Chicago, Il 60637, USA}
\address[6]{Department of Physics and Astronomy,\\ University of California, Los
Angeles, CA 90095-1547, USA}
\address[7]{Kavli IPMU (WPI),\\ University of Tokyo, Kashiwa, Chiba 277-8568, Japan}
\address[8]{Department of Physics and Astronomy,\\
University of Hawaii, Honolulu, HI 96822, USA}

\address[9]{Department of Physics and Astronomy,\\ Vanderbilt University, Nashville TN 37235, USA}

\begin{abstract}

The announcement by the IceCube Collaboration of the
observation of 28 cosmic neutrino candidates has been greeted with a great
deal of justified excitement. The data reported so far depart by
$4.3\sigma$ from the expected atmospheric neutrino background, which
raises the obvious question: ``Where in the Cosmos are these neutrinos
coming from?''  We review the many possibilities which have been
explored in the literature to address this question, including origins
at either Galactic or extragalactic celestial objects. For
completeness, we also briefly discuss new physics processes which may either
explain or  be constrained by IceCube data.

\end{abstract}


\end{frontmatter}


\tableofcontents

\vfill\eject


\section{Introduction} \label{sec:intro}

Neutrinos will serve as unique astronomical messengers. Except for
oscillations induced by transit in a vacuum Higgs field, neutrinos
propagate without interactions between source and Earth, providing
powerful probes of high energy astrophysics.  The neutrino's direction
and energy (modulo the usual red-shifting due to expansion of the
universe) are preserved, and the neutrino's flavor is altered in a
calculable way.  The potential power of neutrino astrophysics has been
discussed in a number of review
articles~\cite{Gaisser:1994yf,Learned:2000sw,Halzen:2002pg,Becker:2007sv,
  Anchordoqui:2009nf}. In addition, the flavor composition of
neutrinos originating at astrophysical sources can serve as a probe of
new physics in the electroweak
sector~\cite{Learned:1994wg,Beacom:2002vi,Beacom:2003eu,Beacom:2003nh,
  Beacom:2003zg,Hooper:2005jp,Anchordoqui:2005gj}. Furthermore, decays
and annihilations of hypothetical dark matter particles accumulated in
Sun are expected to produce a large flux of secondary neutrinos at
energies far above the 1-20~MeV energies of neutrinos produced in
solar burning~\cite{Silk:1985ax,Srednicki:1986vj,
  Halzen:1991kh,Barger:2001ur,Halzen:2005ar,Barger:2010ng,Barger:2011em}.
Observation of such high energy neutrinos coming from the direction of
the Sun would provide ``smoking ice'' for dark matter
hunters~\cite{Aartsen:2012kia}.  However, neutrinos constitute
something of a double-edged sword: they are excellent probes of
astrophysics and particle physics because of their feeble
interactions, but also extremely difficult to detect for the same
reason.

Neutrino (antineutrino) interactions with matter can be reduced to two
categories: {\it (i)}~in charged current (CC) interactions the
neutrino becomes a charged lepton through the exchange of a $W^\pm$
with some particle $X$, $\nu_\alpha \, (\bar \nu_\alpha) + X
\rightarrow l_\alpha^\pm + {\rm anything}$; {\it (ii)} in neutral
current (NC) interactions the neutrino interacts via a $Z$
transferring momentum to jets of hadrons, but producing  a neutrino 
rather than an $l^\pm$ in the final state: 
\mbox{$\nu_\alpha \, (\bar  \nu_\alpha) + X \rightarrow \nu_\alpha \, (\bar \nu_\alpha) + ~{\rm anything}$.}  
Lepton flavor is labeled as $\alpha = e,\, \mu,\,
\tau$ from here on.  The neutrino-nucleon cross section rises roughly
linearly with energy~\cite{Quigg:1986mb,Reno:1987zf,Gandhi:1995tf,Gandhi:1998ri,
Anchordoqui:2006ta,CooperSarkar:2007cv,Jeong:2010za,Block:2010ud,
Connolly:2011vc,Illarionov:2011wc,CooperSarkar:2011pa}.
For neutrino telescopes located on Earth, the detection probability is
modulated by a combination of the neutrino energy $E_\nu$ and the
arrival zenith angle $\theta$.  For $E_\nu \lesssim 10^5~{\rm GeV}$,
most neutrinos pass through the Earth unscattered, and thus in this
energy range the detection probability rises with energy. At about
$10^5$~GeV, the interaction length of neutrinos is roughly equal to
the Earth's diameter, and hence about 80\% (40\%) of $\nu_\mu$ and
$\nu_e$ with $\cos \theta = -1 \, (-0.7)$ are
absorbed~\cite{L'Abbate:2004hv}.  For the case of the  tau neutrino, there
is a subtlety in its propagation through matter due to the short
$\tau$ lifetime.  A $\nu_\tau$ propagating through the Earth can interact to 
generate a $\tau$ lepton which subsequently decays, producing a
$\nu_\tau$ of lower energy, a process referred to as the
``regeneration effect''~\cite{Halzen:1998be} (though this will generally
have negligible consequence for steeply falling spectra).

The rate of interaction of $\nu_e$, $\nu_\mu$, $\nu_\tau$, $\bar
\nu_\mu$, $\bar \nu_\tau$, with electrons is mostly negligible compared to
interactions with nucleons. However, the case of $\bar \nu_e$ is
unique because of resonant scattering, $\bar \nu_e e^- \to W^- \to 
{\rm anything}$, at $E_\nu \simeq 6.3~{\rm PeV}$.
The $W^-$ resonance in this process is commonly referred
to as the Glashow resonance~\cite{Glashow:1960zz}.  The signal for
$\bar \nu_e$ at the Glashow resonance, when normalized to the total
$\nu + \bar \nu$ flux, can be used to differentiate between the two
primary candidates ($p\gamma$ and $pp$ collisions) for
neutrino-producing interactions in optically thin sources of cosmic
rays~\cite{Anchordoqui:2004eb}. In $pp$ collisions the nearly
isotopically neutral mix of pions will create on decay a neutrino
population with the ratio $N_{\nu_\mu}=N_{\bar\nu_\mu} =
2N_{\nu_e}=2N_{\bar\nu_e}.$ On the other hand, in photopion
interactions   
the isotopically asymmetric process 
$p\gamma\rightarrow\Delta^+\rightarrow \pi^+ n$, 
$\pi^+\to\mu^+ \nu_\mu\to e^+\nu_e \nubar_\mu \nu_\mu$,
is the dominant source of neutrinos so that at production, \mbox{$N_{\nu_\mu}=N_{\bar\nu_\mu} = N_{\nu_e} \gg
N_{\bar\nu_e}$.}\footnote{It has been noted that advanced civilizations across
the Galaxy could use a monochromatic signal at the Glashow resonance
for purposes of communication~\cite{Learned:2008gr}.} Note that events at
the Glashow resonance provide the only known physics calibration of neutrino
detectors in this high energy range, always a worrisome problem (witness the
difficulties with the highest energy air showers, as in the Piere Auger Observatory and Telescope Array~\cite{Anchordoqui:2013eqa}).

At PeV energies neutrinos interact with nucleons with a cross section
of about 1~nb ($1~{\rm b} = 10^{-24}~{\rm cm}^2$). Hence, for a
detector medium with a density of about $N_A \simeq 6 \times 10^{23}$
nucleons per ${\rm cm}^3$ we expect only a fraction ${\cal
  O}(10^{-5})$ of PeV neutrinos to interact within 1~km of the medium.
If the medium is transparent, like water or ice, the fast-moving
secondary charged particles created in these interactions can be
observed via the resulting Cherenkov light emission.  Assuming cosmic
ray (CR) sources are optically thin, one can estimate the diffuse flux
of extragalactic neutrinos from the observed cosmic ray flux, since
the relevant particle physics is well-known.  The only wiggle room is
the efficiency of the energy transfer from protons to pions,
$\epsilon_\pi$.  An upper bound on the flux ($\epsilon_\pi = 1$) was
first obtained by Waxman and
Bahcall~\cite{Waxman:1998yy,Bahcall:1999yr}.  For an estimate of
$\epsilon_\pi$ based on our best current knowlege, the diffuse flux of
extragalactic neutrinos would provide ${\cal O}(10^5)$ PeV neutrinos
per year and ${\rm km}^2$. Thus,  observation of a few
extragalactic PeV neutrinos per year requires neutrino telescopes
with active detector volumes on the scale of cubic-kilometers.
IceCube is the first observatory on this scale and we can hope that
the European KM3-NET will soon join the club.

\subsection{Historical Background}

The long road to developing the IceCube experiment has been thoroughly
described in~\cite{Halzen:2007ip,Spiering:2012xe}.  Here we recount
some of the highlights.  Early efforts concentrated on instrumenting
large pre-existing volumes of water to produce giant Cherenkov
detectors.  The first major step from conceptual ideas to large-scale
experimental efforts was taken by the Deep Underwater Muon and
Neutrino Detector (DUMAND) project~\cite{Bosetti:1988kh}. In November
1987, the DUMAND Collaboration measured the muon vertical intensity at
depths ranging between $2 - 4~{\rm km}$ (in intervals of 500~m), with
a prototype string of optical detectors deployed about $30~{\rm km}$
off-shore the  island of Hawaii~\cite{Babson:1989yy}, and set a
limit on high energy showers~\cite{Bolesta:1997}. DUMAND paved the way
for later efforts by pioneering many of the detector technologies in
use today.  This project inspired both the independent development and
deployment of an instrument in the Siberian Lake
Baikal~\cite{Belolaptikov:1997ry}, as well as later efforts to
commission neutrino telescopes in the Mediterranean, NESTOR, NEMO and
ANTARES.

The most developed of the Meditterannean efforts, the ANTARES detector
is deployed at depth of about 2.5~km and has been operating in
its complete configuration since 2008, with 885 photomultiplier tubes
(PMTs) enclosed in Optical Modules (OMs) and distributed in triplets
on 12 detection lines~\cite{Collaboration:2011nsa}.  The Collaboration has
published some physics results, but the apparatus is not large enough
to compete with or
complement the IceCube results, which are the focus of this review (the
geometric volume of ANTARES is $\lesssim 1/1000$ that of IceCube). The
KM3-NET Collaboration is aiming at cubic kilometer scale detector(s)
to be placed in the Mediterranean within the next few years.

In addition to activity in deep underwater neutrino experiments  pursued in the late 1980's, a variety of smaller underground
detectors were active, starting in the early 1960's, but greatly
ramping up in the 1980's largely motivated by the hunt for nucleon
decay. The first large underground water-based Cherenkov detector was
the 10 kiloton IMB (Irvine-Michigan-Brookhaven) water Cherenkov
detector, which began operation in 1982 in a Morton salt mine near
Cleveland, Ohio. It dispatched the $SU(5)$ grand unification prediction
for proton decay and went on to make the first significant
observations of contained neutrino interactions, including the first
hints of muon neutrino oscillations, via the suggestion of a deficit in GeV
atmospheric neutrino fluxes.

Several other experiments were deployed in the 1980s, including the
Baksan (Russian), Frejus (France), LSD (Mont Blanc tunnel), LVD (Gran
Sasso), Homestake Mine (South Dakota, water detector), Soudan
(Minnesota), Kolar Gold Fields (India), and Kamioka (Japan).  All
detected mostly cosmic ray muons and a few neutrinos of typically 100
MeV to 1 GeV.

Most spectacular and unexpected, were the neutrino fireworks (in the 10 MeV range) from
Supernova 1987A, which revealed the first neutrino source ever observed
beyond our solar system~\cite{Hirata:1987hu,Bionta:1987qt} (and until now, the last!).  
These cosmic neutrinos (19 in all) were detected by the IMB and the smaller but more 
sensitive Kamiokande detector, as well as by the Baksan detector. Not only did this 
event confirm theoretical ideas about the mechanism of supernova explosion, but 
it provided a  bound on the neutrino mass and many other 
neutrino properties. This event put much energy into the drive towards neutrino astronomy. 
Unfortunately in order to view a substantial rate of SN one would need to be senstitive 
out to the Virgo Cluster scale (20 Mpc) which requires a gigaton scale detector with 
10 MeV sensitivity (a thousand times lower than IceCube threshold energy). 

In 1998 Super-Kamiokande went on to confirm the long-suspected neutrino oscillation 
phenomenon via muon neutrinos produced in the atmosphere~\cite{Fukuda:1998mi}.

The first telescope on the scale envisaged by the DUMAND Collaboration
was realized by transforming a large volume of the extremely
transparent, deep Antarctic ice into a particle detector, called the
Antarctic Muon and Neutrino Detector Array (AMANDA)~\cite{Halzen:1988wr,Lowder:1991uy}.  During 1993 and
1994, in an exploratory phase, the four-string AMANDA-A array was
deployed and instrumented with 80 PMTs spaced at 10~m intervals from
810 to 1000~m. (The scattering length at that depth turned out to be
too short to allow useful detecton volume.) A deeper array of 10
strings, referred to as AMANDA-B10, was deployed during the austral
summers between 1995 and 1997, to depths between 1500 and 2000~m. The
instrumented volume formed a cylinder with diameter 120~m, viewed by
302 OMs~\cite{Andres:1999hm,Andres:2001ty}. During December 1997 and
January 2000, the detector was expanded with an additional nine outer
strings of OMs. The composite AMANDA-II array of 19 strings and 677
OMs comprising two concentric cylinders with larger diameter of 200~m
became operational in 2000 and continued taking data up to 2009.

\begin{figure}[tpb]
\postscript{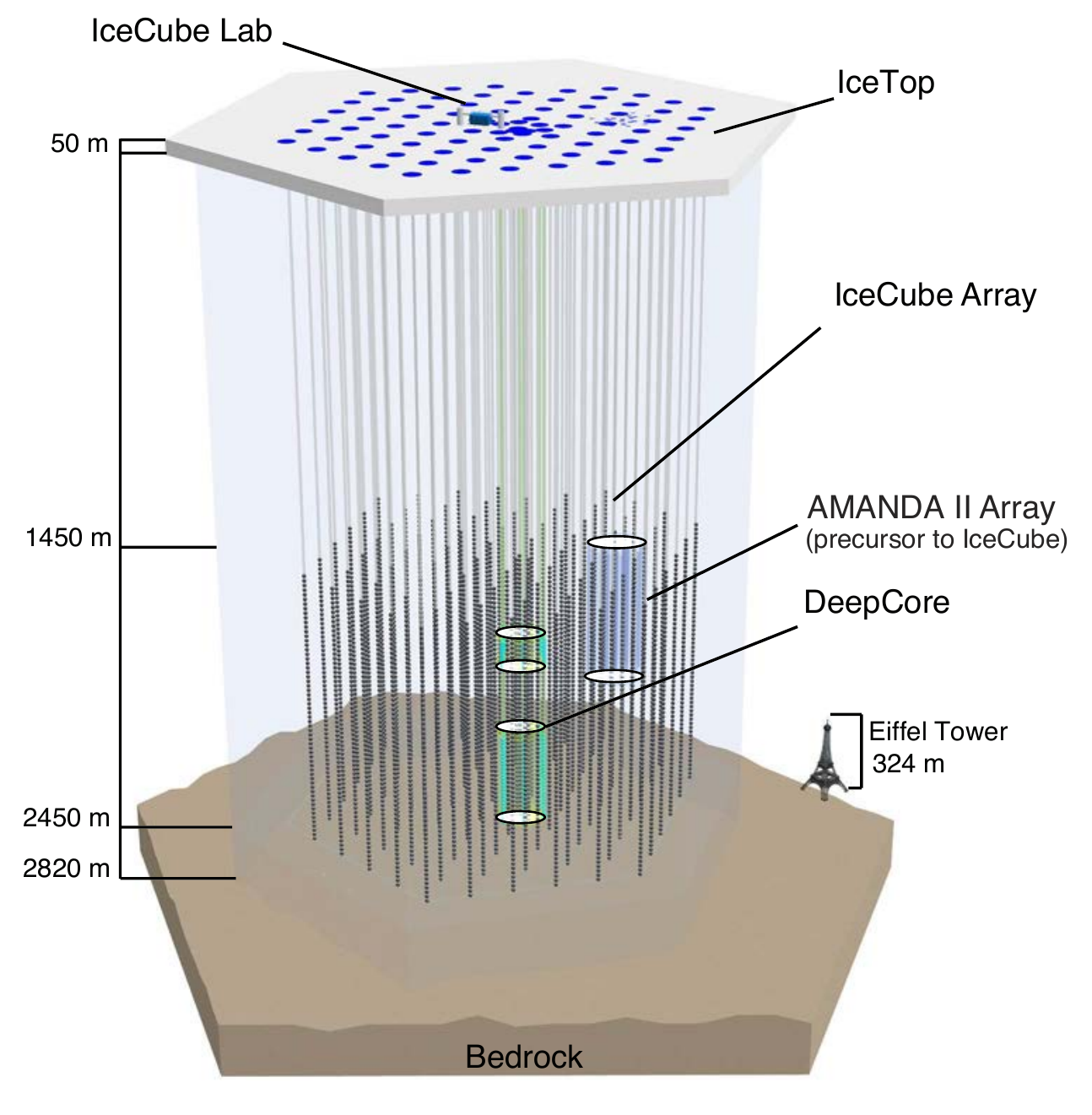}{0.9}
\caption{Schematic of the IceCube instrument, which covers a cubic kilometer of 
  Antarctic glacial ice. It detects neutrinos by observing Cherenkov 
  light from secondary charged particles produced in neutrino-nucleon
  interactions. This light is detected by an array of 5160 DOMs, 
  each of which contains a photomultiplier and readout electronics housed 
  in a clear glass pressure resistant sphere. The DOMs are
  arranged into an array of 86 vertical strings, with 60 DOMs per
  string at depths between 1450 m and 2450 m. Outside of the DeepCore
  low-energy subarray, these DOMs are vertically spaced at 17-meter
  intervals and the strings are on average 125 m apart
  horizontally. The DeepCore subarray fills in the center of the
  detector with a denser array of photomultipliers and provides a
  lower energy threshold of 10 GeV over a fraction of the
  IceCube volume. Figure courtesy of the IceCube Collaboration.}
\label{fig:icecube}
\end{figure}

The follow-up to AMANDA-II, IceCube, finally realized the objective of
instrumenting one cubic kilometer.  IceCube is located near the
Amundsen-Scott station below the surface of the Antarctic ice sheet at
the geographic South Pole, sharing the location of its precursor
observatory~\cite{Achterberg:2006md}.  A sketch of the IceCube
facility is shown in Fig.~\ref{fig:icecube}. The main part of the
detector is the ``InIce'' array of digital optical modules (DOMs)
which detect Cherenkov light~\cite{Abbasi:2008aa}. The DOMs are
attached to km-long supply and read-out cables  -- so-called ``strings''
-- and deployed deep (more than 1.5~km) in the Antarctic ice. Each
string carries 60 DOMs spaced evenly along 1~km.  The full baseline
design of 86 strings was completed in December 2010. There is a region
of dusty ice at about  2000~m, which is not useful. A region
of particularly clear ice is equipped with a denser array of DOMs
attached to six additional strings~\cite{Collaboration:2011ym}. This
``DeepCore'' infill provides a low energy (to ~$10~{\rm GeV}$)
extension of IceCube and increases the sensitivity of indirect dark
matter searches and neutrino oscillations. In addition to the InIce
array, IceCube also possesses an air shower array called ``IceTop''
which comprises 80 stations, each of which consists of two tanks of
water-ice instrumented with 2 DOMs to detect Cherenkov
light~\cite{IceCube:2012nn}. The hybrid observations of air showers in
the InIce and IceTop arrays have mutual benefits, namely significant air
shower background rejection (for neutrino studies) and an improved air
shower muon detection (for CR studies).

\subsection{ The IceCube Detector and Neutrino Detection}

The IceCube event
topologies are classified as cascades, tracks, or combinations of
these. This leads to a zoo of possible signatures and the possibility
to fully disentangle the neutrino flavor composition. The energy and
angular resolution achievable for each event depends on the details of
its topology.  Here we pause to discuss in more detail the various
event topologies.

In a CC event a $\nu_\mu$ produces a muon traveling in nearly the same
direction as the neutrino. Secondary muons range out over kilometers
at $E_\mu \sim 10^3$~GeV, to tens of kilometers at $E_\mu \sim
10^9$~GeV, generating showers along their track by bremsstrahlung,
pair production and photonuclear interactions. All of these are
sources of Cherenkov light. As the energy of the muon degrades along
its track, the energy of the secondary showers diminishes and the
distance from the track over which the associated Cherenkov light can
trigger a PMT becomes smaller. The geometry of the lightpool
surrounding the muon track over which single photo-electron are
produced, for muon of initial energy more than $200~{\rm GeV}$, is
about a kilometer or more long cone with gradually decreasing radius.
Energy is thus determined from range and energy loss rate.  For such a
muon observed over a 1~km path length in the IceCube detector, the
energy resolution is $\Delta (\log_{10} E_\mu) \approx
0.22$~\cite{Abbasi:2012wht}. At energies $E_\mu > 100$~TeV or so,
muons produced inside the instrumented volume will always leave the
detector, allowing a strong lower bound on the neutrino energy, and
depending upon circumstances some energy measurement of muon. The
orientation of the Cherenkov cone reveals the neutrino direction, with
an angular resolution of about $0.7^\circ$~\cite{Ahrens:2002dv}. Muons
created by cosmic ray interactions in the atmosphere constitute the
main background, up to $\sim 100~{\rm TeV}$ energies. (Muons with
energies of PeV shower strongly in the upper ice and cannot penetrate
it to IceCube with substantial energy. A PeV energy muon from any
direction must be due to neutrinos, or something even more
interesting.)

Cascades (or showers of elementary particles) are generated by
neutrino collisions --- $\nu_e\ \mbox{or}\,\,\nu_\tau$ CC
interactions, and all NC interactions --- inside of or near the
detector, and by unseen-muon-generated showers near the
detector. These external showers most frequently will originate from
muon induced pair production, bremsstrahlung or nuclear interactions.
Normally, a reduction of the muon produced shower background is
effected by placing a cut of $10^{4.6}$~GeV on the minimum
reconstructed energy~\cite{Ackermann:2004zw}. Electron neutrinos
deposit 50\% to 80\% of their energy into an electromagnetic shower
initiated by the leading final state electron.  The rest of the energy
goes into the fragments of the target that produce a second (and
nearly co-linear) subdominant shower.

The length of the shower is of orders of meters in ice, which is small
compared to the PMT spacing. As a consequence, the shower results in
roughly a point source of Cherenkov photons projected in some
direction. The optical scattering length in ice (20 m or less) leads
to diffusion of the radiation over a nearly spherical volume, rather
than a conical projection.  Still enough directionality is retained to
reconstruct the shower direction to $15^\circ - 20^{\circ}$ (as
compared to $<1^{\circ}$ for muons).  (We note that this provides an
opportunty for KM3 as compared to IceCube, since in the ocean the
shower directionality can be presumably much better.)

These events trigger the PMTs at the single
photoelectron level over a spherical volume whose radius scales
linearly with the logarithm of shower energy. For ice, the radius is 130~m at
$10^{4}$~GeV and 460~m at $10^{10}$~GeV, {\it i.e.} the shower radius grows
by just over 50\,m per decade in energy.  The measurement of the
radius of the sphere in the lattice of PMTs determines the energy and
renders neutrino-detection-experiments as total energy
calorimeters. The energy resolution is $\Delta (\log_{10} E_\nu) \approx
0.26$~\cite{Abbasi:2011ui}.  (Note that due to this non-linearity in response,
even the unlikely contained interaction by a $10^{10}$~GeV neutrino will not 
saturate a km$^3$ detector volume, a redeeming virtue of the othewise 
annoying optical scattering.)

For $\nu_\tau$'s, CC current interactions produce different signals
depending on the energy. For $\tau$ leptons less energetic than
$10^{6}$~GeV, the shower (hadronic or electromagnetic) from the $\tau$
decay cannot be separated from the hadronic shower of the initial
$\nu_\tau$ interaction. At \mbox{$E_\tau \approx 10^{6}$~GeV}, the $\tau$
range becomes a few hundred meters and the two showers produced may be
easily separated and be identify as a double bang
event~\cite{Learned:1994wg}. At energies $10^7 \lesssim E_\tau/{\rm
  GeV} < 10^{7.5}$, the $\tau$ decay length is comparable to the
instrumented volume. In such cases, one may observe a $\tau$ track
followed by the $\tau$-decay shower (``lollipop topology''), or a
hadronic shower followed by a $\tau$ track which leaves the detector
(``popillol topology'').  At energies $E_\tau > 10^{7.5}$~GeV,  the decay length $\gg
1$~km and $\tau$'s leave only a track like muons. However, a $\tau$
going through the detector at high energies without decaying will not
deposit as much energy in the detector as a comparable-energy muon,
due to the mass difference.  (The direct-pair production process
scales inversely with mass, so it dominates tau-lepton energy
loss~\cite{Becattini:2000fj} resulting in 1/20$^{\rm th}$ the light
produced by a muon.) Such a $\tau$ might then be indistinguishable from a
low energy muon track, and thus might not be flagged as an interesting
event. In summary, the energy range from $10^{6.5} \lesssim E_\nu/{\rm GeV} \lesssim 10^{7.5}$ is
the ``sweet spot'' for $\tau$ detection in IceCube, since here one can
observe all the distinctive topologies.

\begin{figure}[tpb]
\postscript{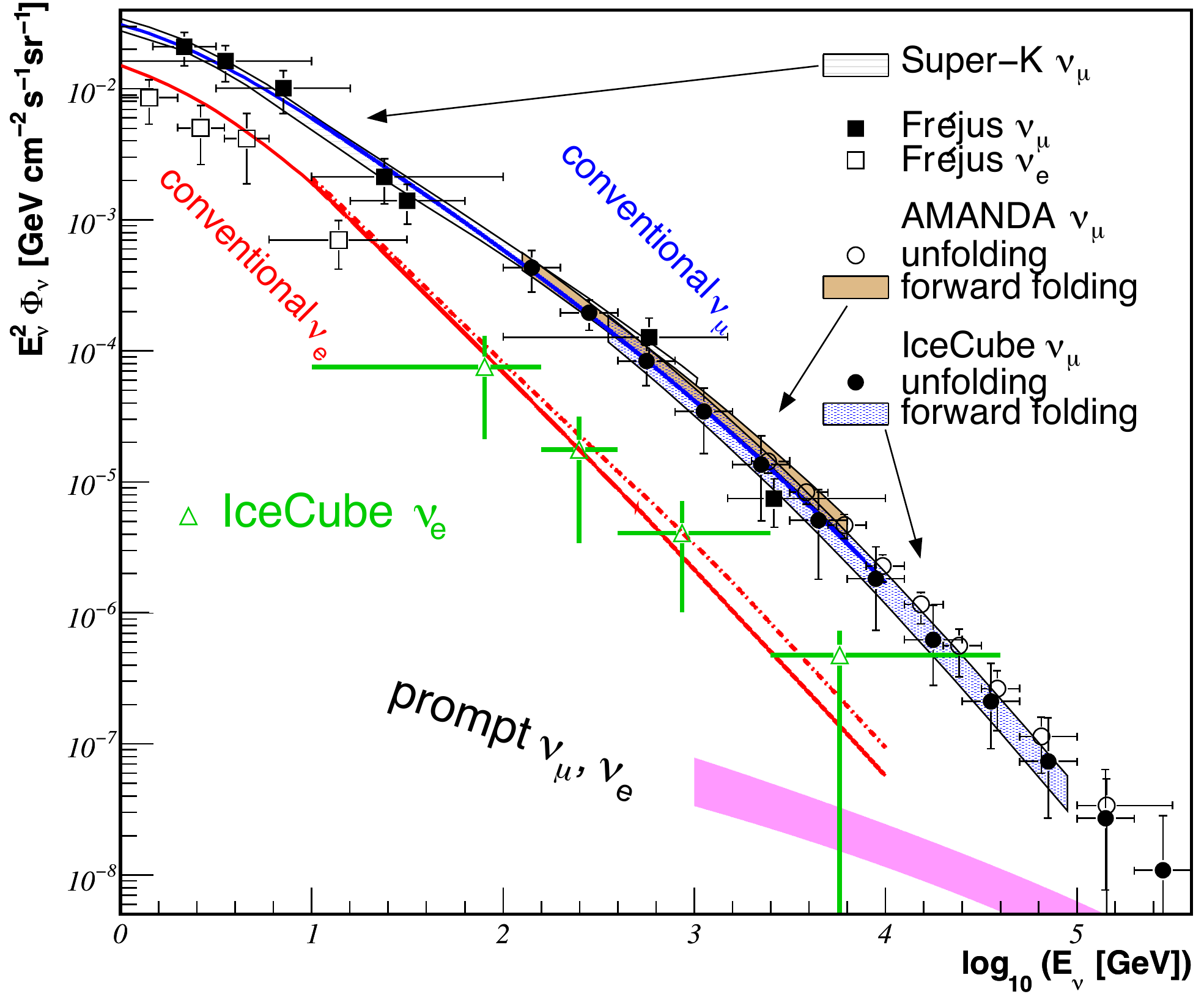}{0.9}
\caption{Atmospheric muon and electron neutrino spectrum as function
  of energy. The open and filled symbols represent measurements of
  various detectors (Super-Kamiokande~\cite{GonzalezGarcia:2006ay},
  Fr\'ejus~\cite{Daum:1994bf}, AMANDA forward folding
  analysis~\cite{Abbasi:2009nfa} and unfolding
  analysis~\cite{Abbasi:2010qv}, IceCube (40 strings) forward folding
  $\nu_\mu$ analysis~\cite{Abbasi:2011jx}, unfolding $\nu_\mu$
  analysis~\cite{Abbasi:2010ie}, and
  $\nu_e$~\cite{Aartsen:2012uu}). Curved lines are theoretical
  predictions of atmospheric fluxes. The conventional $\nu_e$ (red
  solid line) and $\nu_\mu$ (blue solid line) from
  Ref.~\cite{Sanuki:2006yd}, and $\nu_e$ (red dotted line) from
  Ref.~\cite{Barr:2006it}.  The (magenta) band for the prompt flux
  indicates the theoretical uncertainty of charm-induced
  neutrinos~\cite{Enberg:2008te}. From Ref.~\cite{Aartsen:2012uu}. It
  should be noted that this figure contains deceptive spectra
  which do not consist of measured data points at given energies, but
  spectra fitted as a whole, and hence the individual "error bars" are
  misleading at best. The IceCube $\nu_e$ data is an exception.
  Super-Kamiokande has contained events for muons up to about 10 GeV
  and electrons up to about 100 GeV. (Frejus had no events in the
  highest energies in which they report a flux measurement with error
  bars, for example.)  One should note that folding techniques can
  obscure any unexpected distortions of the tested spectra, which are
  typically simple power law distributions.}
  \label{fig:atm}
\end{figure}

When protons and nuclei enter the atmosphere, they collide with the
air molecules and produce all kinds of secondary particles, which in
turn interact, decay or propagate to the ground, depending on their
intrinsic properties and energies. In the GeV range, the most abundant
particles are neutrinos produced in the decay of mesons. Pion decay
dominates the atmospheric neutrino production, $\pi^+ \to \mu^+ \,
\nu_\mu \to e^+ \, \nu_e \, \nu_\mu \, \bar \nu_\mu$ (and the conjugate
process), and determines the neutrino energy spectra up to about
100~GeV. Above this energy, the flux become
increasingly modified by the kaon contribution, which asymptotically
reaches 90\%. In the atmosphere mesons are subject to an
interaction-decay competition.  As a consequence of this, neutrinos
from meson decay have a spectrum that is one power of energy steeper
than the primary cosmic ray spectrum. The muon daughter neutrinos have
a spectrum steeper by two powers of energy, because the muon spectrum
itself is steeper by $1/E$. Electron neutrinos have a differential
spectrum (approximately) $\propto E_\nu^{-4.7}.$ The muon neutrino
spectrum is flatter, $\propto E_\nu^{-3.7}$ up to $10^{5}~{\rm GeV}$,
steepening to $\propto E_\nu^{-4.0}$.  In this energy window, the
flavor ratios are
\mbox{$N_{\nu_e}: N_{\nu_\mu}: N_{\nu_\tau}  \approx 1: 20:0$} and 
the energy spectra are  functions of the zenith angle of the atmospheric 
cascades~\cite{Lipari:1993hd}. This is because mesons in
inclined showers spend more time in tenuous atmosphere where they are
more likely to decay rather than interact. Above about $10^5$~GeV, kaons
are also significantly attenuated before decaying and the ``prompt''
component, arising mainly from very short-lived charmed mesons
($D^\pm,$ $D^0,$ $D_s$ and $\Lambda_c$) dominates the
spectrum~\cite{Zas:1992ci}. Such a prompt neutrino flux
is isotropic with flavor ratios $10:10:1$~\cite{Beacom:2004jb}.   These various spectra
are summarized in Fig.~\ref{fig:atm}. The neutrino flux arising from pion and
kaon decay is reasonably well understood, with an uncertainty in the
range $10\%-20\%$~\cite{Gaisser:2002jj}. The prompt atmospheric neutrino
flux, however, is much less understood, because of uncertainty about
cosmic ray composition and relatively poor knowledge of small-$x$ QCD
processes~\cite{Enberg:2008te}.

The flux of atmospheric neutrinos is a curse and a blessing; the
background of neutrinos produced by cosmic rays in interactions with
air nuclei provides a beam essential for calibrating the detectors and
demonstrating the neutrino measurement technique. It also provides an
opportunity to probe standard neutrino oscillations and those arising
from new physics, such as violation of Lorentz
invariance~\cite{GonzalezGarcia:2005xw,Anchordoqui:2005is}. Over the
next decade, a data set of the order of one million atmospheric
neutrinos will be collected. The statistics will be so large that
some mapping of the Earth's interior denisty profile will be possible 
via neutrino tomography~\cite{GonzalezGarcia:2007gg}. (It is not clear that
this wll permit improvement on the density profile deduced from seismology 
however.)

\subsection{IceCube and Extraterrestrial Neutrinos}

A nearly guaranteed neutrino flux originates from interactions of
ultra-high energy cosmic rays (UHECRs) {\it en route} to
Earth. Ultrahigh energy protons above the photopion production
threshold interact with the cosmic microwave and infrared backgrounds
as they propagate over cosmological distances, the ``GZK''
process~\cite{Greisen:1966jv,Zatsepin:1966jv}.  These interactions
generate pions and neutrons, which decay to produce
neutrinos~\cite{Beresinsky:1969qj}, known as Berezinsky-Zatspein
neutrinos, or the ``BZ'' flux. The accumulation of such neutrinos over
cosmological time is known as the BZ or often the "cosmogenic neutrino
flux"~\cite{Stecker:1978ah,Hill:1983xs,Engel:2001hd,Fodor:2003ph}. Ultra-high
energy nuclei also interact with the cosmic microwave and infrared
backgrounds, undergoing
photodisintegration~\cite{Greisen:1966jv,Zatsepin:1966jv}. The
disassociated nucleons then interact with the cosmic microwave and
infrared backgrounds to produce cosmogenic
neutrinos~\cite{Hooper:2004jc,Ave:2004uj}. While the presence of a
suppression feature in the UHECR spectrum is generally expected for
all compositions, the flux of cosmogenic neutrinos and
$\gamma$ rays~\cite{Gelmini:2007sf,Gelmini:2007jy,Taylor:2008jz}
subsequently produced are both very sensitive to the CR source
model~\cite{Allard:2006mv,Anchordoqui:2007fi,Kotera:2010yn,Ahlers:2012rz}. Indeed
this difference permits information on these fluxes to be used as a
probe of the composition or vice-versa. For instance, an upper limit
on the proton fraction in UHE extragalactic CRs can, in principle, be
inferred from experimental bounds on both the diffuse flux of UHE
neutrinos~\cite{Ahlers:2009rf} and the diffuse flux of UHE
photons~\cite{Hooper:2010ze}. Furthermore, these two messengers
starkly contrast in their subsequent propagation, with UHE neutrinos
freely propagating out to the Hubble-scale whereas UHE $\gamma$ rays
are limited to tens of Mpc distance scales. This difference of scales
highlights the fact that these two messengers can offer complementary
information about the distant and local source distribution
respectively. Moreover, the accompanying output into secondary
electrons and positrons, in particular from Bethe-Heitler pair
production, feeds into electromagnetic cascades from the cosmic
microwave background (CMB) and intergalactic magnetic fields. This
leads to the accumulation of $\gamma$ rays in the energy range ${\rm
  GeV} \lesssim E_\gamma \lesssim {\rm TeV}$. The observed diffuse
$\gamma$ ray flux by \textit{Fermi}-LAT~\cite{Abdo:2010nz} hence
provides a constraint on the total energy injected into such cascades
over the Universe's entire history and can be translated into upper
limits on the cosmic diffuse flux of photons and
neutrinos~\cite{Berezinsky:1975zz,Berezinsky:2010xa,Ahlers:2010fw}.

By devising a search dedicated to finding these cosmogenic neutrinos,
the two (now three) highest energy neutrinos ever observed were quite recently
uncovered~\cite{Aartsen:2013bka}. We will review the possible origins
of these events, as well as more recently detected lower energy
neutrinos from cosmic sources~\cite{Aartsen:2013pza,FrancisHalzenfortheIceCube:2013uva}. 
The layout of the review is as follows. In Sec.~\ref{section-2} we will discuss
the published IceCube data, including the characteristics of the
energy spectrum, the arrival directions, and the role of atmospheric
prompt neutrinos. In Sec.~\ref{section-3} we discuss the consequences
of the neutrino observations for theories of Galactic neutrino
production. In Sec.~\ref{section-4} we consider potential
extragalactic sources, including active galactic nuclei (AGNs),
gamma-ray bursts (GRBs), starburst galaxies (SBGs), and newly-born pulsars. In
Sec.~\ref{section-5} we turn to beyond Standard Model physics,
including the production of neutrinos from heavy particle decay, and
the relevance of this to dark matter. We will also discuss possible
enhancement of the neutrino-nucleon cross section and the consequence
of certain new physics processes for neutrino oscillations. Finally,
in Sec.~\ref{section-6} we make a few observations on the consequences
of the overall picture discussed herein.

\section{Evidence for Extraterrestrial High Energy Neutrinos}
\label{section-2}

\begin{figure}[tbp]
\begin{minipage}[t]{0.52\textwidth}
\postscript{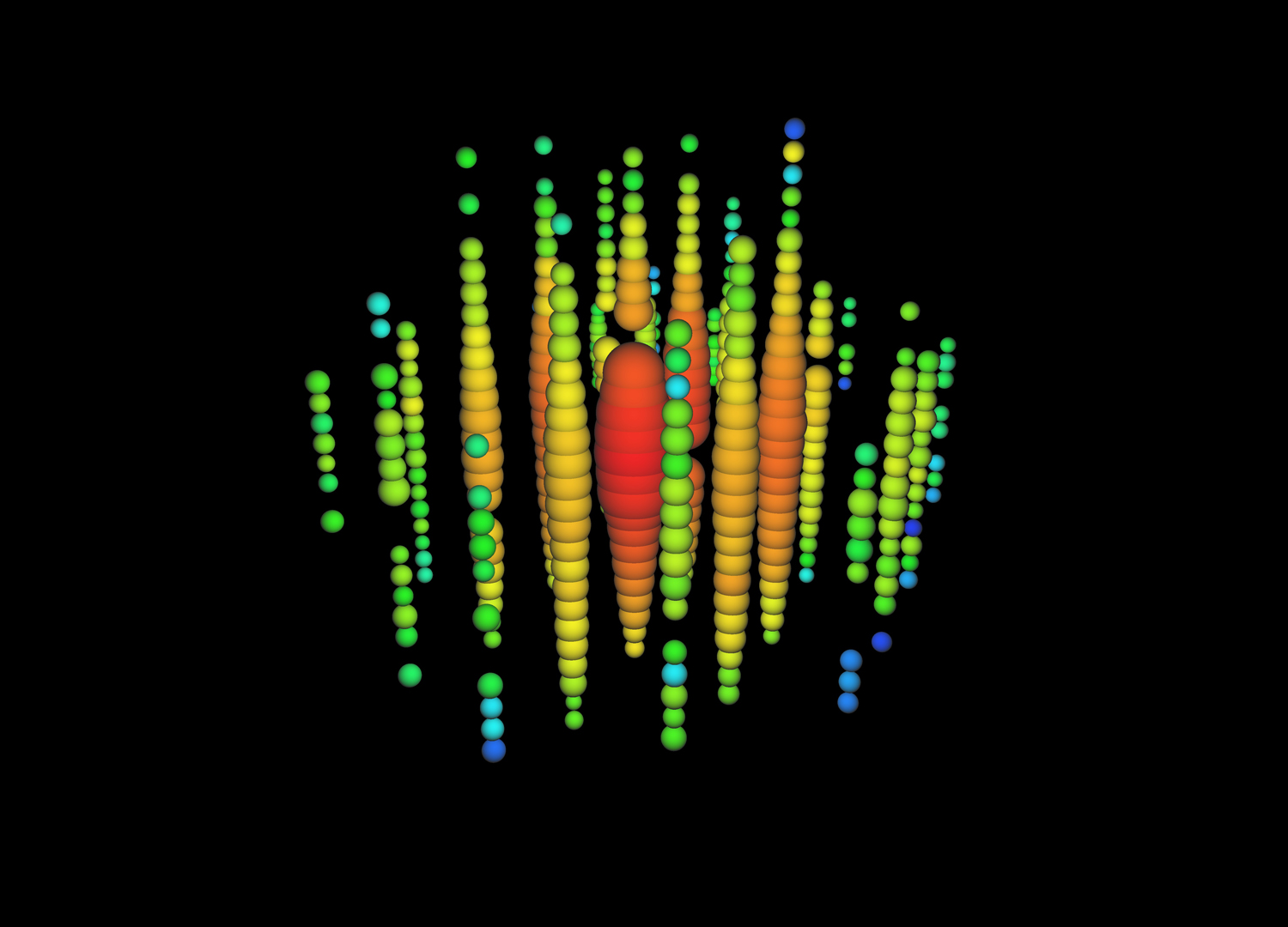}{0.99}
\end{minipage}
\hfill
\begin{minipage}[t]{0.47\textwidth}
\postscript{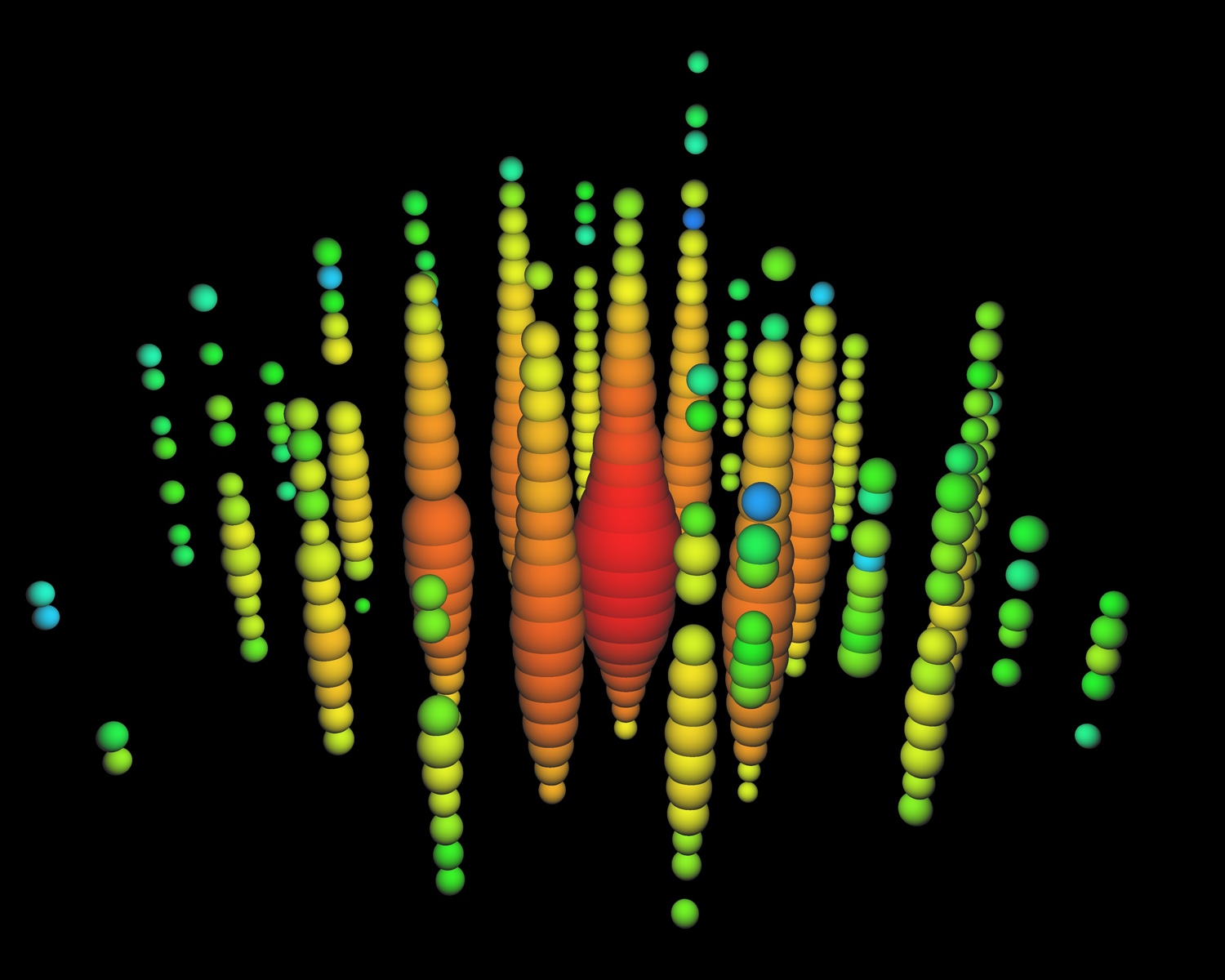}{0.99}
\end{minipage}
\caption{The two highest energy neutrino events reported by the
  IceCube Collaboration. The left panel corresponds to the event
  called Bert that ocurred in August 2011, whereas the right panel
  shows the event in January 2012, called Ernie. Each sphere represents a
  DOM. Colors represent the arrival times of the photons where red
  indicates early and blue late times. The size of the spheres is a
  measure for the recorded number of photo-electrons. Figure courtesy
  of the IceCube Collaboration.}
    \label{fig:eventview} 
\end{figure}

In April 2013 the IceCube Collaboration published an observation of
two $\sim$~1 PeV neutrinos, with a p-value 2.8$\sigma$ beyond the
hypothesis that these events were atmospherically
generated~\cite{Aartsen:2013bka}.  These two candidates were found in
a search for events with a significant energy deposition as expected
for cosmogenic neutrinos. These two events
are the highest energy neutrino candidates as yet reported. As can be 
seen in Fig.~\ref{fig:eventview} the
events exhibit a ``cascade'' morphology consistent with that expected
to result from CC interactions of electron neutrinos,
low-energy tau neutrinos, and NC interactions for all
three flavors.

New results were presented in May 2013 at the IceCube Particle
Astrophysics Symposium (IPA
2013)~\cite{Kopper:2013ipa,Naoko:2013ipa,Whitehorn:2013ipa}.  In a new
search protocol, down-going events were selected based on the
requirement that they display a vertex contained within the
instrumented ice volume, effectively employing the edges of the
IceCube detector as a veto for down-going muons.  Since atmospheric
neutrinos are produced by the same parent mesons which generate the
shower muons, imposing this veto also provides a partial self-veto of
the accompanying down-going atmospheric neutrino background, as
discussed in~\cite{Schonert:2008is}.  This technique is particularly
effective for energies $E_\nu > 1$~TeV, where the boost is sufficient
to ensure that the shower muons and neutrinos follow nearly identical
trajectories.  The new analysis, published in November 2013, revealed
an additional 26 neutrino candidates depositing ``electromagnetic
equivalent energies'' ranging from about 30~TeV up to
250~TeV~\cite{Aartsen:2013pza}. The main properties of these events,
which were observed between May 2010 to May 2012, are given in
Table~\ref{tab:events}.\footnote{
The energy given in Table~\ref{tab:events} is equal to the neutrino 
energy for $\nu_e$ CC events, within experimental 
uncertainties, and is otherwise a lower limit on the neutrino energy 
due to exiting muons or neutrinos.  Errors on energy and the angle include both
statistical and systematic effects.  Systematic uncertainties on
directions for shower-like events were determined on an individual
basis; track systematic uncertainties here are equal to $1^\circ$. 
The arrival directions are given in equatorial coordinates, right 
ascension (R.A.) and declination (Dec.). The topologies of all these 
events are shown in Ref.~\cite{Aartsen:2013pza}.}

\begin{table}
\begin{center}
\begin{tabular}{|c|c|c|c|c|c|c|}
\hline \hline
   & Dep. Energy & Time  & Decl. & R.A. & Med. Angular    &   Event   \\
ID & (TeV)            & (MJD) & (deg.)      & (deg.)          & Error (deg.) & Type\\ \hline 
\hline \\ [-2.0 ex]
1 & $47.6 \,^{+6.5}_{-5.4}$ & 55351.3222110 & $-1.8$ & $35.2$ & $16.3$ & Shower \\
\hline \\ [-2.0 ex]
2 & $117 \,^{+15}_{-15}$ & 55351.4659612 & $-28.0$ & $282.6$ & $25.4$ & Shower \\
\hline \\ [-2.0 ex]
3 & $78.7 \,^{+10.8}_{-8.7}$ & 55451.0707415 & $-31.2$ & $127.9$ & $\lesssim 1.4$ & Track \\
\hline \\ [-2.0 ex]
4 & $165 \,^{+20}_{-15}$ & 55477.3930911 & $-51.2$ & $169.5$ & $7.1$ & Shower \\
\hline \\ [-2.0 ex]
5 & $71.4 \,^{+9.0}_{-9.0}$ & 55512.5516214 & $-0.4$ & $110.6$ & $\lesssim 1.2$ & Track \\
\hline \\ [-2.0 ex]
6 & $28.4 \,^{+2.7}_{-2.5}$ & 55567.6388084 & $-27.2$ & $133.9$ & $9.8$ & Shower \\
\hline \\ [-2.0 ex]
7 & $34.3 \,^{+3.5}_{-4.3}$ & 55571.2585307 & $-45.1$ & $15.6$ & $24.1$ & Shower \\
\hline \\ [-2.0 ex]
8 & $32.6 \,^{+10.3}_{-11.1}$ & 55608.8201277 & $-21.2$ & $182.4$ & $\lesssim 1.3$ & Track \\
\hline \\ [-2.0 ex]
9 & $63.2 \,^{+7.1}_{-8.0}$ & 55685.6629638 & $33.6$ & $151.3$ & $16.5$ & Shower \\
\hline \\ [-2.0 ex]
10 & $97.2 \,^{+10.4}_{-12.4}$ & 55695.2730442 & $-29.4$ & $5.0$ & $8.1$ & Shower \\
\hline \\ [-2.0 ex]
11 & $88.4 \,^{+12.5}_{-10.7}$ & 55714.5909268 & $-8.9$ & $155.3$ & $16.7$ & Shower \\
\hline \\ [-2.0 ex]
12 & $104 \,^{+13}_{-13}$ & 55739.4411227 & $-52.8$ & $296.1$ & $9.8$ & Shower \\
\hline \\ [-2.0 ex]
13 & $253 \,^{+26}_{-22}$ & 55756.1129755 & $40.3$ & $67.9$ & $\lesssim 1.2$ & Track \\
\hline \\ [-2.0 ex]
14 & $1041 \,^{+132}_{-144}$ & 55782.5161816 & $-27.9$ & $265.6$ & $13.2$ & Shower \\
\hline \\ [-2.0 ex]
15 & $57.5 \,^{+8.3}_{-7.8}$ & 55783.1854172 & $-49.7$ & $287.3$ & $19.7$ & Shower \\
\hline \\ [-2.0 ex]
16 & $30.6 \,^{+3.6}_{-3.5}$ & 55798.6271191 & $-22.6$ & $192.1$ & $19.4$ & Shower \\
\hline \\ [-2.0 ex]
17 & $200 \,^{+27}_{-27}$ & 55800.3755444 & $14.5$ & $247.4$ & $11.6$ & Shower \\
\hline \\ [-2.0 ex]
18 & $31.5 \,^{+4.6}_{-3.3}$ & 55923.5318175 & $-24.8$ & $345.6$ & $\lesssim 1.3$ & Track \\
\hline \\ [-2.0 ex]
19 & $71.5 \,^{+7.0}_{-7.2}$ & 55925.7958570 & $-59.7$ & $76.9$ & $9.7$ & Shower \\
\hline \\ [-2.0 ex]
20 & $1141 \,^{+143}_{-133}$ & 55929.3986232 & $-67.2$ & $38.3$ & $10.7$ & Shower \\
\hline \\ [-2.0 ex]
21 & $30.2 \,^{+3.5}_{-3.3}$ & 55936.5416440 & $-24.0$ & $9.0$ & $20.9$ & Shower \\
\hline \\ [-2.0 ex]
22 & $220 \,^{+21}_{-24}$ & 55941.9757760 & $-22.1$ & $293.7$ & $12.1$ & Shower \\
\hline \\ [-2.0 ex]
23 & $82.2 \,^{+8.6}_{-8.4}$ & 55949.5693177 & $-13.2$ & $208.7$ & $\lesssim 1.9$ & Track \\
\hline \\ [-2.0 ex]
24 & $30.5 \,^{+3.2}_{-2.6}$ & 55950.8474887 & $-15.1$ & $282.2$ & $15.5$ & Shower \\
\hline \\ [-2.0 ex]
25 & $33.5 \,^{+4.9}_{-5.0}$ & 55966.7422457 & $-14.5$ & $286.0$ & $46.3$ & Shower \\
\hline \\ [-2.0 ex]
26 & $210 \,^{+29}_{-26}$ & 55979.2551738 & $22.7$ & $143.4$ & $11.8$ & Shower \\
\hline \\ [-2.0 ex]
27 & $60.2 \,^{+5.6}_{-5.6}$ & 56008.6845606 & $-12.6$ & $121.7$ & $6.6$ & Shower \\
\hline \\ [-2.0 ex]
28 & $46.1 \,^{+5.7}_{-4.4}$ & 56048.5704171 & $-71.5$ & $164.8$ & $\lesssim 1.3$ & Track \\
\hline 
\hline
\end{tabular}
\end{center}
\caption{
  Properties of the 28 events.
  Shown are the deposited electromagnetic-equivalent energy (the energy deposited by the events in IceCube assuming all light was made in electromagnetic showers) as well as the arrival time and direction of each event and its topology (track or shower-like). The events are ordered according to the Modified Julian Date (MJD).}
\label{tab:events}
\end{table}

\begin{figure}[tbp]
\postscript{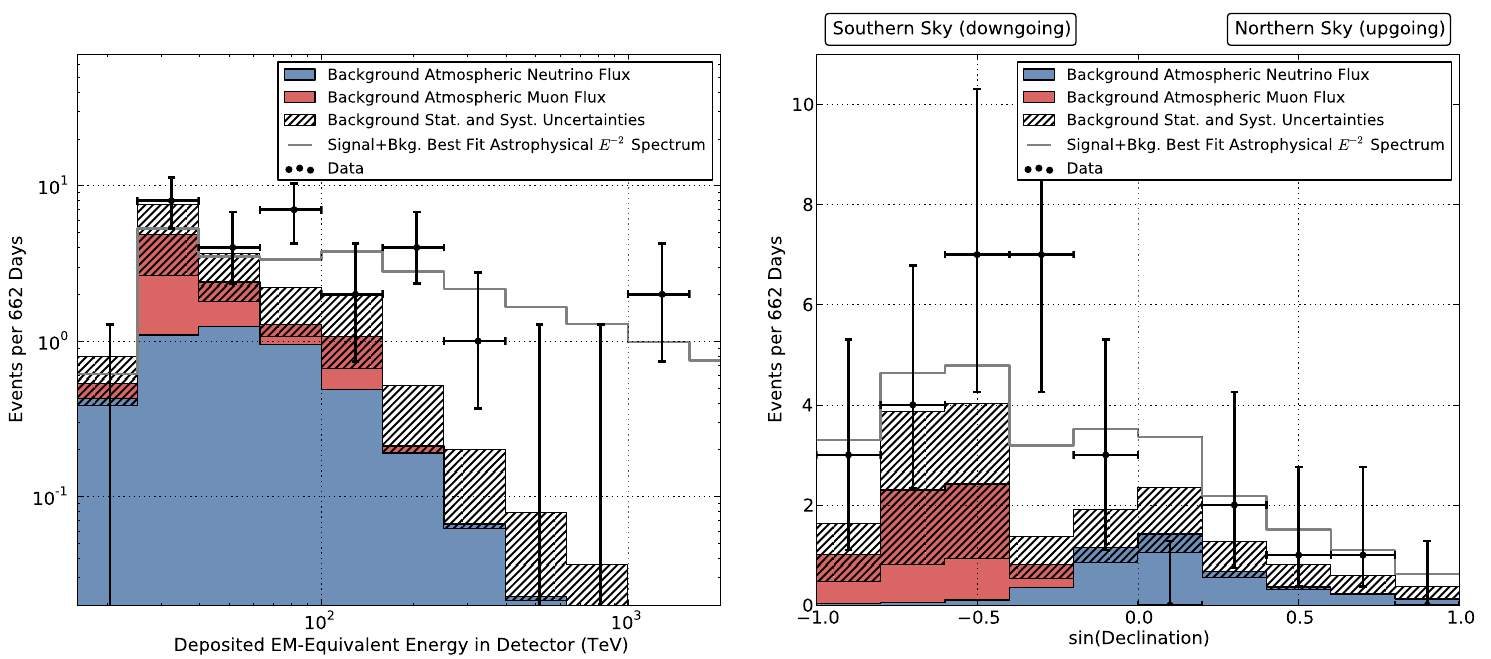}{0.98}
\caption{Distribution of the deposited energies (left) and declination
  angles (right) of the IceCube observed events compared to model
  predictions. Energies plotted are in-detector visible energies,
  which are lower limits on the neutrino energy. Note that deposited
  energy spectra are always harder than the spectrum of the neutrinos
  that produced them due to the neutrino cross-section increasing with
  energy. The expected rate of atmospheric neutrinos is based on
  northern hemisphere muon neutrino observations at lower
  energies. The estimated distribution of the background from
  atmospheric muons is shown in red. Due to lack of statistics from
  data far above the cut threshold, the shape of the distributions
  from muons in this figure has been determined using Monte Carlo
  simulations with total rate normalized to the estimate obtained from
  the in-data control sample. Combined statistical and systematic
  uncertainties on the sum of backgrounds are indicated with a hatched
  area. The gray line shows the best-fit canonical $E^{-2}$
  astrophysical spectrum with all-flavor normalization (1:1:1) of
  $E_\nu^2 \Phi_\nu^{\rm total} (E_\nu) = 3.6 \times 10^{-8}~{\rm GeV}
  \, {\rm cm}^{-2} \, {\rm s}^{-1} \, {\rm sr}^{-1}$ and a spectral
  cutoff of 2~PeV derived in~\cite{Aartsen:2013bka}. (An $E_\nu^{-2}$
  spectrum is used here as a reference, as this spectral index is
  expected for canonical firs-order Fermi shock acceleration. In
  reality, this index may be somewhat larger or smaller.) From
  Ref.~\cite{Aartsen:2013pza}.}
\label{fig:snowmass}
\end{figure}

These events, together with atmospheric neutrino background
expectations, are displayed in Fig.~\ref{fig:snowmass}. The left panel
shows the distribution of electromagnetic (EM) equivalent energy.  At
first glance, one may notice a gap between 250~TeV and the 2 highest
energy events~\cite{He:2013zpa}. Keep in mind, however, that the lower
energy events contain track topologies, which, as discussed before,
represent only a lower bound on the neutrino energy. For example, the
highest energy event in the search for $\nu_\mu$ performed using data
collected when IceCube was running in its 59-string configuration (May
2009 to May 2010) is most likely originated from a neutrino of energy
$E_\nu \sim 0.5 - 1~{\rm PeV}$, producing a muon that passed through
the detector with an energy $E_\mu \approx 400~{\rm TeV}$~\cite{Aartsen:2013eka}. 

Thus at present statistics are not sufficient to determine whether 
the suggestive gap in event energies represents a real structure in
the spectrum~\cite{Anchordoqui:2013qsi}.  Seven of the events show
visible evidence of a muon track, and the remainder are consistent
with cascades induced by $\nu_e$'s or $\nu_\tau$'s (or their antiparticles 
or neutral current events). The quoted background estimate from atmospheric
neutrinos is $10.6^{+4.6}_{-3.9}$. Taken together, the total sample of
28 events departs from the atmospherically-generated neutrino
hypothesis by $4.3\sigma$.\footnote{See however~\cite{Lipari:2013taa}
for the possible significance of prompt neutrinos.}

Interpreting these results in terms of popular astrophysical models
appears to be challenging.  First of all, if the neutrino flux is
indeed a Fermi-shock flux falling as an unbroken $E_\nu^{-2}$
power-law spectrum~\cite{Fermi:1949ee} would lead to about 8-9 events
above 1~PeV, which thus far are not observed.  This null result at
high-energy may be indicative of a cutoff in the spectrum at
$1.6^{+1.5}_{-0.4}$~PeV~\cite{Whitehorn:2013ipa}.  (But note the newly
reported Big Bird
event~\cite{SpencerR.KleinfortheIceCube:2013vca,Halzen:2013dva}, with
$E_\nu \gtrsim 2 ~{\rm PeV}$ which will raise this cutoff somewhat.)
On the other hand, it may be possible to maintain consistency with the
data with a steeper but still unbroken $E_\nu^{-\Gamma}$ spectrum,
with $\Gamma > 2$.

\begin{figure}
\postscript{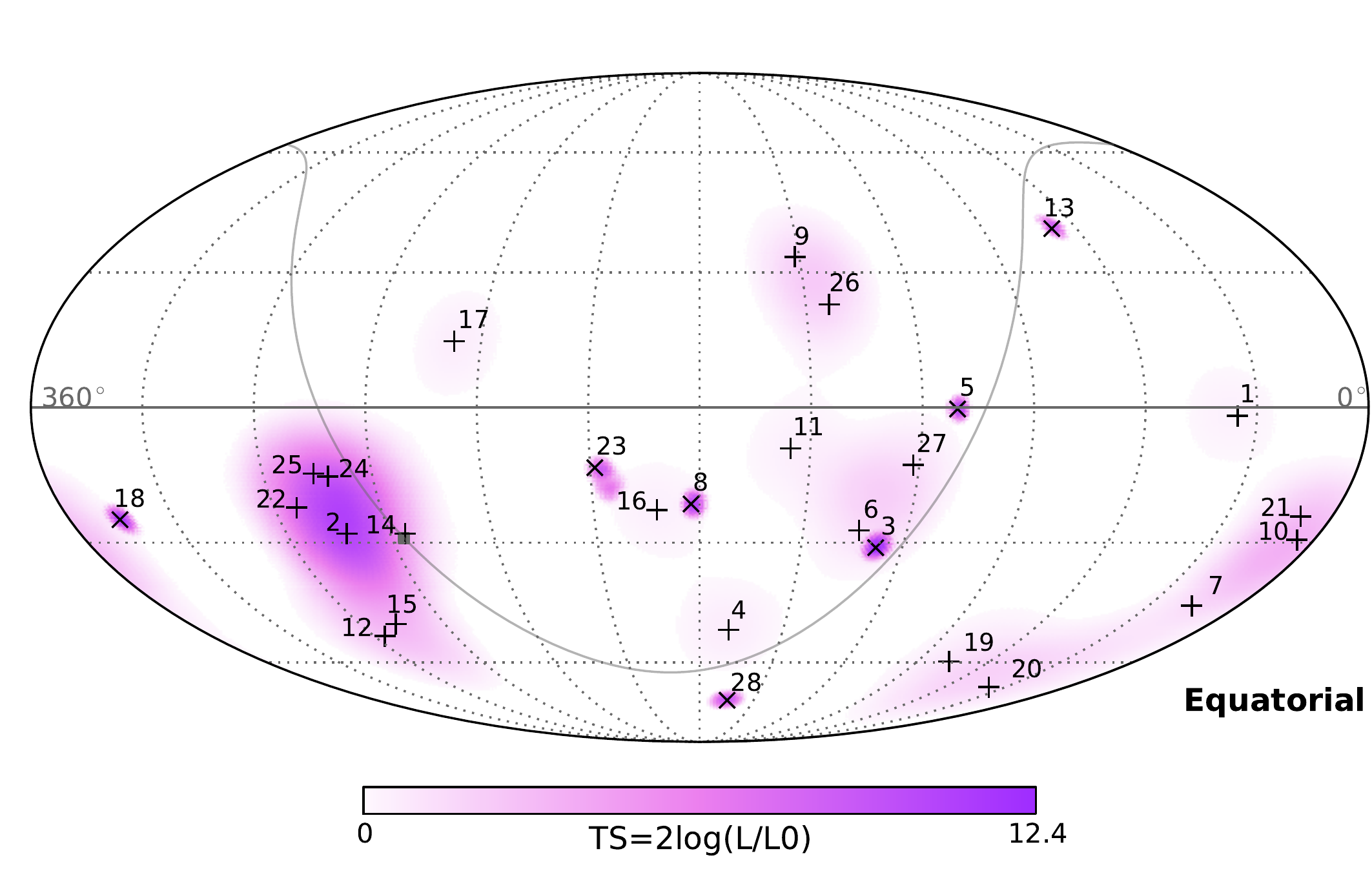}{0.98}
\caption{IceCube skymap in equatorial coordinates of the Test Statistic value
  (TS) from the maximum likelihood point-source analysis.  The most
  significant cluster consists of five events---all showers and
  including the second-highest energy event in the sample---with a
  final significance of 8\%.  This is not sufficient to identify any
  neutrino sources from the clustering study.  The galactic plane is
  shown as a gray line with the galactic center denoted as a filled
  gray square.  Best-fit locations of individual events (listed in
  Table~\ref{tab:events}) are indicated with vertical crosses ($+$)
  for showers and angled crosses ($\times$) for muon tracks.  From
  Ref.~\cite{Aartsen:2013pza}.}
\label{fig:skymap}
\end{figure}

The arrival directions of the 28 neutrinos are shown in
Fig~\ref{fig:skymap}.  The IceCube angular resolution for shower
events is poor, $15^\circ$ to $20^\circ$, so firm conclusions are
elusive at present. The largest concentration of events is near the
Galactic center~\cite{Razzaque:2013uoa}, consisting of 7 shower events
(with a p-value of 8\%~\cite{Naoko:2013ipa}). It is further tempting
to observe that one of the highest energy events, \#14 at 1 PeV,
points directly towards the Galactic Center.  It has also been noted
that the possible clustering could be associated to the Norma arm of
the Galaxy~\cite{Neronov:2013lza}. While these interpertations are
interesting, it is worth reiteration that at present statistics are
limited and we have seen many incorrect suggestions of source
association in the cosmic rays~\cite{Sigl:2000sn,Torres:2003ee,Finley:2003ur}.

Concerning the issue of possible structure in the neutrino 
spectrum (either a gap or a cutoff), let us examine the 
consistenty of a single power law over the entire energy range,
with no cutoff. We consider the hypothesis that the cosmic neutrino
flux per flavor, averaged over all three flavors,
follows an unbroken power law of the form
\begin{equation}
\Phi_\nu (E_\nu) \equiv \frac{dF_\nu}{d\Omega dA dt dE_\nu} = \Phi_0 \ \left(\frac{E_\nu}{1~{\rm GeV}} \right)^{-\Gamma}\,,
\label{eqn:flux}
\end{equation}
for a factor of several or more above the highest energies so far observed.
Then we ask ``What value(s) of the spectral index $\Gamma$ are consistent with the recent IceCube observations?''
We partition the observations into three bins:
\begin{itemize}
\item 26 events from 50~TeV to 1~PeV, which includes the $\sim 10$ atmospheric background events;
\item 2 events from 1~PeV to 2~PeV;
\item zero events above 2~PeV, say from 2~PeV to 10~PeV, with a background of zero events.
\end{itemize}
This choice of binning was selected a priori for the following reason.
The 2 highest energy events exhibit shower topologies. For cascade
events, the IceCube Collaboration has a sensitive method for
determination of the energy resolution, $\Delta (\log_{10} E_\nu) \approx
0.26$, so we place these two events in 1 bin. The low energy bin
contains both background as well as a number of events exhibiting
track topologies.  For track events, the $\nu_\mu$ energy may be 5
times higher than the deposited energy.  We therefore conservatively
group all of these lower energy events into a single bin.

\begin{figure}[tbp]
\postscript{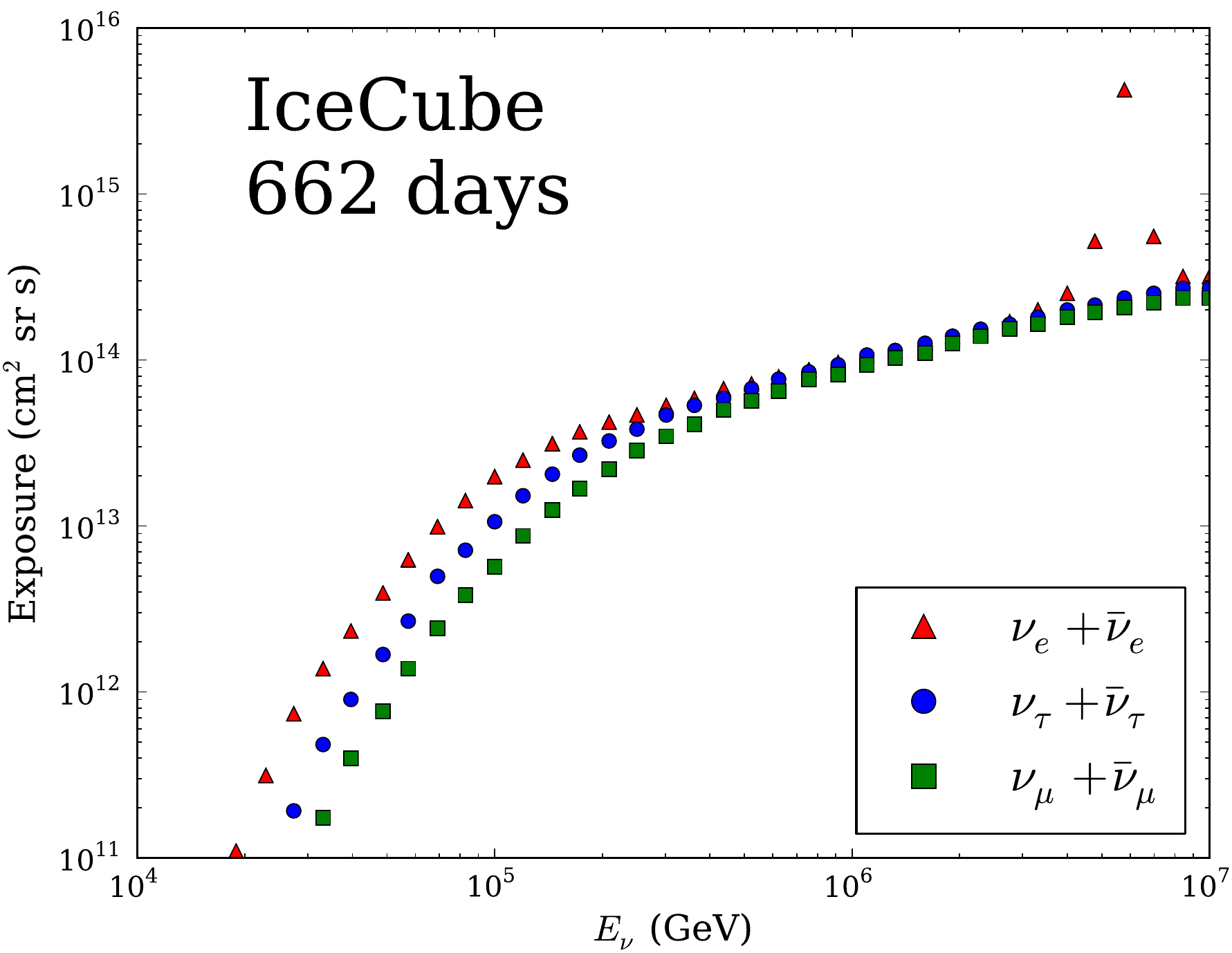}{0.9}
\caption{IceCube exposure for 662 days of data collection, for
  contained events.. The sharp-peaked structure for $\bar \nu_e$ at
  $10^{6.8}~{\rm GeV}$ is due to the Glashow resonance. One should
  note the relatively smaller ``exposure'' for muon events below100
  TeV. Taken from Ref.~\cite{Anchordoqui:2013qsi}.}
\label{fig:exposure}
\end{figure}

For various spectral indices from 2.0 to 2.8, we fit the neutrino flux to each of these three bins,
by integrating over the energy span of the bin.
A key point is that we employ IceCube's energy-dependent, flavor-dependent exposure functions
for the 662 days of observation time reported thus far.
The IceCube exposures are shown in Fig.~\ref{fig:exposure}.

\begin{table}[tb]
\caption{Normalization $\Phi_0$ for the ``low energy'' ($E<1$~PeV) and ``high energy'' (1-2~PeV) bins ,
and normalization upper limits for the ``null'' bin (2-10~PeV) at 68 \%C.L. ($\Phi^{\rm max}_{68}$)
and 90\%C.L. ($\Phi^{\rm max}_{90}$) in units of $({\rm GeV}\cdot{\rm cm}^{2}\cdot{\rm s}\cdot{\rm sr})^{-1}$,
for various spectral indices, $\Gamma$.
\label{tab:cls}}
\centering
\begin{tabular}{ |c|c|c|c|c|}
\hline \hline \\ [-2.0 ex]
$\Gamma$~&~$\Phi_0^{E_\nu < 1{\rm PeV}}$~&$\Phi_0^{1 {\rm PeV} < E_\nu < 2
  {\rm PeV}}$  &~$\Phi^{\rm max}_{68}$~&~$\Phi^{\rm max}_{90}$~\\
\hline \\ [-2.0 ex]
~~2.0~~ & $1.66 \times 10^{-8}$ & $9.50 \times 10^{-9}$ & $3.94 \times
10^{-9}$ & $7.44 \times 10^{-9}$ \\
\hline \\ [-2.0 ex]
2.1 & $5.70 \times 10^{-8}$ & $3.91 \times 10^{-8}$ & $1.84 \times
10^{-8}$ & $3.49 \times 10^{-8}$ \\
\hline \\ [-2.0 ex]
2.2 & $1.95 \times 10^{-7}$ & $1.61 \times 10^{-7}$ & $8.62 \times
10^{-8}$ & $1.63 \times 10^{-7}$ \\
\hline \\ [-2.0 ex]
2.3 & $6.63 \times 10^{-7}$ & $6.62 \times 10^{-7}$ & $4.02 \times
10^{-7}$ & $7.61 \times 10^{-7}$ \\
\hline \\ [-2.0 ex]
2.4  & $2.24 \times 10^{-6}$ & $2.72 \times 10^{-6}$  & $1.88 \times
10^{-6}$ & $3.55 \times 10^{-6}$\\
\hline \\ [-2.0 ex]
2.5 & $7.54 \times 10^{-6}$ & $1.12 \times 10^{-5}$ & $8.73 \times 10^{-6}$ & $1.65 \times$ $10^{-5}$ \\
\hline \\ [-2.0 ex]
2.6 & $2.52 \times 10^-5$ & $4.59 \times 10^{-5}$ & $4.06 \times
10^{-5}$ & $7.68 \times 10^{-5}$\\
\hline \\ [-2.0 ex]
2.7 & $8.39 \times 10^{-5}$ & $1.88 \times 10^{-4}$ & $1.88 \times 10^{-4}$ & $3.56 \times 10^{-4}$\\
\hline \\ [-2.0 ex]
2.8 & $2.78 \times 10^{-4}$ & $7.71 \times 10^{-4}$& $8.73\times 10^{-4}$  & $1.65 \times 10^{-3}$ \\
\hline
\hline
\end{tabular}
\end{table}

The results of the fit are summarized in Table~\ref{tab:cls}.  Column two (three)
shows the fitted flux normalization $\Phi_0$ for the first (second)
bin.  The null, third bin requires more explanation: According to the
statistics of small numbers~\cite{Feldman:1997qc}, any flux yielding
more than 1.29 (2.44) events in the null 2-10~PeV range of bin three,
is excluded at 68\% C.L. (90\% C.L.).  Accordingly, columns four and five
show the maximum flux normalizations allowed by the null bin three, at
the 68\% and 90\%~C.L.'s.

Under the assumption of a single power-law across the three energy
bins, consistency requires that the maximum flux normalization
determined by bin three must exceed the flux normalizations from bins
one and two.  Moreover, the fitted normalizations from bins one and
two should be the same, or nearly so.  In terms of the Table columns,
if flux numbers from columns two or three exceed the maximums of
columns 4 and 5, then the fit is ruled out at 68\%  and
90\%  C.L.
Thus, Table~\ref{tab:cls} reveals that spectral indices shallower than
2.3 are inconsistent with the data at 90\% C.L. or more, while indices
shallower than 2.7 are inconsistent at 68\% C.L..  Note also that for
$\Gamma = 2.3$, and only for $\Gamma = 2.3$, are the normalizations
from bins one ($E_\nu < 1$~PeV) and two ($1~{\rm PeV} < E_\nu < 2~{\rm
  PeV}$) quite consistent with each other, and therefore with an
unbroken power law.  The overall consistency of the $\Gamma=2.3$ power
law across all three bins is at roughly the 90\%, $1.5\sigma$ level.
We therefore choose $\Gamma = 2.3$ as our reference value for the
unbroken power law cutoff-free hypothesis. It is worth reiterating
here that consistency with a single power low does not exclude a
cutoff in the spectrum; rather, given current statistics it is {\em
  possible} to characterize the spectrum with a single power law.

To assess the effect of uncertainties on our reference spectral index, we note that a $1\sigma$ upper fluctuation
of the background is 15.2 events. The consequence
of such a fluctuation on the hypothesis of a single power law with a
high-energy cutoff would be to favor a spectral index close to $\Gamma
= 2$,  
with a normalization for the all-neutrino flux $\Phi_0^{\rm total} = 3 \,
  \Phi_0 =  2.85
\times 10^{-8}~{\rm GeV}^{-1} \, {\rm cm}^{-2} \, {\rm s}^{-1} \, {\rm sr}^{-1}$.  
Note that this normalization is about 20\% smaller the
normalization quoted in Fig.~\ref{fig:snowmass}. The reason for this
difference arises from the particular selection critera applied in the
analysis corresponding to Fig. 3 of reference~\cite{Aartsen:2013bka} compared the
selection in the analysis just discussed. In~\cite{Aartsen:2013bka} the exposure is
computed only for shower topologies, whereas in the analysis discussed
herein  both shower and tracks are included in the exposure calculation. From
a comparison of Fig. 3 in reference~\cite{Aartsen:2013bka} to 
Fig.~\ref{fig:exposure} one can see that
taking account of both CC and NC interactions in the case of muons,
increases the number muon neutrino events by about a factor 5,
compared to considering NC interactions alone.

To close this section, we consider some aspects of the connection
among CRs, neutrinos and $\gamma$ rays.  We anticipate that if the
neutrino spectrum ultimately turns out to be dominated by Galactic
sources, the lack of observed CR anisotropy will require a soft
neutrino spectrum of $\Gamma \approx 2.3$. This will be discussed
further in Sec.~\ref{section-3}.  If cosmic neutrinos are primarily of
extragalactic origin, then the 100~GeV gamma ray flux observed by
\textit{Fermi}-LAT constrains the normalization at PeV energies at injection,
which in turn demands a neutrino spectral index $\Gamma < 2.1$~\cite{Murase:2013rfa}.

\section{Galactic Models}
\label{section-3}

Above about 10~GeV, the CR energy spectrum is observed to fall roughly
as a power law; the flux decreases nearly three orders of magnitude
per energy decade until eventually suffering a strong suppression near
$10^{10.7}~{\rm
  GeV}$~\cite{Abbasi:2007sv,Abraham:2008ru,Abraham:2010mj}. Close
examination reveals several other spectral features. A steepening of
the spectrum from $J(E) \propto E^{-2.67\pm 0.07}$ to $E^{-3.07 \pm
  0.11}$ occurring at an energy $E_{\rm knee} \approx 10^{6.5}~{\rm
  GeV}$ is known in cosmic vernacular as the
``knee''~\cite{Hoerandel:2002yg,Blasi:2011fi}.  A less prominent
``second knee'', corresponding to a further softening $J (E) \propto
E^{-3.52 \pm 0.19}$ appears above $10^{8.5}~{\rm
  GeV}$~\cite{AbuZayyad:2000ay}. At $E_{\rm ankle} \approx
10^{9.5}~{\rm GeV}$ a pronounced hardening of the spectrum becomes
evident, generating the so-called ``ankle''
feature~\cite{Bird:1993yi,Abbasi:2005ni}. Given that the CR spectrum
exhibits breaks at the knee and second knee, we should ask whether it
is plausible for the proton injection spectrum to be charactized by a
single index over the energy range of interest. In this section we
first discuss the connection between the neutrino spectrum and
structures in the cosmic ray spectrum. In the process, we have to
consider consistency with source power requirements as well as other
multimessenger constraints including observations of TeV gamma rays
and bounds at higher energies. We will also discuss the neutrino
“hot-spot” near the Galactic Center, and assuming it is not a
statistical fluke, we make predictions for future observations by
IceCube and ANTARES.

\subsection{Shape of the Source Spectrum} \label{sec:b}

Relativistic charged particles produced in our Galaxy are likely to be
confined by the Galactic magnetic field, $|\bm{B}| \sim
3~\mu {\rm G}$. The Larmor radius of a charged particle in a magnetic field
is
\begin{equation}
r_{\rm L} =
\frac{E}{Z \, e \, |\bm{B}|} \simeq \frac{1.08 \times 10^{-9}} {Z}~
\frac{E_{\rm GeV}}{B_{\mu{\rm G}}} ~{\rm kpc} \,,
\end{equation}
where $E_{\rm GeV} \equiv E/{\rm GeV}$ and $B_{\mu{\rm G}} \equiv B/\mu{\rm G}$. 
The quantity $E/Ze$ is termed the ``rigidity'' of the particle.
Particles may only leak from the Galaxy if their gyroradius is
comparable to the size of the Galaxy~\cite{Cocconi}.
As a zeroth order approximation, in which we pretend the Galaxy
has a spherical homogeneous halo, this leakage energy corresponds to
\begin{equation}
E \ga  Z \, e\, B\, R_H \simeq 1.5 \times 10^{10} \, Z \, B_{\mu{\rm G}} \,
\left (\frac{R_H}{15~{\rm kpc}} \right)~{\rm GeV} \, ,
\end{equation}
where $R_H$ is the radius of the halo.  The majority of CRs possess a
considerably lower energy than this and are thus trapped in the
Galaxy. It is important to note here that the $ \bm{B}$ field
comprises approximately equal contributions from a ``regular''
component $B_{\rm reg}$ (that is, the galactic plane has field lines
that run parallel to the spiral arms) and an irregular component
$B_{\rm rand}$ generated by turbulent motions in the interstellar
medium~\cite{Stanev:1996qj,Jansson:2012rt}. The motion of trapped
cosmic rays can be reasonably approximated as a diffusive process
controlled by the turbulent components of the magnetic field.

Neglecting the effects of convection, feed-down from fragmentation of
heavier nuclei, and any energy losses, the CR transport in the Galaxy
can be described by steady-state diffusion equation, in which the
current $\bm{j}$ is related to the CR density $n_{\rm CR}$ through~\cite{Ginzburg:1964}
\begin{equation}
\bm{\nabla}  \bm{\cdot}  \bm{j} \equiv -  \nabla_i  D_{ij}  \nabla_j  n_{\rm CR}  = Q ,
\label{diffusion-eq}
\end{equation}
where  $Q$ is the generation rate of primary CRs and $D_{ij}$
is the cosmic ray diffusion tensor, with components
\begin{equation}
D_{ij} = (D_\parallel - D_\perp) b_i b_j + D_\perp \delta_{ij} + D_A \epsilon_{ijk} b_k,
\end{equation}
where $b_i = B_{{\rm reg}, i}/B_{\rm reg}$ is a unit vector along the
regular Galactic magnetic field, $\delta_{ij}$ is the Kronecker delta,
and $\epsilon_{ijk}$ is the Levi-Civita fully antisymmetric
tensor.  The symmetric terms of $D_{ij}$ contain
the diffusion coefficients parallel (field-aligned) and perpendicular
(transverse), $D_\parallel$ and $D_\perp$, which describe diffusion
due to small-scale turbulent fluctuations. The antisymmetric (Hall)
diffusion coefficient $D_A$ is responsible for macroscopic drift
currents. The anisotropy vector $\bm {\delta}$ is given by
\begin{equation}
\delta_i = \frac{3 \  j_i }{n_{\rm CR}  \, c} \, .
\end{equation}
The diffusion coefficients and their energy dependence are primarily
determined by the level of turbulence in the interstellar medium.
Under the assumption that the regular magnetic field is directed in
the azimuthal direction and that both the Galaxy and the CR sources
can be considered to have cylindrical symmetry, one finds that
$D_\parallel$ plays no role in the diffusion equation.  In most
interesting cases the turbulent spectrum of the random magnetic field
is described by a power-law, yielding $D_\perp \propto (E/Z)^\delta$
and $D_A \propto E/Z$, with $\delta = 1/3$ for a Kolmogorov
spectrum~\cite{Kolmogorov} and $\delta = 1/2$ for a Kraichnan
hydromagnetic spectrum~\cite{Kraichnan}.

There are two broad categories of explanatory models for the knee: one
ascribes the structure {\em mainly} to properties of the source(s)
such as different acceleration mechanisms or variations in
acceleration efficiency with
energy~\cite{Fichtel:1986kn,Biermann:1993,Biermann:1995qy,Erlykin:1997bs,Kobayakawa:2000nq,Hillas:2005cs,Erlykin:2005xn,Hillas:2006ye};
the second broad category attributes the break at the knee to the
details of the Galactic magnetic field and the resulting rigidity
dependent leakage from the
Galaxy~\cite{Syrovatskii:1971,Ptuskin:1993,Candia:2002we,Candia:2003dk}.
In order to determine which type of model is more likely to be viable, it
is crucial to employ both data on the energy spectrum and composition
as well as the anisotropy around and above the knee.

Many of the salient features of the two models can be visualized in terms of a
``leaky box,'' in which CRs propagate freely in the Galaxy, contained by the
magnetic field but with some probability to escape which is constant in time~\cite{Gaisser:1990vg}.
For a homogeneous volume $V$ (enclosed by the surface $S$), after averaging over that
volume, gives for Eq.~(\ref{diffusion-eq})
\begin{eqnarray}
\overline{-  \nabla_i  D_{ij}  \nabla_j  n_{\rm CR} } & = & - \int_V
d^3x \, \nabla_i  D_{ij}  \nabla_j  n_{\rm CR} / V \nonumber \\
& = & \int_S d^2x \ \hat n_i  D_{ij}  \nabla_j
  n_{\rm CR}/ V \nonumber \\
& = & \int_S d^2 x \  v_{\rm esc} \,  n_{\rm CR} /V \,,
\end{eqnarray}
where $\bm{\hat n}$ is the unit vector orthogonal to $S$. Assuming
that the position where the particles escape through $S$ is independent of
$\bm{x}$, which is the leaky-box ansatz, then 
\begin{eqnarray}
\overline{-  \nabla_i  D_{ij}  \nabla_j  n_{\rm CR} } & \sim & \frac{v_{\rm
  esc} \ n_{\rm CR} }{x_{\rm esc}} = \frac{n_{\rm CR}}{\tau_{\rm
  esc}} \,,
\end{eqnarray}
where $x_{\rm esc}$ is the characteristic escape distance, $\tau_{\rm
  esc}$ is the characteristic escape time, and $V \sim S x_{\rm esc}$. 
 Averaging over $V$ on the right-hand side of
Eq.~(\ref{diffusion-eq}) gives $\overline Q$ .  Thus, under the 
homogeneity assumption we have derived the leaky-box equation (without
energy loss/gain or decay terms)
\begin{equation}
n_{\rm CR} (E) \equiv \frac{4\pi}{c} J (E) \, \approx \overline{Q(E)}  \
\tau_{\rm esc} (E/Z) \, .
\label{leaky}
\end{equation}
There is a subtle point here with regard to boundary conditions. In
the diffusion model, the boundary condition is that $n_{\rm CR}$ drops
to zero everywhere on the border $S$ (Dirichlet boundary condition).
In the leaky-box ansatz a strong reflection is assumed at the border
for particles below some energy threshold, {\it i.e.}, the gradient of
$n_{\rm CR}$ drops to zero at any point on $S$ (Neumann boundary
condition). For cosmic ray particles that do escape, we have $|\bm{j}| >
j_{\rm esc} = v_{\rm esc} \, n_{\rm CR}$. These different boundary
conditions are associated to different Sturm-Liouville operators. In
this approach the leaky-box limit can be considered as a 
mathematical limit (however physical or unphysical that limit may be) employed to solve the differential equation (\ref{diffusion-eq}) at any point inside $S$.

The basic model for the investigation of cosmic ray propagation in the
Galaxy is the flat halo diffusion model~\cite{Ginzburg:1976dj}. The
model has a simple geometry which reflects, however, the most
essential features of the real system. It is assumed that the system
has the shape of a cylinder with a radius $R_H \sim 15~{\rm kpc}$ and
half height $H = 4~{\rm kpc}$. The cosmic ray sources are distributed
within the inner disk having characteristic thickness $h \sim 1~{\rm
  kpc}$ and radius $R_G \sim 10~{\rm kpc}$. For $E < Z E_{\rm knee}$,
the escape time from the Galaxy as a function of energy is solidly
parametrized by~\cite{Parizot:2004wh,DeMarco:2007eh}
\begin{equation}
\tau_{\rm esc} (E/Z) \simeq \frac{H^2}{6 D_\perp(E)} \propto (E/Z)^{-\delta} \, .
\end{equation}

\begin{figure}[tbp]
\postscript{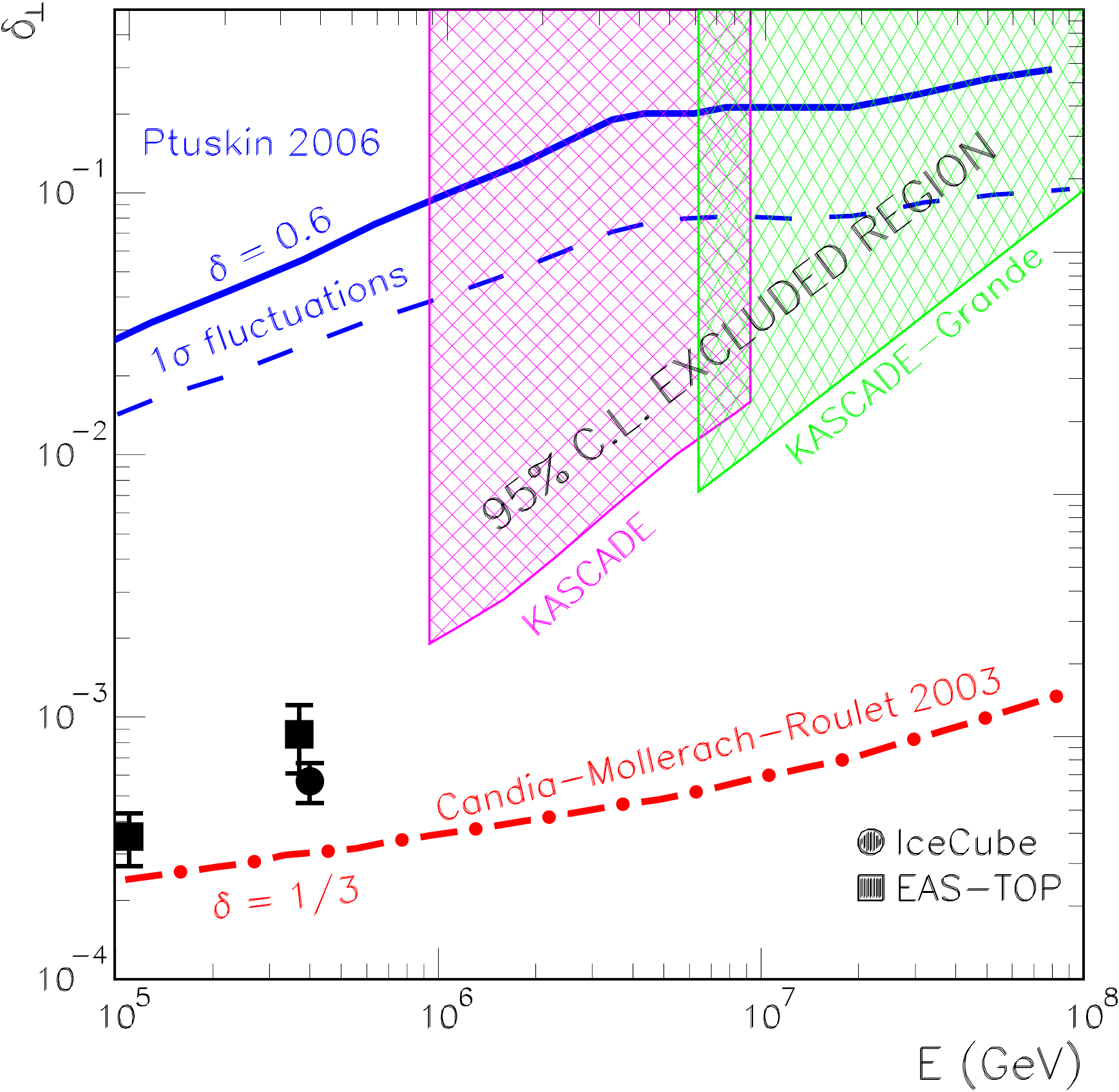}{0.8}
\caption{Comparison of the effect of different injection spectral
  indices on the anisotropy amplitude in the direction of the
  equatorial plane, $\delta_\perp$. The solid line indicates the
  average value for $\delta = 0.6$~\cite{Ptuskin:2006xz}, whereas the
  dot-dashed line corresponds to the Kolmogorov value $\delta =
  1/3$~\cite{Candia:2003dk} .  In order to compare with existing
  observations, the anisotropy amplitude has been projected into the
  equatorial direction~\cite{Abreu:2011ve}, as described
  in~\ref{AI}. Measurements of the first harmonic amplitude (corrected
  by the mean value of the cosine of the event declinations) by
  EAS-TOP~\cite{Aglietta:2009mu} and IceCube~\cite{Abbasi:2011zka} are
  shown for comparison. The shaded regions are excluded by null
  results of searches by KASCADE~\cite{Antoni:2003jm} and
  KASCADE-Grande~\cite{Stumpert:2008zz} collaborations. The dashed
  line indicates the $1\sigma$ downward fluctuation for $\delta =
  0.6$.}
 \label{fig:anisotropy}
\end{figure}

Hereafter, we consider the particular case of a source term whose energy
dependence is taken as a power-law, $\overline{Q(E)} \propto
E^{-\alpha}$. An effective way to determine the rigidity
behavior is to uncover the spectrum of secondary nuclei.  Fits to the
energy dependence of secondary to primary ratios yield
$\delta=0.6$~\cite{Gupta:1989,Engelmann:1990,Swordy:1993dz}. For a
source index $\alpha \simeq 2.07$, which is close to the prediction of
Fermi shock acceleration~\cite{Fermi:1949ee}, inclusion of propagation effects reproduces
the observed spectrum. However, as shown in Fig.~\ref{fig:anisotropy},
$\delta = 0.6$ results in an excesively large anisotropy which is
inconsistent with observations. Consistency with anisotropy can be
achieved by adopting a Kolmogorov index, $\delta =
1/3$~\cite{Biermann:1995qy,Candia:2003dk}. The apparent conflict with
the secondary to primary composition analyses can be alleviated
through small variations of the energy dependnence of the spallation
cross sections, or variation in the matter distribution in the
Galaxy~\cite{Biermann:1995qy}.  This hypothesis implies a steeper
source spectrum, $\alpha \simeq 2.34$, which agrees remarkably well
with the fit of an unbroken power law to IceCube data, as discussed
herein.

For rigidity $E/Z > 3~{\rm GeV}$, the secondary to primary ratios and the
abundances of unstable isotopes can be best fit by choosing
\begin{equation}
D_\perp (E) = 10^{28} \ D_{28} \left(\frac{E_{\rm GeV}/Z}{3~{\rm GeV}} 
\right)^\delta 
\end{equation}
with $D_{28}/H_{\rm kpc} = 1.33$ for $\delta = 1/3$ and $D_{28}/H_{\rm kpc}
= 0.55$ for $\delta = 0.6$~\cite{Blasi:2011fm}. For Kolmogorv difussion, this leads to  
\begin{equation}
\tau_{\rm esc} \approx 2 \times 10^7 \, \left(\frac{E_{\rm GeV}}{Z} \right)^{1/3} ~{\rm yr} \,,
\label{tau-escape}
\end{equation}
which is in good agreement with the CR confinement time derived from
the observed abundance of the radioactive
$^{10}$Be~\cite{Garcia-Munoz}.

A model accomodating the single power law hypothesis can be concocted 
by assuming cosmic ray leakage is
dominated by Kolmogorov diffusion, $\tau \propto (E/Z)^{-1/3}$, for $E
< Z E_{\rm knee}$, with increasing leakege due to decreasing trapping
efficiency with rising energy, $\tau \propto (E/Z)^{-1}$ for $E \gg Z
E_{\rm knee}$~\cite{Candia:2003dk}. The knee is etched into the spectrum by a transiton
from diffusion to drift motion, while the second knee results from a subsequent
transiton to quasirectilinear motion.  Each CR nucleus is
affected by drifts at $E \simeq Z E_{\rm knee}$, resulting in a
progressive steepening of the CR spectrum. Since the lighter
components 
are strongly suppressed above $10^8~{\rm GeV}$, we are left with an iron
dominated spectrum which progressively steepens until the overall
spectrum becomes $J(E) \propto E^{-2.67 + \Delta\delta}$, 
where $\Delta\delta = -1 -(-1/3)=-2/3$,
in agreement with observation of the second knee~\cite{AbuZayyad:2000ay}.
Such a power law index is also in agreement with upper limits on flux anisotropies. 

In closing, we note that detailed Monte Carlo
simulations~\cite{Blasi:2011fm} demonstrate that the leaky box
approximation (in which anisotropy depends in a simple way on energy)
is a dramatic oversimplification that only represents an average over
a large ensemble of possible source distributions in space and time.
Significantly, while some realizations for $\delta = 1/3$ can
accommodate all existing observations, a value $\delta =0.6$ yields an
anisotropy which is systematically larger than what is observed at any
energy.  From these arguments, we posit that a proton injection
spectrum of $\alpha = 2.34$ is favored, and consequently a similar
injection index for neutrinos.

\subsection{Consistency with Upper Limits on the Diffuse $\gamma$ Ray
    Flux} \label{sec:photons}

  Since neutrinos are produced by $\pi^{\pm}$ decays at the same unshielded
  sources where $\gamma$ rays are produced by $\pi^0$ decays, and since
  neither is deflected or attenuated (assuming the photons are of
  Galactic origin), coordinated observations of these cosmic
  messengers will allow a new way of exploring the highest-energy
  Galactic sources~\cite{Anchordoqui:2013qsi,Gupta:2013xfa}.  At
  present we are dealing only with bounds on PeV photons, which
  constrain the source injection of photons, neutrinos, and charged
  particles.

  In this section, we address the impact of experimental bounds on the
  diffuse $\bm{\gamma}$ ray flux, for the hypothesis of a predominantly
  Galactic origin of the IceCube neutrino excess.  From experimental
  bounds on the photon fraction we can acquire limits on the integral
  $\bm{\gamma}$ ray flux in regimes spanning one or more decades in
  energy~\cite{Matthews:1990zp,Chantell:1997gs,Schatz:2003aw,Aartsen:2012gka}. It
  is worth keeping in mind that this very lax binning means that we
  cannot strictly place a bound on the differential photon flux
  without making some assumption about the photon energy spectrum,
  though future improvements in experimental methods could nudge us
  closer to a bound on (or observation of) a differential flux.  In
  this section, we therefore consider the relationship between the
  neutrino flux and the integral photon flux, compare this with
  experimental bounds, and draw some conclusions regarding how
  strongly the photon observations constrain the neutrino sources. 

Inelastic $pp$ collisions lead to roughly equal numbers of $\pi^0$'s,
$\pi^+$'s, and $\pi^-$'s, hence one expects two photons, two $\nu_e$'s,
and four $\nu_\mu$'s per $\pi^0$.  On average, the photons carry one-half of
the energy of the pion.
The average  neutrino energy from the direct pion decay  is
$\langle E_{\nu_\mu} \rangle^\pi = (1-r)\,E_\pi/2 \simeq 0.22\,E_\pi$
and that of the muon is $\langle E_{\mu} \rangle^\pi = (1+r)\,E_\pi/2 \simeq
0.78\,E_\pi$, where $r$ is the ratio
of muon to the pion mass squared. Now, taking the $\nu_\mu$ from muon decay
to have 1/3 the energy of the muon, the average energy of the $\nu_\mu$ from
muon decay is $\langle E_{\nu_\mu} \rangle^\mu =(1+r)E_\pi/6=0.26 \, E_\pi$.
This gives a total $\nu_\mu$ energy per charged pion
 $ \langle E_{\nu_\mu} \rangle \simeq 0.48 \, E_\pi$,
with a total $\langle E_{\nu_\mu} \rangle^{\rm total} = 0.96 \langle
E_\gamma \rangle$ for each triplet of
$\pi^+,$ $\pi^-,$ and  $\pi^0$ produced. For simplicity, we hereafter
consider that all neutrinos carry one-quarter of the energy of the pion.

The total number of $\gamma$ rays in the energy interval $(E_1/2,\, E_2/2)$ is equal
to the total number of charged pions in the interval $(E_1,\, E_2)$ and twice the number
of neutral pions in the same energy interval,
\begin{equation}
\int_{E_1/2}^{E_2/2} \frac{dN_\gamma}{dE_{\gamma}} dE_\gamma = 2
\int_{E_1}^{E_2} \frac{dN_{\pi^0}}{dE_{\pi}} dE_\pi =
2 N_{\pi^0}\,\, .
\label{po}
\end{equation}
Additionally, since $N_{\pi^\pm} = 2 \,N_{\pi^0}$, the number of $\nu_\mu$ in
the energy interval  $(E_1/4,\, E_2/4)$ scales as
\begin{equation}
\int_{E_1/4}^{E_2/4} \frac{dN_{\nu_\mu}}{dE_{\nu}} dE_\nu = 2
\int_{E_1}^{E_2} \frac{dN_{\pi^\pm}}{dE_{\pi}} dE_\pi =
2 N_{\pi^\pm}\,.
\label{pu}
\end{equation}
Now, taking $d/dE_2$ on each side of Eqs.~(\ref{po}) and (\ref{pu}) leads to
\begin{equation}
\left. \frac{1}{2} \frac{dN_\gamma}{dE_\gamma} \right|_{E_\gamma = E_2/2} = \left. 2\,
\frac{dN_\pi^0}{dE_\pi} \right|_{E_2} \,\,\,{\rm and} \,\,\,
\left. \frac{1}{4} \frac{dN_{\nu_\mu}}{dE_{\nu}} \right|_{E_\nu = E_2/4}
= 2 \left. \frac{dN_\pi^\pm}{dE_\pi} \right|_{E_2} \,,
\end{equation}
respectively. The
energy-bins $dE$ scale with these fractions, and we arrive at
\begin{eqnarray}
\left. \frac{dN_\gamma }{dt d\Epho} \right|_{\Epho=\Epi/2}  & = &
  \left.  4\,\frac{dN_{\pi}}{dt d\Epi} \right|_{\Epi}\,, \nonumber \\
\left. \frac{dN_{\nu_e}}{dt d E_\nu} \right|_{E_\nu = \Epi/4}  & = &
  \left.  8\,\frac{dN_{\pi}}{dt d\Epi} \right|_{\Epi}\,, \label{pumitavalles}\\
\left. \frac{dN_{\nu_\mu}}{dt dE_\nu} \right|_{E_\nu= \Epi/4} & = &
\left.    16\,\frac{dN_{\pi}}{dt d\Epi}\right|_{\Epi}\,, \nonumber
\end{eqnarray}
for the total fluxes at the source, where $\pi$ denotes any one of the
three pion charge-states and $dt$ is the time differential. 

During propagation TeV-PeV $\gamma$ rays are absorbed in radiation
backgrounds, with interaction length $\lambda_{\gamma \gamma}
(E_\gamma)$. From Eq.~(\ref{pumitavalles}), we see that neutrinos are produced at
sources with a flavor ratio of 1:2:0.  After propagating over large
distances, however, oscillations convert this ratio to approximately
1:1:1  (for details, see~\ref{AII}). From these observations, one
finds a nearly identical flux for each of the three neutrino flavors,
which relates to the $\gamma$ ray counterpart
according to~\cite{Anchordoqui:2004eu}
\begin{equation}
e^{-\frac{r}{\lambda_{\gamma \gamma}}} \, \left. \frac{dF_{\nu_\alpha}}{dAdtdE_\nu} \right|_{E_\nu= \Epho/2} =
\left. 2\frac{d\Fpho}{dA dtd\Epho}\right|_{\Epho}\, \,  .
\label{fnu}
\end{equation}

Now that we have a relation between the neutrino and photon differential fluxes,
we can check whether existing experimental bounds on the CR photon fraction at
various energies leave room for the possibility that the IceCube excess is
generated in optically thin sources in the Galaxy. To do this, we consider
Eq.~(\ref{fnu}) for a distribution of sources, and write the integral flux of
photons above some minimum energy $E_\gamma^{\rm min}$ in terms of the neutrino
flux
\begin{equation}
\int_{E_\gamma^{\rm min}} \frac{dF_\gamma}{d\Omega dA dt dE_\gamma}
dE_\gamma = \frac{1}{2} \,  \int_{E_\gamma^{\rm min}/2}  \ \sum_i e^{-\frac{r_i}{\lambda_{\gamma
      \gamma}}} \  \frac{dF_{\nu_\alpha}}{d\Omega dA dt dE_\nu} dE_\nu \,,
\end{equation}
where
\begin{equation}
\frac{dF_{\nu_\alpha}}{d\Omega dA dt dE_\nu} \simeq 6.62 \times 10^{-7} \left(\frac{E_\nu}{\rm GeV}\right)^{-2.3}~({\rm GeV}\cdot{\rm cm}^{2}\cdot{\rm s}\cdot{\rm sr})^{-1}
\end{equation}
is the cosmic neutrino flux per flavor, averaged over all three
flavors, according to the best fit of an unbroken power law to IceCube data~\cite{Anchordoqui:2013qsi}. 
The CASA-MIA 90\% C.L. upper limits on the integral
$\gamma$ ray flux, $I_\gamma$, for
\begin{equation}
\frac{E_\gamma^{\rm min}}{{\rm GeV}} = 3.30 \times 10^5,\ 7.75 \times 10^5,\ 2.450 \times 10^6 \,,
\label{tres}
\end{equation}
are
\begin{equation}
\frac{I_\gamma}{{\rm cm}^{-2} \ {\rm s}^{-1} \ {\rm  sr}^{-1}} < 1.0 \times 10^{-13}, \, 2.6 \times 10^{-14}, \ 2.1 \times 10^{-15}\,,
\label{cuatro}
\end{equation}
respectively~\cite{Chantell:1997gs}. Under the assumption that there is no photon absorption, the integral photon flux (in units of photons ${\rm cm}^{-2} \ {\rm s}^{-1} \ {\rm  sr}^{-1}$) above the energies specified in (\ref{tres}) are
\begin{equation}
\int_{E_\gamma^{\rm min}} \frac{dF_\gamma}{d\Omega dA dt dE_\gamma} dE_\gamma = 4.2 \times 10^{-14}, \ 1.4 \times 10^{-14}, \ 3.1 \times 10^{-15} \, .
\label{cinco}
\end{equation}
We can see that for the first two energies the predicted fluxes are
comfortably below the 90\% C.L. upper limits. For the highest energy
the predicted integral photon flux slightly exceeds the 90\%
C.L. bound. However, {\it this does not necessarily signify the
  Galactic origin for the IceCube flux is excluded at the 90\% C.L.}
First of all, in this energy regime photon absorption starts to play
an important role, as the mean free path of PeV photons in the CMB is about 10~kpc. Secondly we do not know
the maximum neutrino energy achieved, and hence the maximum
photon energy is not certain.

To more strictly comply with the CASA-MIA bound we would need about a
30\% flux reduction. To see whether this is plausible we {\it (i)}
conduct a calculation of the flux at the edge of the Galactic disk
with an estimate of absorption effects and {\it (ii)} quantify the
importance of assumption about the maximum photon energy.

Consider the case where the observer $O$ is at the edge of the
Galactic disk of radius $R_G$. Denote the vector from $O$ to the center $C$ of the
galaxy by $\bm{R}_G$, from $C$ to the source $S_i$ by $\bm{r_i}^{\prime}$,
and from $O$ to $S_i$, by $\bm{
  r_i}$.  Then \be \bm{
  r_i} = \bm{ R}_G +
\bm{ r_i}^{\prime}  \, . \ee The energy-weighted neutrino flux from the Galactic
source distribution with normal
incidence at $O$ is \bea
E_\nu \frac{dF_{\nu_\alpha}}{dA dt dE_\nu} & = &  \frac{1}{4\pi}\
\sum_i \frac{P_i}{ {r_i}^2} \nonumber \\
&=& \frac{1}{4\pi}\ \sum_i\frac{P_i}{R_G^2 + 2 R_G\ r_i' \cos\theta_i '
  +r_i'^2} \,, \eea where $P_i$ is power output of source $i$ and
$\theta_i '$ is the angle subtended by $\bm{
  r_i}^{\prime}$ and $\bm{R}_G$. Assuming
equal power for all sources, and thus equal power density per unit
area of the disk, we convert the sum to an integral
\bea
E_\nu \frac{dF_{\nu_\alpha}}{dA dt dE_\nu} &=&\frac{1}{4\pi}
\frac{P}{\pi R_G^2}\ \int_0^{r'_{\rm max}}  r' dr' \int_0^{2\pi}
d\theta' \ \frac{1}{R_G^2 + 2r' R_G\cos\theta' + r'^2} \nonumber \\
&=&\frac{1}{4\pi} \frac{P}{\pi R_G^2}\ \frac{1}{2} \int_0^{ {r'_{\rm
      max}}^2}  dr'^2  \frac{2\pi}{R_G^2-r'^2} \nonumber \\
&=&\frac{P}{4\pi R_G^2}\ln \frac{1} {1 - (r'_{\rm max}/R_G)^2}.
\label{nueve}
\eea The
divergence is avoided by cutting off the integral for sources closer
than within a ring of radius $h$, the thickness of the disk at the position of the observer,
so that $r'_{\rm max} = R_G-h$. Note that this cut is unlikely to remove sources,
which are expected to be clustered close to the Galactic center.
After this regularization, the
energy-weighted neutrino flux at Earth becomes \be E_\nu
\frac{dF_{\nu_\alpha}}{dA dt dE_\nu} = \frac{P}{4\pi R_G^2} \ln\ \frac{1}{\tau
  (2 - \tau)} \,, 
\label{ventiseis}
\ee where $\tau \equiv h/R_G.$ For $h/R_G = 0.1,$ we get
\begin{equation}
E_\nu \frac{dF_{\nu_\alpha}}{dA dt dE_\nu} = 1.66 \frac{P}{4\pi R_G^2} \, .
\end{equation}

We now estimate the photon flux for $E_\gamma^{\rm min} > 1$~PeV.
The $\gamma$ absorption mean free path on the CMB is about 10~kpc,
roughly the same as the distance of Earth from our Galactic center.
Consequently, a simple assumption is justified, that photons traveling distances larger than $R_G$ are
completely absorbed.  This absorption suppresses contributions from sources at distances greater than 10~kpc,
to $0.2 \, P/(4 \pi R_G^2)$ (a result that was computed by changing the integration
range $(0, 2\pi) \to (-\pi/2, \pi/2)$ in the angular integral of (\ref{nueve}) ). For a
Galactic disk thickness of 1~kpc, we find a 12\% reduction in the photon
flux.

Now we explore the effect of varying the maximum energy cutoff. If we
set the upper limit of integration in (\ref{cinco}) to
\begin{equation}
\frac{E_\gamma^{\rm max}}{\rm PeV} = 6, \ 7, \ 8  \,,
\end{equation}
we obtain
\begin{equation}
\int_{E_\gamma^{\rm min} }^{E_\gamma^{\rm max}} \frac{dF_\gamma}{d\Omega dA dt dE_\gamma} dE_\gamma = 2.1 \times 10^{-15},\ 2.3 \times 10^{-15}, \ 2.4 \times 10^{-15} \, .
\end{equation}
{}From these results we can see there are several possible ways to
comply with the CASA-MIA upper bound at the highest energy bin of (\ref{cuatro}). 
For example, even without absorption, $E_\gamma^{\rm max} = 6~{\rm PeV}$ is consistent with the
CASA-MIA bound; if the cutoff is at $E_\gamma^{\rm max} = 8~{\rm PeV}$,
$\gamma$ ray absorption on the CMB provides enough additional suppression of the
photon flux to be consistent with data.

\begin{figure}[tbp]
\postscript{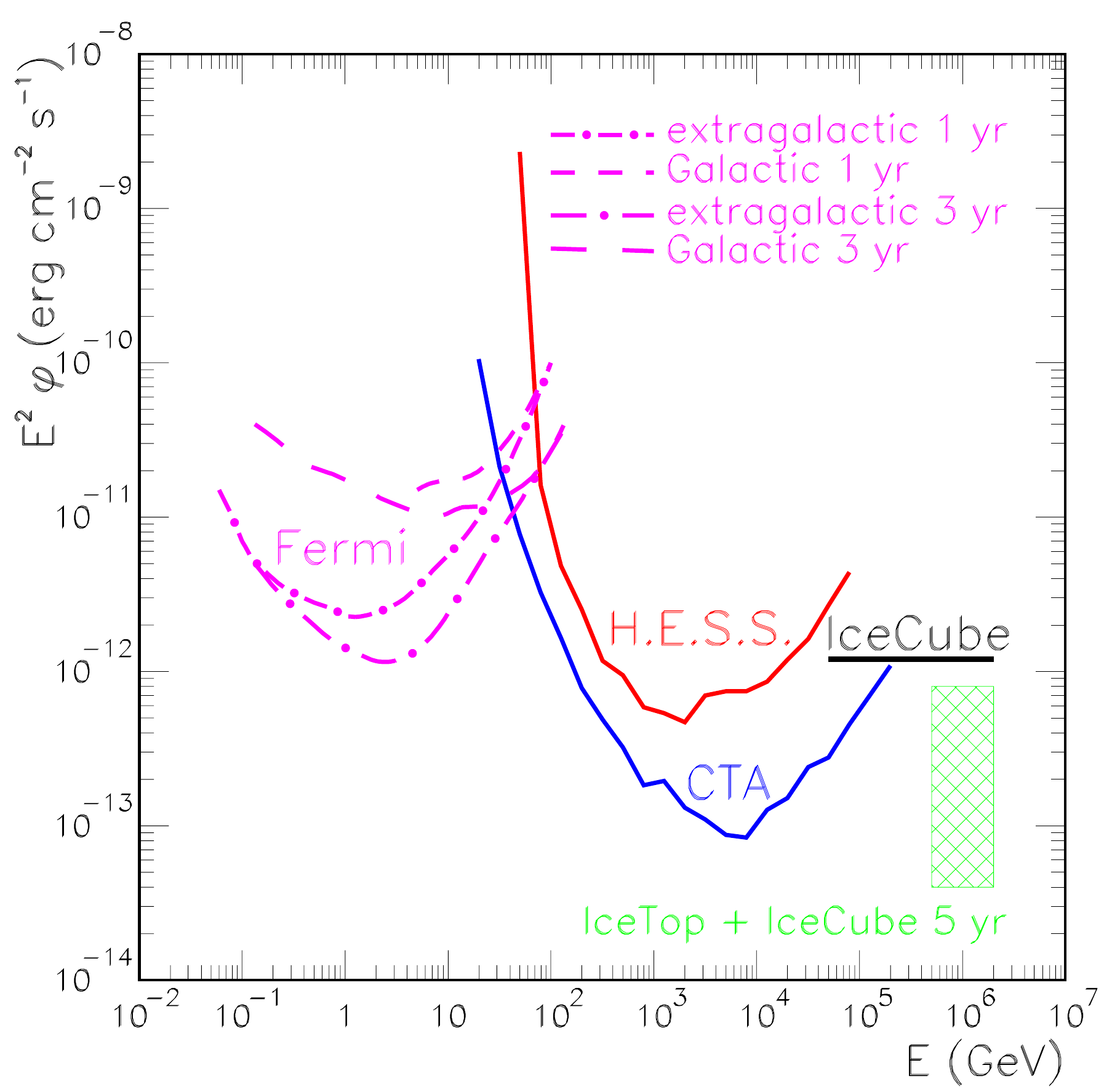}{0.8}
\caption{The solid horizontal line shows the directional (eight degree
  solid angle) cosmic neutrino flux (all flavors) observed by IceCube
  (for details, see Table~\ref{tab:cls-GC1}).  We also show $\gamma$ ray point source
  sensitivities of CTA (for 50 hours observation time in configuration
  I)~\cite{Bernlohr:2012we}, H.E.S.S. (also for 50 hours of
  observation time), Fermi for 1~yr~\cite{fermi1yr} and 3
  yr~\cite{fermi3yr}, and IceTop + IceCube for 5 yr (declinations
  $-67^\circ \lesssim \delta \lesssim -
  57^\circ$)~\cite{Aartsen:2012gka}. (The H.E.S.S. sensitivity is
  estimated by considering a sub-array of CTA~\cite{Bernlohr:2012we}
  which has the same configuration as the HESS array, namely 4
  telescopes of 12~m diameter at a separation of 120~m.)}
 \label{fig:jheap}
\end{figure}

In summary, current data still allow sufficient plausible wiggle room
for consistency with a Galactic origin of the IceCube flux even if the
sources are optically thin.  In fact, the highest fluctuation in the
IceCube $\gamma$ ray map~\cite{Aartsen:2012gka} is in the direction of
one of the PeV neutrino events~\cite{Ahlers:2013xia}.  It is also
worth noting that sources which are optically thin up to $E_\gamma
\sim 100~{\rm TeV}$, may not be optically thin above $E_\gamma \sim
100~{\rm TeV}$, suggesting that the importance of photon bounds in
establishing the origin of IceCube events should be considered with
some caution.  In Fig.~\ref{fig:jheap} we compare the IceCube
sensitivity to neutrino point sources with the $\gamma$ ray sensitivity
of several current and pending experiments. Note that if the source
emissivity is rougnly the same for photons and neutrinos, then the Cherenkov Telescope Array (CTA)
will be able to observe the associated $\gamma$ ray flux at IceCube energies

\subsection{Waxman-Bahcall Energetics}

Next we turn to the question of what the Galactic power-law model
developed above would imply regarding the average efficiency of
transferring proton energy to charged pions.  Assume that the source  
spectral index of  CRs in the range 0.1 - 100~PeV is $\Gamma$ from here on. In the spirit of~\cite{Ahlers:2005sn}, we define
the two constants
\begin{equation}
C_{\rm CR}^p(\Gamma) \equiv \frac{dF_{\rm CR}^p}{dE\,dA\,dt}\,E^\Gamma\,,
\ \
{\rm and\ \ }
C_{\nu}(\Gamma)  \equiv  \frac{dF_{\nu}}{dE\,dA\,dt}\,E^\Gamma\,,
\end{equation}
where $C_\nu = 4\pi \Phi_0^{\rm total} \, {\rm GeV}^\Gamma$ and
$\Phi_0^{\rm total} = 3 \Phi_0$, given our assumption of flavor
equilibration.  In conventional notation, we next define $\epspi$ to
be the ratio of CR power (energy/time) emitted in charged pions to
that in the parent nucleons.  We also need $\epsilon_\nu$, defined as
the fractional energy in neutrinos per single charged pion decay.  If
the pion decay chain is complete
($\pi^\pm\rarr e \, \nue \, \numu \, \numubar$), then $\epsilon_\nu \simeq 3/4$,
whereas if the pion decay chain is terminated in the source region by energy loss
of the relatively long-lived muon, then $\epsilon_\nu \simeq 1/4$.
Comparing the energy produced in charged pions at the source to the
neutrino energy detected at Earth, one gets the energy conservation
relation
\begin{equation}
\epsilon_\nu\,\epspi \int_{E_1}^{E_2} \frac{dF_{\rm CR}^p}{dE\,dA\,dt}\;E dE =
	\int_{E_{\nu 1}}^{E_{\nu
            2}}\,\frac{dF_{\nu}}{dE_\nu\,dA\,dt}\;E_\nu dE_\nu\,, \nonumber
\end{equation}
where $E_{\nu 1}=\frac{{E_1}}{16}$, and $E_{\nu 2}=\frac{{E_2}}{16}$;
these integrals may be done analytically to yield (for $\Gamma\neq 2$)
\begin{equation}
\epsilon_\nu \; \epspi \; C_{\rm CR}^p   \; \frac{E_1^{2-\Gamma} -
  E_2^{2-\Gamma}}{\Gamma-2} =  \frac{\left(\frac{E_1\,}{16}
  \right)^{2-\Gamma} -
  \left(\frac{E_2\,}{16}\right)^{2-\Gamma}}{\Gamma-2} \; C_\nu \, .
\end{equation}
Then, solving for $\epspi$ we arrive at
\beq{epspi}
\epspi = \left(\frac{1}{16} \right)^{2-\Gamma} \,
\frac{C_\nu(\Gamma)}{\epsilon_\nu\,C_{\rm CR}^p(\Gamma)} \,.
\eeq
The numerology for $C_\nu$ is given in Table~\ref{tab:cls}.
For the favored spectral index $\Gamma=2.3$, we have
\begin{equation}
C_\nu (2.3) = 12 \pi\times 6.6\times 10^{-7} \,{\rm GeV}^{2.3}\,({\rm GeV\, s\, cm}^2)^{-1}\,.
\label{Cnu2.3}
\end{equation}
The constant $C_{\rm CR}^p(2.3)$ is related to the
injection power  of CR protons, $d\eps_{\rm CR}^p/dt$, as follows:
\bea
\frac{d\eps_{\rm CR}^p}{dt}[E_1, E_2] &=& A \int_{E_1}^{E_2} \frac{dF_{\rm CR}^p}{dE\,dA\,dt}\;E\,dE \nn\\
 &=& A\int_{E_1}^{E_2} \left( \frac{dF_{\rm CR}^p}{dE\,dA\,dt}\;E^{\Gamma}\right) E^{(1-\Gamma)} dE  \nn\\
 &=& A\,C_{\rm CR}^p \,\frac{ \left( {E_1}^{(2-\Gamma)}-{E_2}^{(2-\Gamma)}\right)}{\Gamma-2}\,,
\label{Ccr2eps}
\eea
where $A = 4 \pi r^2$ is an appropriately weighted surface area for
the arriving cosmic ray or neutrino flux.  In~\cite{Gaisser:1994yf},
$A$ is set equal to $4 \pi R_G^2\equiv A_0$.  However, keeping in
mind that $\langle r^{-2} \rangle$ diverges as $\ln(R/2r_{\min})$,
with $r_{\min}$ being the distance to the nearest source, $A^{-1}$ can
easily be a factor of 2 larger than $A_0^{-1}$.  Two independent
arguments support such an enhancement.  The first is to simply note
that a local void radius of 0.7~kpc gives $A_0/A = 2$.  The second is
to note that the thin-disk approximation breaks down at a small
distance $z$ of order of the disk height, leading to a similar
estimate of integration cutoff and resulting enhancement
factor; the second argument was previously discussed in the
text preceeding Eq.~(\ref{ventiseis}).

 Inverting (\ref{Ccr2eps}) and using the fact that
${E_2}^{(2-\Gamma)} \ll {E_1}^{(2-\Gamma)}$, we get the conversion
\beq{eps2Ccr}
C_{\rm CR}^p =  \frac{d\eps_{\rm CR}^p}{dt} [E_1,E_2] \ \frac{(\Gamma-2)\,E_1^{(\Gamma-2)}}{A}\  .
\eeq
How, and how well, is $d\eps_{\rm CR}^p/dt$ known?  The assumption
underlying the leaky box model is that  the energy density
in CRs observed locally is typical of other regions of the
Galactic disk. If so, the total power required to maintain the cosmic
radiation in equilibrium can be obtained by integrating the generation
rate of primary CRs over energy and space. Using (\ref{leaky}), we obtain
\begin{equation}
\frac{d\epsilon_{\rm CR}}{dt} = \int d^3x \int Q(E) \: dE = V_G \frac{4\pi}{c}
\int \frac{J(E)}{ \tau(E/Z)}  dE \, ,
\end{equation}
where $V_G \sim 10^{67}~{\rm cm}^3$ is the Galactic disk
volume~\cite{Gaisser:2005tu}. For $E_{\rm knee} < E < E_{\rm ankle}$,
we conservatively assume that the trapping time in the Galaxy scales
with energy as in (\ref{tau-escape}). (Note that an evolution into
quasirectilinear motion would increase the power allowence.) In this
case the power budget required to fill in the spectrum from the knee
to the ankle is found to be $d\epsilon_{\rm CR}/dt \simeq 2 \times
10^{39}~{\rm erg/s}$~\cite{Gaisser:2006xs}.

We also note that recent data from KASCADE-Grande~\cite{::2013dga}
indicate that at $\sim 30~{\rm PeV}$ the flux of protons is about an
order of magnitude smaller than the all-species CR flux. Taken at face
value, this implies that the fraction of the power budget allocated to
nucleons of energy $E_p$ which do not escape the Galaxy is about $0.1$
of the all-species power.  However, light elements possess higher
magnetic rigidity and are therefore more likely to escape the Galaxy.
From the functional form of $\tau(E/Z)$ above, we estimate the
survival probability for protons at $30~{\rm PeV}$ to be 46\% of that
at $E_{\rm knee}$.  This leads to a value for the proton fraction of
total flux at injection ($\zeta$) of $\zeta = 0.1 / 0.46 = 0.22$.  In
the following discussion, we will consider a wide range for
$\bar\zeta\equiv\zeta A_0/A$, with $0.22 \lesssim \bar\zeta \lesssim
0.44$ seemingly the most realistic range.

Then, we find for $C_{\rm CR}^p$ the particular result
\beq{Ccr2.3}
C_{\rm CR}^p(2.3) = \frac{0.3\times (0.1\,{\rm PeV})^{0.3} \times 2 \bar\zeta \times 10^{39}{\rm erg/s}}{4\pi(10\,{\rm kpc})^2}\,.
\eeq
Finally, inserting Eqs.~\rf{Cnu2.3} and \rf{Ccr2.3} into \rf{epspi}, we get
\beq{epspi2.3} \epspi (2.3) = \left(\frac{1}{16} \right)^{-0.3}
\frac{C_\nu(2.3)}{\epsilon_\nu\, C_{\rm CR}^p(2.3)} =
\frac{0.055}{\bar\zeta \, \epsilon_\nu}\,, \eeq where in the final
expression, we have set $\Gamma$ equal to the favored value of
2.3.  Substituting in Eq.~(\ref{epspi2.3}) $\epsilon_\nu = \frac{3}{4}$ and
$\frac{1}{4}$ for the complete and damped pion decay chain,
respectively, we finally arrive at $\epspi(2.3) = 0.073/\zeta$ and $0.22/\zeta$.\footnote{Though the damped pion decay chain does not yield flavor
  equipartition on Earth, any deviation from $\Phi_0^{\rm total}$
  falls in the range of uncertainty.}  

\begin{figure}[tbp]
\postscript{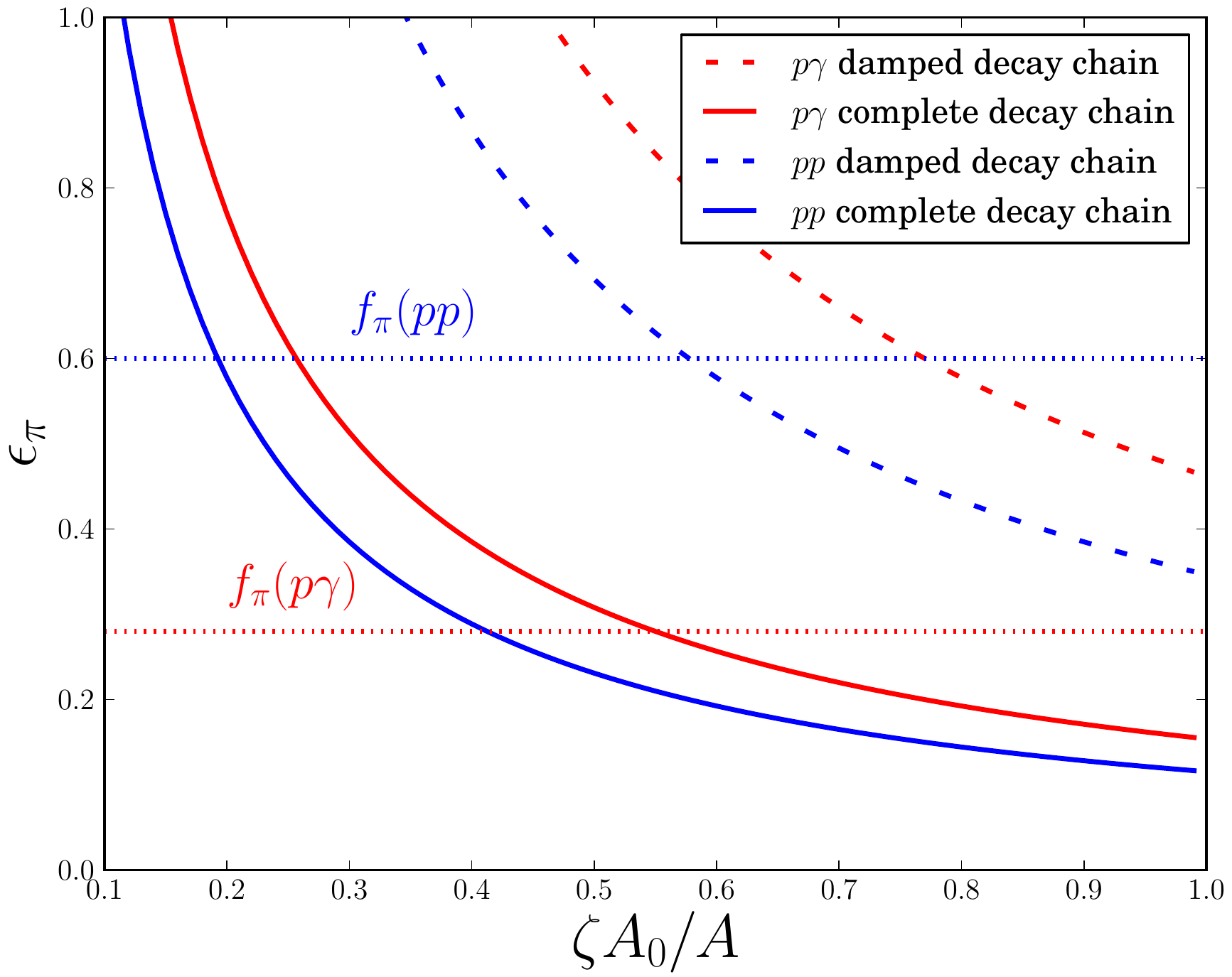}{0.62}
\caption{Total pion energy fractions of parent proton, for favored
  spectral index $\Gamma=2.3$. The average inelasticity $f_\pi$ for
  $pp$ and $p\gamma$ collisions is also shown for comparison. From
  Ref.~\cite{Anchordoqui:2013qsi}.}
\label{fig:epsilonpi}
\end{figure}

If  neutrinos are produced in $pp$ collisions, one can interpret
$\epspi$ in terms of the efficiency of transferring proton energy to
all three pion species, $\epsilon_\pi$, by simply scaling $\epspi$ by
$\frac{3}{2}$ to yield $\epsilon_\pi = 0.12 /\zeta$ and
$\epsilon_\pi = 0.33/ \zeta$ for the complete and damped
chains, respectively. Alternatively, if  neutrinos are produced in
$p\gamma$ collisions, we scale by 2, yielding
$\epsilon_\pi = 0.15/ \zeta$ and $\epsilon_\pi = 0.44/
\zeta$ for the complete and damped pion chains,
respectively.\footnote{Resonant $p\gamma$ interactions produce twice as many neutral
pions as charged pions. Direct pion production via virtual meson
exchange contributes about 20\% to the total cross section, almost
exclusively producing $\pi^+$. Hence, $p\gamma$ interactions produce
roughly equal numbers of $\pi^+$ and $\pi^0$.}  We show $\epsilon_\pi (\zeta A_0/A)$ for all four cases in
Fig.~\ref{fig:epsilonpi}.

In $pp$ collisions, hadronic models predict that $f_\pi \sim 0.6$ of
the ``beam'' proton energy is channeled into
pions~\cite{Frichter:1997wh}.  Since the value of $\epsilon_\pi$
reflects both the inelasticity as well as the fraction of protons
which escape the source without producing pions, we expect
$\epsilon_\pi$ to be smaller than $f_\pi$.  This turns out to be the
case for a complete pion decay chain if $\zeta A_0/A> 0.19$.  Note,
however, that the incomplete pion decay chain requires a considerably
larger fraction, $\zeta A_0/A> 0.59$, which pushes the realm of
plausibility. For $p\gamma$ interactions, $f_\pi \sim
0.28$~\cite{Stecker:1968uc}, thereby excluding the incomplete decay
chain hypothesis for this case.  On the other hand, the complete decay
chain appears to be allowed only for $\zeta A_0/A> 0.56$.

We have seen that the energetics of the Galaxy are sufficient to
explain the neutrino flux, assuming that the neutrinos are produced in $pp$
interactions. Potential Galactic neutrino sources have long been
considered, see
{\it e.g.}~\cite{Levinson:2001as,Distefano:2002qw,AlvarezMuniz:2002tn,Anchordoqui:2002xu,Amato:2003kw,Anchordoqui:2003vc,Kistler:2006hp,Torres:2006ub,Anchordoqui:2006pe,Anchordoqui:2006pb,Beacom:2007yu,Kappes:2006fg,Lunardini:2011br};
some recent refinements to previous models in light of new IceCube
data are discussed in~\cite{Joshi:2013aua,Gonzalez-Garcia:2013iha}.
As an example, we discuss in the next section the Galactic Center as
one of the more likely neutrino engines.

\subsection{Clustering at the Center of the Milky Way?}

\begin{figure}[ht]
\postscript{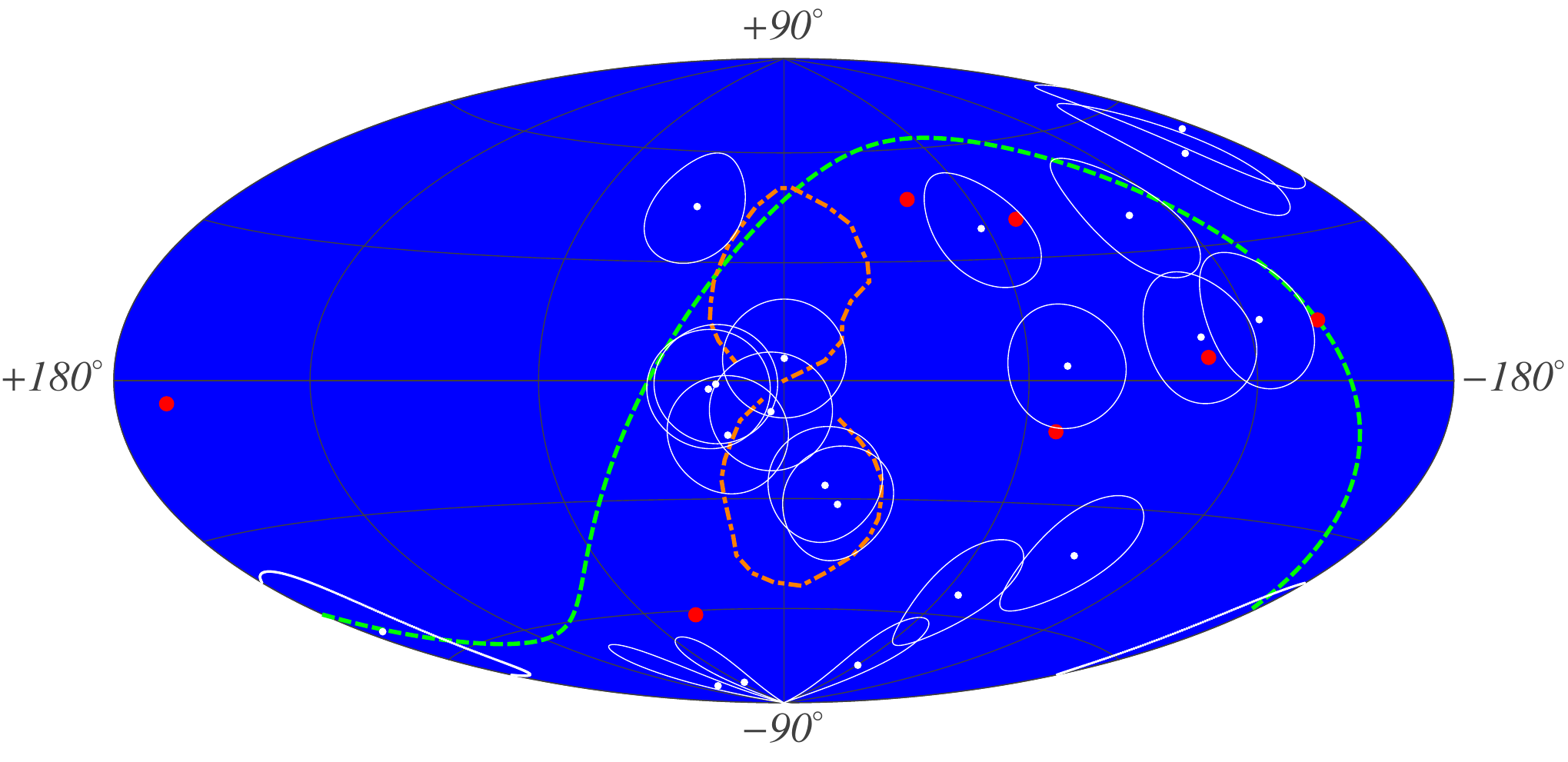}{0.75}
\caption{IceCube neutrino events in Galactic coordinates. The 21
  shower-like events are shown with $15^\circ$ error circles around
  the approximate positions (small white points) reported by the IceCube Collaboration. The 7 track-like events are shown as larger red
  points. Also shown are the boundaries of the Fermi bubbles
  (dot-dashed line) and the Equatorial plane (dashed line). From
  Ref.~\cite{Razzaque:2013uoa}.}
\label{fig:icevents}
\end{figure}

We previously discussed the distribution of arrival directions. Let us
revisit Fig.~\ref{fig:skymap}, where we see that the largest
concentration of events is near the Galactic Center. In
Fig.~\ref{fig:icevents} we reproduce the skymap with all the events,
considering a shower average reconstruction angular uncertainty of
$15^\circ$. We can see that 5 shower-like events fall near
the Galactic center, one of which is the second-highest energy event
in the data, with an additional 3 shower events~\cite{Lunardini:2013gva} consistent with the
extent of the \textit{Fermi} bubbles~\cite{Su:2010qj}.  Curiously, there are no
track-like events in this region, although statistics are very limited
at present and the selection efficiency for cascades is higher than it
is for tracks below $10^{6}~{\rm GeV}$ (see Fig.~\ref{fig:exposure}).

In the spirit of Ref.~\cite{Razzaque:2013uoa},  we define  the ``directional'' cosmic neutrino flux per flavor, averaged over all flavors, as
\begin{equation}
\varphi_\nu (E_\nu) \equiv \frac{dF_{\nu_\alpha}}{dA dt dE_\nu} \simeq
\varphi_0 \left(\frac{E_\nu}{\rm GeV}\right)^{-\Gamma} \, .
\end{equation}
Next, we compute the normalizations for an $8^\circ$ circular window encompassing the Galactic
Center,  {\it i.e.} solid angle $\Omega_{8^\circ} = 2 \pi (1 - \cos 8^\circ) =
0.06~{\rm sr}$, with the results shown in Table~\ref{tab:cls-GC1}.  We find that an unbroken spectrum with $\Gamma = 2$ and a cutoff
at $\sim 2~{\rm PeV}$ yields the observed number of high and low
energy events in this region, with normalizations differing only by
12\%. If we are willing to tolerate consistency of an empty  bin at the 90\% C.L., then we do not require a cutoff.

\begin{table}[tb]
\caption{Normalization $\varphi_0$ for 4 events in the ``low energy''
  ($E<1$~PeV) bin and 1 event in the ``high energy'' (1-2~PeV) bin,
and normalization upper limits for the ``null'' bin (2-10~PeV) at 68\%C.L. ($\varphi^{\rm max}_{68}$)
and 90\%C.L. ($\varphi^{\rm max}_{90}$) in units of $({\rm GeV}\cdot{\rm cm}^{2}\cdot{\rm s})^{-1}$,
for various spectral indices, $\Gamma$, and $\Omega_{8^\circ} =
0.06~{\rm sr}$.
\label{tab:cls-GC1}}
\centering
\begin{tabular}{ |c|c|c|c|c|}
\hline \hline \\ [-2.0 ex]
$\Gamma$~&~$\varphi_0^{E_\nu < 1{\rm PeV}}$~&$\varphi_0^{1 {\rm PeV} < E_\nu < 2
  {\rm PeV}}$  &~$\varphi^{\rm max}_{68}$~&~$\varphi^{\rm max}_{90}$~\\
\hline \\ [-2.0 ex]
~~2.0~~ & $2.49 \times 10^{-10}$ & $2.85 \times 10^{-10}$ & $2.36
\times 10^{-10}$ & $4.64 \times 10^{-10}$ \\
\hline \\ [-2.0 ex]
2.1 & $8.55 \times 10^{-10}$ & $1.17 \times 10^{-9}$ & $1.10 \times
10^{-9}$ & $2.09 \times 10^{-9}$ \\
\hline \\ [-2.0 ex]
2.2 & $2.92 \times 10^{-9}$ & $4.83 \times 10^{-9}$ & $5.17 \times
10^{-9}$ & $9.78 \times 10^{-9}$ \\
\hline \\ [-2.0 ex]
2.3 & $9.94 \times 10^{-9}$ & $1.99\times 10^{-8}$ & $2.41 \times
10^{-8}$ & $4.57 \times 10^{-8}$ \\
\hline \\ [-2.0 ex]
2.4  & $3.36 \times 10^{-8}$ & $8.16 \times 10^{-8}$  & $1.13 \times
10^{-7}$ & $2.01 \times 10^{-7}$\\
\hline \\ [-2.0 ex]
2.5 & $1.31 \times 10^{-7}$ & $3.36 \times 10^{-7}$ & $5.24 \times
10^{-7}$ & $9.90 \times 10^{-7}$ \\
\hline
\hline
\end{tabular}
\end{table}
\begin{table}[tb]
\caption{Normalization $\varphi_0$ for 6 events in the ``low energy''
  ($E<1$~PeV) bin and 1 event in the ``high energy'' (1-2~PeV) bin,
and normalization upper limits for the ``null'' bin (2-10~PeV) at 68\%C.L. ($\varphi^{\rm max}_{68}$)
and 90\%C.L. ($\varphi^{\rm max}_{90}$) in units of $({\rm GeV}\cdot{\rm cm}^{2}\cdot{\rm s})^{-1}$,
for various spectral indices, $\Gamma$, and $\Omega_{20^\circ} =
0.38~{\rm sr}$.
\label{tab:cls-GC2}}
\centering
\begin{tabular}{ |c|c|c|c|c|}
\hline \hline \\ [-2.0 ex]
$\Gamma$~&~$\varphi_0^{E_\nu < 1{\rm PeV}}$~&$\varphi_0^{1 {\rm PeV} < E_\nu < 2
  {\rm PeV}}$  &~$\varphi^{\rm max}_{68}$~&~$\varphi^{\rm max}_{90}$~\\
\hline \\ [-2.0 ex]
~~2.0~~ & $2.36 \times 10^{-9}$ & $1.80 \times 10^{-9}$ & $1.50
\times 10^{-9}$ & $2.82 \times 10^{-9}$ \\
\hline \\ [-2.0 ex]
2.1 & $8.12 \times 10^{-9}$ & $7.45 \times 10^{-9}$ &  $6.99
\times 10^{-9}$ & $1.33 \times 10^{-8}$ \\
\hline \\ [-2.0 ex]
2.2 & $2.78 \times 10^{-8}$ & $3.06 \times 10^{-8}$ &  $3.27
\times 10^{-8}$ & $6.19 \times 10^{-8}$ \\
\hline \\ [-2.0 ex]
2.3 & $9.45 \times 10^{-8}$ & $1.26\times 10^{-7}$ & $1.53 \times
10^{-7}$ & $2.89 \times 10^{-7}$ \\
\hline \\ [-2.0 ex]
2.4  & $3.19 \times 10^{-7}$ & $5.17 \times$ $10^{-7}$  &  $7.14
\times 10^{-7}$ & $1.27 \times 10^{-6}$\\
\hline \\ [-2.0 ex]
2.5 & $1.07 \times 10^{-6}$ & $2.13 \times 10^{-6}$ & $3.31 \times
10^{-6}$ & $ 6.27 \times 10^{-6}$ \\
\hline
\hline
\end{tabular}

\end{table}

Using  data collected from 2007 to 2010  the ANTARES Collaboration
performed a time integrated search for point sources of cosmic neutrinos~\cite{AdrianMartinez:2012rp}.
No statistically significant signal has been found and upper limits on
the neutrino flux have been obtained. Assuming an $E_\nu^{-2}$
spectrum,  with a $20^\circ$ circular window around the source, the
Collaboration reported an upper limit from the direction of the Galactic Center,
\begin{equation}
E_\nu^2 \ \varphi_\nu^{\rm total} (E_\nu) < 3.8 \times 10^{-8}~{\rm GeV} \, {\rm cm}^{-2}\, {\rm
  s}^{-1} \,,
\end{equation}
at the 90\% C.L. Comparison with the normalizations given in
Table~\ref{tab:cls-GC1}, for a spectrum  $\propto E_\nu^{-2}$, shows that
the  flux required to explain IceCube data is safely two orders of
magnitude below the current ANTARES bound.

In closing we make a speculative observation, to be digested with a
dash of salt. If we take the cascade angular resolution of IceCube to be
closer to $20^\circ$ than the fiducial $15^\circ$, then 6 events are
consistent with a common origin near the Galactic Center.  The
relevant normalizations are summarized in Table~\ref{tab:cls-GC2}. For
spectral indices $\Gamma \geq 2.2$, the data do not demand a cutoff at
the highest energies. For $\Gamma = 2.2$, the normalizations in the
first two bins differ by 10\%, whereas for $\Gamma =2.3$, there is a
25\% difference. To
cross-check whether these scenarii are consistent with ANTARES
observations, we must translate the $E_\nu$-square-weighted bound reported by
ANTARES into integral limits. For $\Gamma = 2.3$, we obtain
\begin{equation}
\varphi_\nu^{\rm total} (E_\nu > 50~{\rm TeV}) < 7.6 \times 10^{-13}~{\rm cm}^{-2} \, {\rm s}^{-1} \,,
\end{equation}
at the 90\% C.L. 

Using the normalization in the
low energy bin for the all flavor neutrino flux,
\begin{equation}
\varphi_0^{\rm total} = 2.8 \times 10^{-7}~({\rm GeV}\cdot{\rm cm}^{2}\cdot{\rm
  s})^{-1}\,,
\end{equation}
we obtain the integrated flux required to reproduced the IceCube data
\begin{equation}
\varphi_\nu^{\rm total} (E_\nu > 50~{\rm TeV}) = 1.7 \times 10^{-13}~{\rm cm}^{-2} \, {\rm s}^{-1} \,,
\end{equation}
which is still a factor of $\approx 4$
below  the bound of ANTARES.  Interestingly the
H.E.S.S. Collaboration has reported a point-like source at the
Galactic center with a spectral index $2.21 \pm 0.09 \pm
0.15$~\cite{Aharonian:2004wa}, which is, within errors, consistent
with $\Gamma = 2.2 - 2.3$.

\section{Extragalactic Models}
\label{section-4}

In many respects, extragalactic cosmic ray accelerators provide the most natural sources for the extraterrestrial neutrinos observed by IceCube. In particular, interactions of high energy and UHECRs with energetic photons can generate charged pions, and thus neutrinos in their decays. The flux of neutrinos produced through
the photo-meson interactions of cosmic ray protons can be directly tied to the cosmic ray
injection rate~\cite{Waxman:1998yy}:
\begin{eqnarray}
E^2_{\nu} \Phi_\nu (E_\nu)
&\approx& \frac{3}{8} \epsilon_\pi \xi_Z \, t_{\rm{H}}\, \frac{c}{4\pi}\,E^2_{\rm{CR}} 
\frac{d\dot{N}_{\rm{CR}}}{dE_{\rm{CR}}},\nonumber \\ 
& \approx & 2.3
  \times 10^{-8}\,\epsilon_\pi\,\xi_Z\, \rm{GeV}\,
  \rm{cm}^{-2}\,\rm{s}^{-1}\,\rm{sr}^{-1},
  \label{wbflux}
\end{eqnarray}
where $t_{\rm{H}}$ is the Hubble time and $\epsilon_\pi$ is the
fraction of the energy which is injected in protons lost to photo-meson
interactions.\footnote{A similar argument applies to the case of cosmic ray nuclei~\cite{Anchordoqui:2007tn}.}  The factor of 3/8 comes from the fact that, near the threshold for pion production, roughly half of the pions produced in photo-meson interactions are neutral and do not
generate neutrinos, and three quarters of the energy of charged pion
decays ($\pi^+ \rightarrow \mu^+ \nu_{\mu} \rightarrow e^+ \nu_e
\nu_{\mu} \bar{\nu}_{\mu}$) go into neutrinos. The quantity $\xi_Z$ accounts for the effects of redshift dependent source evolution ($\xi_Z=1$ in the case of no evolution, and $\xi_Z\approx$ 5.75 for sources distributed according to the star formation rate, for example), and $d\epsilon_{\rm CR}/dt \equiv E^2_{\rm CR} d \dot N_{\rm CR}/d E_{\rm CR}$ is the (cosmologically) local energy injection rate of cosmic rays. In a target consisting of ice, the flux given in Eq.~(\ref{wbflux}) is predicted to yield a rate of approximately $14 \times \, \epsilon_{\pi} \, \xi_Z$ showers per km$^3$ per year with energies above 1 PeV~\cite{Cholis:2012kq}. This simple calculation illustrates that a generic class of extragalactic cosmic ray sources with $\epsilon_{\pi} \, \xi_Z$ of order unity would be expected to produce a flux of neutrinos approximately equal to that observed by IceCube. This is highly suggestive of a connection between the the observed neutrinos and the extragalactic sources of the high energy cosmic ray spectrum. 

In the remainder of this section, we review a number of specific
classes of extragalactic sources that could potentially be responsible
for the neutrinos observed by IceCube, including GRBs, AGN,  SBGs, and
newly-born pulsars.

\subsection{Gamma-Ray Bursts}

GRBs constitute one of the most promising sources of high and UHECRs,
and may be capable of accelerating protons to energies as high as
$\sim$$10^{20}$ eV~\cite{Milgrom:1995um, Waxman:1995vg, Vietri:1997st,
  Wick:2003ex}.  Furthermore, as their name implies, gamma-ray burst
fireballs contain high densities of $\gamma$ rays, enabling the efficient
production of neutrinos via the photo-meson interactions of high
energy protons~\cite{Waxman:1997ti, Meszaros:2006rc}.

Typical GRBs exhibit a broken power-law spectrum of the form: 
$dN_{\gamma}/dE_{\gamma} \propto E_{\gamma}^{-2}$ for $E_{\gamma}\ga 0.1-1$ 
MeV and $dN_{\gamma}/dE_{\gamma} \propto E_{\gamma}^{-1}$ at lower
energies~\cite{Band:1993eg}.  The radiation pressure resulting from the very high optical depth of GRB 
fireballs leads to their ultra-relativistic expansion, accelerating the plasma to Lorentz 
factors on the order of $\Gamma \sim 10^2-10^3$. In order for proton-photon collisions 
in this environment to exceed the threshold for pion production, the proton must have 
an energy that, in the observer's frame, meets the following condition~\cite{Waxman:1997ti, Dermer:2003zv, Guetta:2003wi}:
\begin{equation}
E_p \ga 40 \,{\rm PeV} \,\bigg(\frac{\Gamma}{300}\bigg)^2 \, \bigg(\frac{0.3 \, {\rm MeV}}{E_{\gamma}}\bigg)\, \bigg(\frac{1}{1+z}\bigg)^2,
\label{grbc1}
\end{equation}
where $z$ is the redshift of the burst. In any falling spectrum of
high-energy protons, such interactions will predominantly take place
near this threshold. After taking into account that only about 1/5 of
the proton's energy goes into the charged pion produced in such an
interaction, and that each neutrino carries away only about a quarter
of the charged pion's energy, this leads to the production of
neutrinos of characteristic energy:
\begin{equation}
E_{\nu} \sim 2 \,{\rm PeV} \,\bigg(\frac{\Gamma}{300}\bigg)^2 \, \bigg(\frac{0.3 \, {\rm MeV}}{E_{\gamma}}\bigg)\, \bigg(\frac{1}{1+z}\bigg)^2.
\label{grbc2}
\end{equation}
Thus for protons interacting with photons near the observed spectral
break, the resulting neutrinos will have energies near that of the two
most energetic events reported by IceCube.

Most GRBs are observed to have maximum isotropic luminosities in the
range of $L_{\rm max} \sim 10^{51}-10^{53}$ erg/s.\footnote{As the
  observed hard X ray and $\gamma$ ray luminosity is synchrotron
  emission from internal shocks in the relativistic
  fireball~\cite{Rees:1994nw}, this emission will be relativistically
  beamed to within an opening angle on the order of $\theta \sim
  1/\Gamma$. The isotropic equivalent luminosity is related to the true
  luminosity by: $L_{\rm iso} = L_{\rm true}/(1-\cos\theta)$. The
  total isotropic energy emitted is $E_{\rm iso} \simeq L_{\rm
    iso}^{\rm max} \tau_{\rm dur}$. $L_{\rm iso}^{\rm max} \equiv
  L_{\rm max}$ (for simplicity in the remaining text) is the maximum
  isotropic equivalent luminosity.  The duration timescale (as
  observed in hard X rays and $\gamma$ rays), is taken to be
  $\tau_{\rm dur} = 2$ sec for high luminosity GRBs and and $\tau_{\rm
    dur} = 50$ sec for the low luminosity sample (see
  Ref.~\cite{Cholis:2012kq, Goldstein:2012uf, Paciesas:2012vs}).}
This class of ``high luminosity'' GRBs is further divided into short
and long duration bursts with observed timescales of $0.1 - 1$ and $10 - 100$
seconds, respectively, with the majority of observed bursts being of
long duration~\cite{Zhang:2012jr, Goldstein:2012uf,
  Paciesas:2012vs}. In addition, another population of low luminosity
GRBs with $L_{\rm max} \sim 10^{47}$ erg/s has been
suggested~\cite{Liang:2006ci} (see also Ref.~\cite{Lv:2010bz}).  These
low luminosity GRBs, which are potentially much more numerous than
their high luminosity counterparts, generally exhibit smooth light
curves, wider emission cones and durations that are typically in the
range $50 - 1000$~s~\cite{Murase:2006mm, Gupta:2006jm, Coward2006,
  Cobb:2006cu, Pian:2006pr, Daigne:2007qz, Bromberg:2011fm}. At
present, there is considerable variation in the GRB luminosity
functions, $\Phi(L)$, appearing in the
literature~\cite{Gupta:2006jm,Liang:2006ci,Wanderman:2009es}. The
redshift distribution of GRBs, $R_{\rm GRB}(z)$, is generally assumed
to approximately follow the star formation
rate~\cite{Porciani:2000ag}, with high luminosity GRBs occurring at a
local rate of $\sim$1~Gpc$^{-3}$\,yr$^{-1}$ and low luminosity GRBs at
a rate of 230~\cite{Soderberg:2006vh} to 5000~\cite{Soderberg:2005vp}
Gpc$^{-3}$\,yr$^{-1}$.

For a given luminosity function and redshift distribution, one can calculate the diffuse flux of neutrinos or photons at the 
location of the Earth from the population of all GRBs:
\begin{eqnarray}
\Phi_{\nu (\gamma)} = \int_{0}^{z_{\rm max}} dz \int_{L_{\rm min}}^{L_{\rm max}} dL  \;
\Phi(L) \frac{R_{\rm GRB}(z)}{1+z} \frac{4\pi D_{L}(z)^{2}}{(1+z)^2} 
 \frac{c}{H_{0} \sqrt{\Omega_{\Lambda} + \Omega_{M}(1+z)^3}} \ \varphi_{\nu (\gamma)} \, ,
\label{eq:Nuflux}
\end{eqnarray}
where $D_L$ is the luminosity distance
and $\varphi_{\nu (\gamma)}$ refers to the observable neutrino/photon fluence from
an individual GRB located at comoving distance, $D(z)$:
\begin{equation}
\varphi_{\nu (\gamma)} = \frac{dN_{\nu (\gamma)}}{dE_{\nu (\gamma)}^{\rm inj}} \frac{1+z}{4\pi D(z)^{2}}. 
\label{eq:NuObsSpect}
\end{equation}
$dN_{\nu (\gamma)}/dE_{\nu (\gamma)}^{\rm inj}$ is the equivalent injection neutrino/photon spectrum. 

The spectrum of neutrinos (and anti-neutrinos) at injection can be approximated by a doubly broken power-law~\cite{Waxman:1998yy}:
\begin{equation}
\frac{dN_{\nu}}{dE_\nu^{\rm inj}} \propto \left\{ \begin{array}{ll} 
& \left(\frac{E_{\nu}^{\rm inj}}{E_{1}}\right)^{-1} \; \textrm{for} \; E_{\nu}^{\rm inj} \leq E_{1}   \\
&\left(\frac{E_{\nu}^{\rm inj}}{E_{1}}\right)^{-2} \; \textrm{for} \; E_{1} \leq E_{\nu}^{\rm inj} \leq E_{2}  \\
&\left(\frac{E_{2}}{E_{1}}\right)^{-2} \times \left(\frac{E_{\nu}^{\rm inj}}{E_{2}}\right)^{-3} \; \textrm{for} \; E_{\nu}^{\rm inj} \geq E_{2} 
\end{array}\right. \ .
\label{eq:NuInjSpect}
\end{equation}
The first of these spectral features (at $E_{\nu}=E_1$) corresponds to the pion production threshold 
for scattering off photons at the observed break in the gamma ray spectrum of GRBs, while the higher 
energy break (at $E_{\nu}=E_2$) appears as a result of the synchrotron cooling of muons and pions. 
The exact locations of these breaks differs between individual GRBs  due to differences in the strengths 
of the fireballs' magnetic and radiation fields.
Deviations from the standard Waxman-Bahcall spectrum reproducible by the $\Delta^{+}$ resonance 
can result from additional neutrino production modes~\cite{Baerwald:2010fk}. Yet, such modifications of 
the spectrum do not affect the diffuse neutrino flux by more than a
factor of 2. (See however~\cite{Hummer:2011ms,Li:2011ah,He:2012tq}.) In fact the most important uncertainty is 
the amount of the burst's internal energy that goes into accelerating protons to energies of $\sim 10^{16}$ eV 
and above.  Refs.~\cite{Zhang:2006uj, Racusin:2011jf} have suggested that the energy into accelerated protons may be a 
factor of $\sim$10 higher than that into accelerated electrons. Of the energy in protons, $\sim$$1 - 10 \%$ is expected 
to go  into neutrinos, with significant variation from burst-to-burst~\cite{Guetta:2003wi}. 
It is the averaged value for the overall population of GRBs that is the most significant uncertainty yet. 

\begin{figure}[!]
\postscript{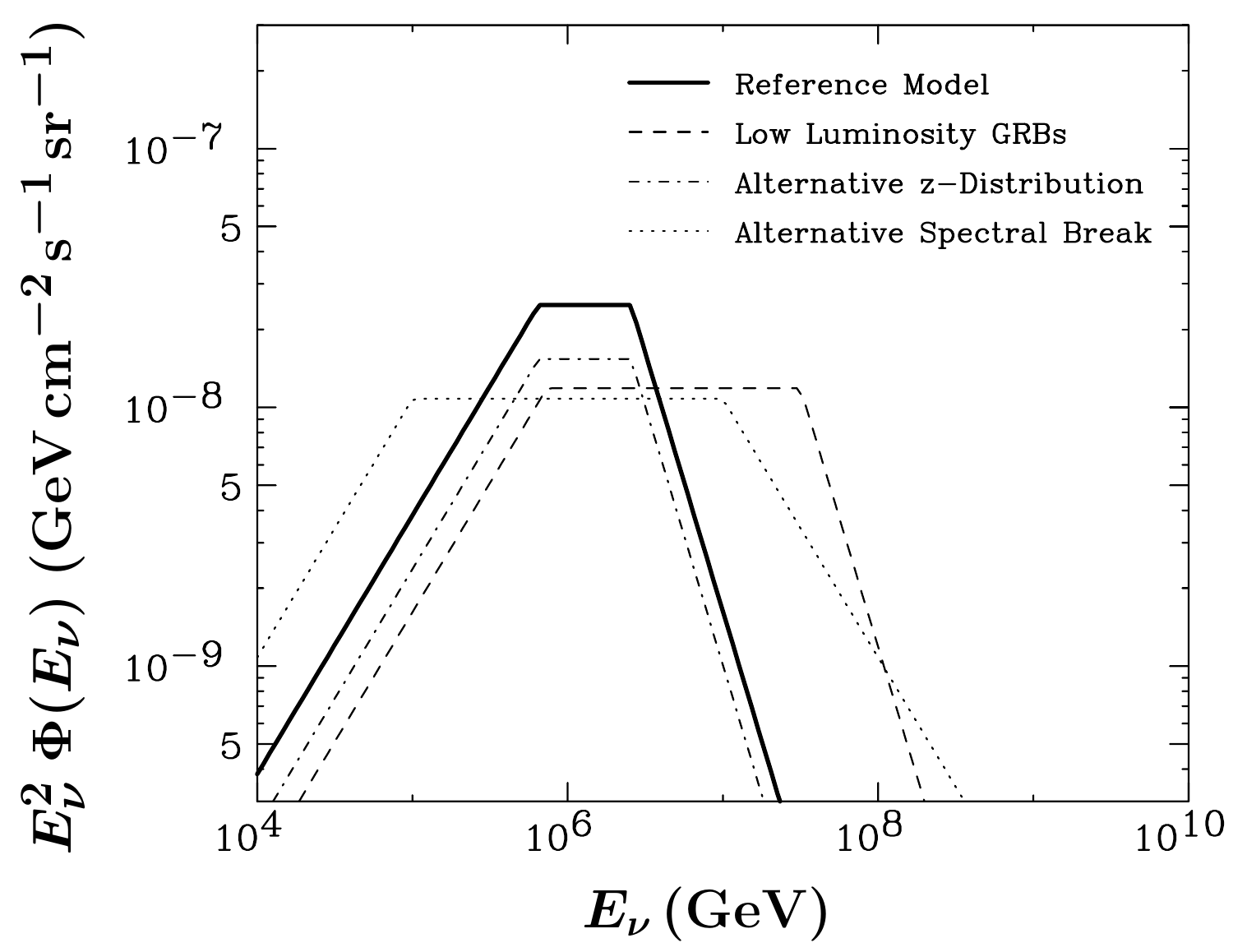}{0.8}
\caption{The contribution of GRBs to the diffuse neutrino (plus anti-neutrino) spectrum. 
Results are shown for high luminosity (solid) and low luminosity (dashed) GRBs, calculated using default 
parameters, and for high luminosity GRB models with a suppressed high redshift distribution (dot-dash) 
and alternative spectral characteristics (dots). Each of these models yields a rate of PeV events which is 
comparable to that implied by the two most energetic events reported by IceCube. Plot taken from Ref.~\cite{Cholis:2012kq}.}
\label{fig:GRBvariations}
\end{figure}

In Fig.~\ref{fig:GRBvariations}, taken from Ref.~\cite{Cholis:2012kq}, the total diffuse  flux of neutrinos and anti-neutrinos 
from GRBs is shown for default parameter choices, and for some representative variations of these parameters.
The solid and dashed lines represent the predicted flux from high and low luminosity GRBs, respectively, each assumed to evolve according to the rate of star formation.
The dot-dashed line represents the contribution from high luminosity GRBs with a redshift distribution that is suppressed above $z=3$. The dotted line shows the flux from high luminosity GRB with alternative choices 
for the parameters leading to the location of the spectral breaks. 
These variations show a fairly wide range of assumptions for GRBs; expected to generate fluxes of PeV neutrinos that are similar to that implied by IceCube's two most energetic events. 
As also suggested by Ref.~\cite{Liu:2012pf}, luminosity functions with steeper slopes on the low luminosity end, or
redshift distributions with higher rates at high redshifts, favor the existence of more dim and untriggered GRBs
in the X rays and $\gamma$ rays.  

Additionally, ultra-long GRBs with $\gamma$ ray luminosities in the
range of $10^{49} - 10^{51}$ erg s$^{-1}$ and durations of $\sim
10^{4}$ s (associated with larger progenitors, such as Wolf-Rayet
stars with radius $R \sim 0.6-3.0 R_{\odot}$) have been proposed for the
source of IceCube's observed neutrinos~\cite{Murase:2013ffa} (see
also, Refs.~\cite{Liu:2012pf, Vieyro:2013lxa} on Pop.~III GRBs at high redshifts).  The
calculations of Ref.~\cite{Murase:2013ffa} suggest that such sources
could produce a neutrino flux of $\sim 10^{-9}$ GeV cm$^{-2}$ s$^{-1}$
sr$^{-1}$ from successful jets and of $\sim 10^{-8}$ GeV cm$^{-2}$
s$^{-1}$ sr$^{-1}$ from choked jets.  For such a case, the spectral
slope is expected to steepen above a few PeV, and an associated
multi-TeV neutrino signal is also predicted \cite{Murase:2013ffa}.
 
By taking into account the times and/or directions of known GRBs, it
is possible to conduct a nearly background free search for neutrinos
originating from such sources. Recently, the IceCube Collaboration has
applied such a strategy, and used the results to derive a stringent
upper limit on the flux of high energy neutrinos from observed
GRBs~\cite{Abbasi:2012zw}. Under standard astrophysical assumptions,
this limit implies that GRBs cannot be the only sources of the highest
energy ($>10^{18}$ eV) cosmic rays (see also,
Ref.~\cite{Ahlers:2011jj}).\footnote{Of course this constraint can be
  evaded if cosmic rays escape from GRBs without
scattering~\cite{Baerwald:2013pu}.} The events reported by IceCube, however,
could still originate from GRBs if either, {\it (i)} a greater
fraction than expected of the high energy neutrinos from GRBs
originate from bursts which are not sufficiently luminous to be
observed by $\gamma$ ray or X ray observatories~\cite{Cholis:2012kq,Liu:2012pf,Murase:2013ffa,Fraija:2013cha}, or {\it
  (ii)} a significant fraction of the $10^{16}-10^{18}$ eV cosmic ray
spectrum originates from GRBs, while most of the $>10^{18}$~eV cosmic
ray spectrum originates from other
sources~\cite{Cholis:2012kq,Li:2012gf,Winter:2013cla}.  Having an
alternative source for the CRs above $E >10^{18}$ eV is attractive
from the perspective of the cosmic ray spectrum's chemical
composition.  Measurements from the Pierre Auger Observatory of the
depth of shower maxima and its variation suggest that the highest
energy cosmic rays are largely of heavy chemical composition (closer
in mass to iron nuclei than protons), while the composition becomes
steadily lighter at lower energies, appearing to be dominated by
protons at $10^{18}$ eV~\cite{Abraham:2010yv,Abreu:2013env}.  As ultra
high-energy nuclei accelerated in a GRB are expected to be entirely
disintegrated into individual nucleons before escaping the
fireball~\cite{Anchordoqui:2007tn}, the possibility that GRBs provide
most of the CRs below $\sim 10^{18}$~eV, but that another class of
sources provide the bulk of the highest energy (heavy nuclei) CRs, is
a well motivated one.

The fact that the events reported by IceCube do not correlate in time
with any known GRBs does not necessarily rule out the hypothesis that
these events originate from this class of sources. Many GRBs, while in the field-of-view of
either the \textit{Swift} Burst Alert Telescope (BAT) and the
\textit{Fermi} Gamma-ray Burst Monitor (GBM), may still go
undetected if they are of sufficiently low luminosity, or are
sufficiently distant.  Given the fluence sensitivity of
\textit{Swift}'s BAT and \textit{Fermi}'s GBM ($\sim$$1\times 10^{-8}$~erg cm$^{-2}$\,s$^{-1}$ and $\sim 2\times 10^{-8}$~erg
cm$^{-2}$\,s$^{-1}$ at 20~keV, respectively), one can estimate how
distant a GRB of a given luminosity could be and still trigger these
detectors. These experiments should be capable of detecting
essentially all high luminosity GRBs ($L \ga 10^{51}$~erg/s) within
their fields-of-view out to a distance of about 8~Gpc $(z\approx
5)$. Thus the observed collection of high luminosity GRBs is fairly
complete (within the given fields-of-view). In contrast, low
luminosity GRBs ($L\sim 10^{47}$ erg/s) are likely to be detected only
within a radius of $\sim$$100$ Mpc, suggesting that the vast majority
of the diffuse neutrino flux from low luminosity GRBs will be
uncorrelated in time or direction with any observed $\gamma$ ray or
X ray signals~\cite{Cholis:2012kq}.

\subsection{Active Galactic Nuclei}

The kinematics of high-energy neutrino production in AGNs is similar
to that of GRBs, with protons accelerated in the cores of AGN close to
the accretion disk~\cite{Kazanas:1985ud}.  Yet there are important
differences; the Lorentz factors of AGN jets are significantly lower
than those of GRB shocks, with values of $\Gamma\sim 30$ rather than
$\sim 300$~\cite{Atoyan:2001ey, Mannheim:1993jg, Mannheim:1998fs}.  As
a result, $\sim 10-100$~PeV protons can exceed the threshold for pion
production much more easily, requiring only the presence of $\sim$keV
photons (rather than the $\sim$100 keV photons required in GRBs).

In GRBs, the observed photon spectral break ($\sim 0.1 - 1$~MeV) leads
to a break at $\sim$1 PeV in the neutrino spectrum. Thus we may expect
the first detections of GRB neutrinos to appear at around this energy
scale. In contrast, AGN do not typically exhibit a spectral peak at
keV energies, but instead in the ultra-violet, typically at around
$\sim$10 eV. This leads one to expect the neutrino spectrum to peak
EeV energies, much higher than that from GRBs. There is a considerable
degree of model dependence in this conclusion, however, deriving in
large part from uncertainties in the spectrum of the target radiation
fields.  In addition, there is significant uncertainty in the magnitude
of the diffuse neutrino flux associated to the number density of AGN~\cite{Stecker:2013fxa}.

\begin{figure}[!]
\postscript{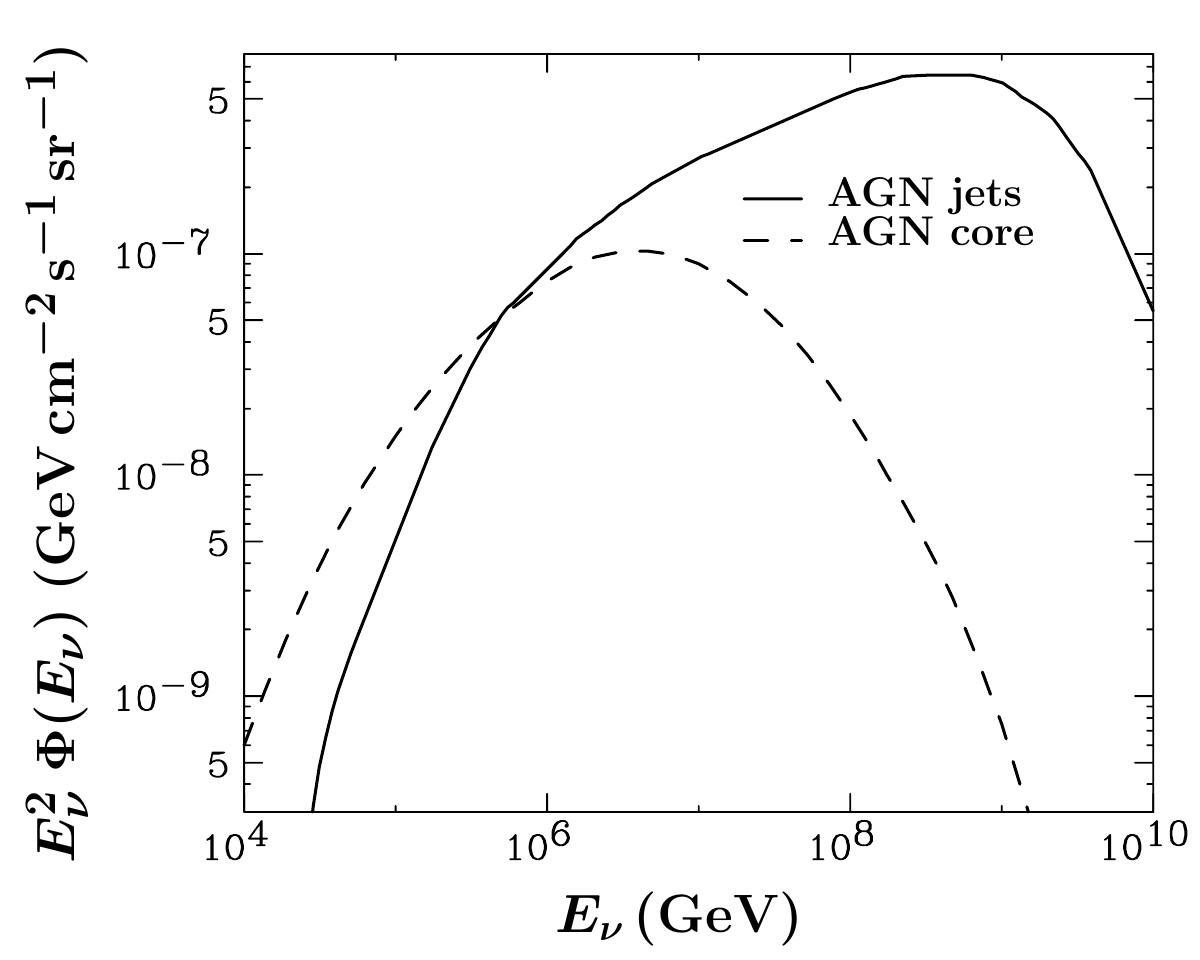}{0.8}
\caption{The contribution of AGNs to the diffuse neutrino (plus anti-neutrino) flux. Results are shown for the energy spectrum of the AGN core model~\cite{Stecker:1991vm,Stecker:2005hn} and the maximal neutrino intensity from AGN jets~\cite{Mannheim:1998wp}. Plot from Ref.~\cite{Cholis:2012kq}.}
\label{fig:AGN}
\end{figure}

In Fig.~\ref{fig:AGN} we show a comparison of two canonical models for the
diffuse neutrino emission from AGN cores and from optically thick AGN jets~\cite{Cholis:2012kq}. In both cases, the diffuse neutrino
flux has been normalized to that observed at $\sim 1$ PeV by IceCube.
In the model of Ref.~\cite{Mannheim:1998wp}, the scattering of
ultra-high energy protons with ultra-violet radiation leads to a
neutrino spectrum which peaks at EeV energies. For this spectral
shape, most showers initiated within IceCube's volume will be of
energy 20 PeV or greater.  Assuming that IceCube's existing data does
not contain a sizable number of enormous (non-contained) showers in
this energy range~\cite{Aartsen:2013pza}, this AGN model will likely
not be able to account for the reported events. In contrast, the model
of Ref.~\cite{Stecker:1991vm,Stecker:2005hn} predicts a neutrino
spectrum from AGN which peaks at a much lower energy of a few PeV, not
unlike the predictions for GRBs.  This is in large part due to an
assumed high density of ambient X rays present around the AGN.

While neutrino emission from known GRBs can be efficiently constrained by 
searching in the time window around the occurrence of a given burst, such a background-free 
strategy is not possible for AGN. As a result, it will be much more difficult to definitively test the 
hypothesis that these neutrinos originate from AGN. 

Due to the ability of AGN jets to accelerate CRs up to the EeV scale,
there are additional constraints that can be imposed on such
model. First, the injected CR spectrum and composition must be
consistent with the observations of CRs at very high
energies~\cite{Apel:2012rm, IceCube:2013wda, Berezhnev:2012vq,
  Garyaka:2008gs, Amenomori:2008aa}. Secondly, the electromagnetic
cascades generated in CR production and propagation cannot lead to a
$\gamma$ ray flux in excess of the extragalactic background measured
by \textit{Fermi}-LAT~\cite{Abdo:2010nz}.  In
Ref.~\cite{Kalashev:2013vba} it was shown that in order to explain the
observed flux of $\sim$~PeV neutrinos without overproducing neutrinos
at even higher energies, and satisfying CR and gamma ray constraints,
AGN jets should emit anywhere between $2 \times 10^{39}$ erg s$^{-1}$
Mpc$^{-3}$ and $7 \times 10^{40}$ erg s$^{-1}$ Mpc$^{-3}$ (see also,
Ref.~\cite{Kistler:2013my}).  This implies an individual AGN
luminosity of approximately $L \sim 10^{44}$ erg
s$^{-1}$~\cite{Kalashev:2013vba}.\footnote{In
  Ref.~\cite{Treister:2009kw} the observed volume density of AGN with
  luminosity in the X rays $L_{X}>10^{43}$ erg s$^{-1}$ has been
  measured to be $\simeq 10^{-5}$ Mpc$^{-3}$ for $z>0.5$.  This
  density refers to the observable AGN, and does not account for
  distant AGN with their jets pointing at an angle far from our line-
  of-sight.} In the next subsection we discuss in more detail the
energetics of the AGN jet model.

\subsection{Blazars}
\label{section:blazars}

Blazars are AGNs with a relativistic jet pointing in the general
direction of Earth.  They are very bright gamma ray sources observed
over a broad range of energies.  Powered by a central engine believed
to be a supermassive black hole, a blazar jet is capable of
accelerating electrons, protons, and nuclei to very high energies.
Detailed numerical simulations support the possibility of accelerating
protons up to $E_{\rm p,max}\sim
10^{17}$~eV~\cite{Sironi:2010rb,Sironi:2013ri}, and even higher
energies are possible under some exceptional conditions, such as
alignment of magnetic fields in the internal shocks.  It is unclear
how much AGNs contribute at the highest end of observed UHECR
spectrum, which extends well above $10^{19}$~eV.  Contributions of
unusual supernova explosions, GRBs, and, possibly, nuclei from nearby
sources remain viable possibilities for explaining UHECR at energies
above
$10^{18}$~eV~\cite{Gaisser:2013bla,Anchordoqui:1999cu,Aloisio:2009sj,Calvez:2010uh,Taylor:2011ta}. However,
there is little doubt that AGNs can produce substantial fluxes of
cosmic rays at least up to the ''ankle.''

There is growing evidence that intergalactic cascades initiated by
line-of-sight interactions of cosmic rays produced by AGNs are
responsible for the highest-energy gamma rays observed from
blazars~\cite{Essey:2009zg,Essey:2009ju,Essey:2010er,Essey:2011wv,Murase:2011cy,Razzaque:2011jc,Prosekin:2012ne,Aharonian:2012fu,Zheng:2013lza,Kalashev:2013vba,Takami:2013gfa,Essey:2013kma,Inoue:2013vpa}.
As long as the intergalactic magnetic fields (IGMFs) are in the range
$10^{-17} {\rm G} \lesssim B\lesssim 3\times
10^{-14}$~G~\cite{Essey:2010nd}, the spectra of distant blazars are
explained remarkably well with secondary photons from such
cascades~\cite{Essey:2009zg,Essey:2009ju,Essey:2010er}.  In the
absence of cosmic ray contribution (for example, if one assumes large
IGMFs), the observed spectra from distant blazars should be much
softer because of the gamma ray interactions with extragalactic
background light.  Models for hard intrinsic spectra of $\gamma$ rays
can be constructed~\cite{Stecker:2007zj,Lefa:2011xh,Dermer:2011uu},
but neither source-intrinsic features, nor selection effects can
explain the observed anomaly in the broad range of energies and
redshifts for which the data are available~\cite{Horns:2012fx}.
Furthermore, there is a growing list of remarkably distant TeV
sources, such as a VHE blazar PKS~1424+240: the recent measurement of
its redshift $z \ge 0.6035$~\cite{Furniss:2013kv} puts it at an
optical depth $\tau >5$ for the highest energy gamma rays observed by
VERITAS~\cite{Acciari:2009ft}.  At such an optical depth, all the
primary gamma rays should be attenuated and filtered out, while
proton-induced secondary component agrees with the observed spectrum
of PKS~1424+240 for redshifts $0.6 \le z \le
1.3$~\cite{Essey:2013kma}.  There is a tantalizing possibility that
new axion-like particles exist and couple to photons with a coupling
that is large enough to allow for gamma ray
conversions~\cite{De_Angelis:2007dy,Simet:2007sa,Horns:2012kw}.  This
effect could reduce the effective opacity of the universe
dramatically, as could some forms of Lorentz-invariance
violation~\cite{Kifune:1999ex}.  However, the natural ease with which
secondary photons from cosmic ray interactions reproduce the data
makes the explanation based on cosmic rays very appealing.
Furthermore, the lack of time variability of the most distant blazars
at energies above TeV is in agreement with this hypothesis, which
predicts that the shortest variability time scales for $z \gtrsim
0.15$ and $E\gtrsim 1\ {\rm TeV}$ should be greater than
$(0.1-10^3)$~years, depending on the model
parameters~\cite{Prosekin:2012ne}.

\begin{figure}[tbp]
\begin{minipage}[t]{0.49\textwidth}
\postscript{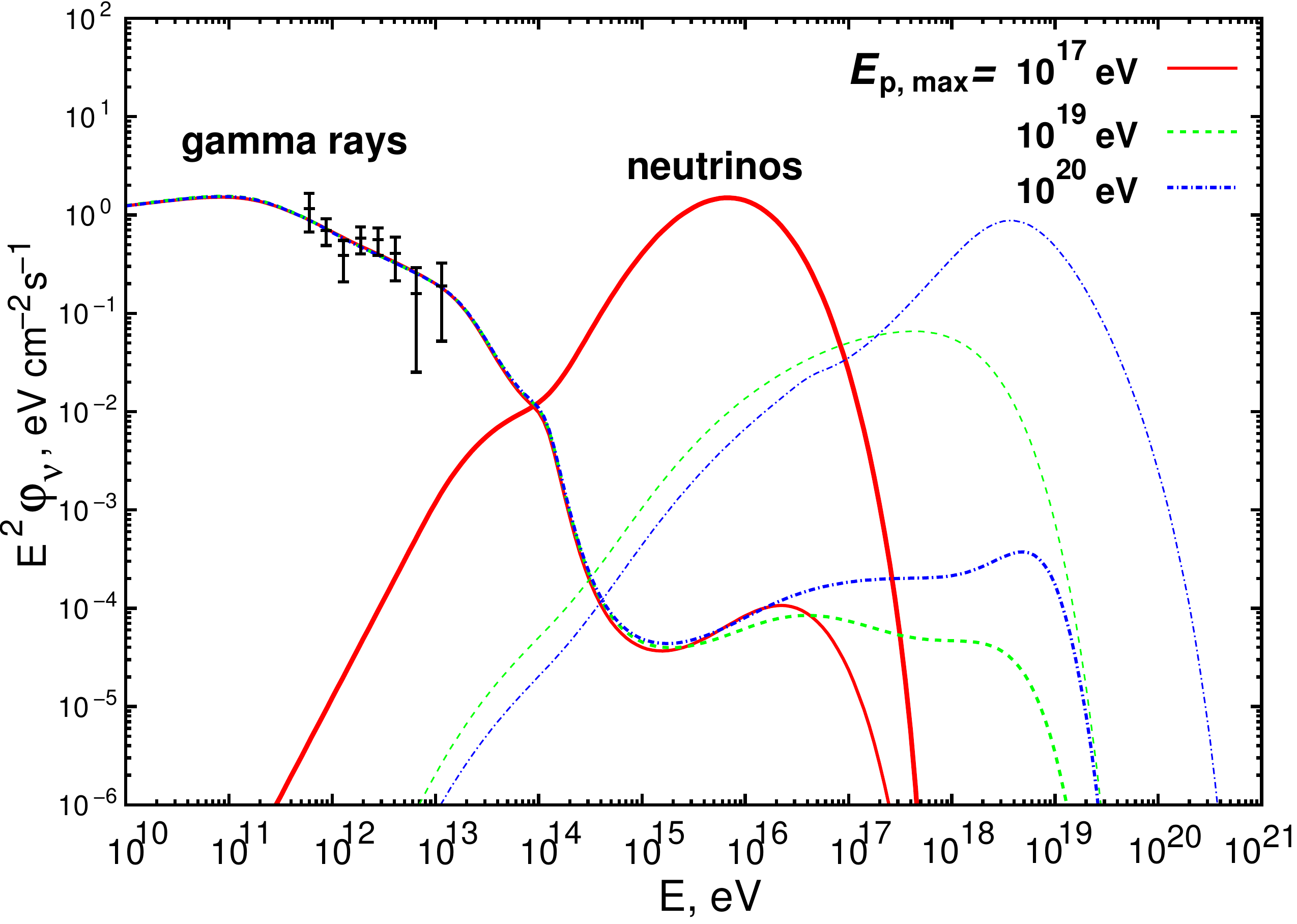}{0.99}
\end{minipage}
\hfill
\begin{minipage}[t]{0.49\textwidth}
\postscript{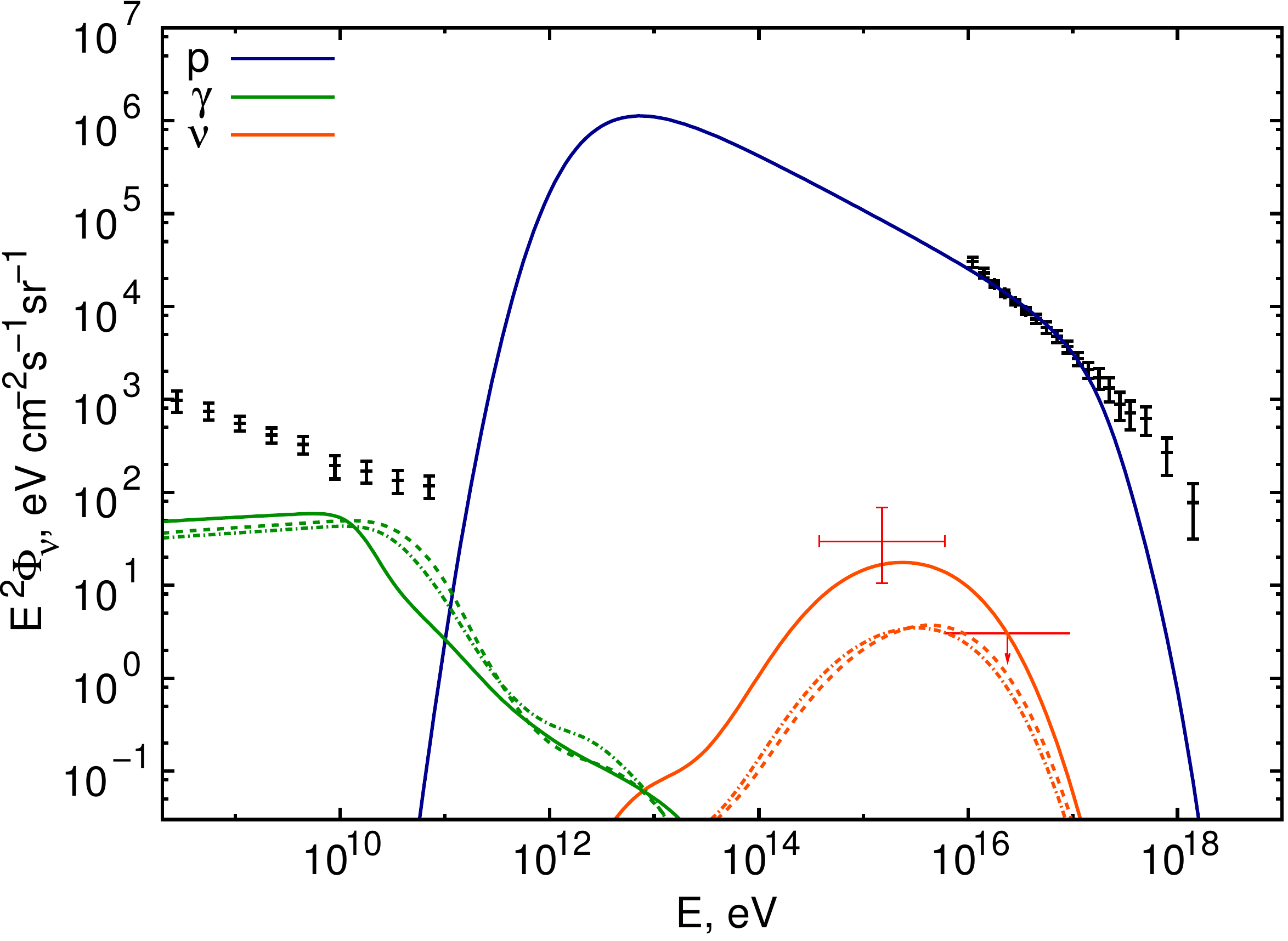}{0.99}
\end{minipage}
\caption{{\bf Left:}~A peaked neutrino spectrum accompanies
  secondary gamma rays produced in line-of-sight interactions of
  cosmic rays emitted by
  blazars~\cite{Essey:2009zg,Essey:2009ju,Essey:2010er}. The lowest
  position of the peak is at 1~PeV.  Assuming that a distribution of
  AGN with respect to maximal proton energy $E_{p, {\rm max}}$ is a
  decreasing function of $E_{p, {\rm max}}$, as implied by numerical
  simulations~\cite{Sironi:2010rb,Sironi:2013ri}, the diffuse neutrino
  spectrum shown in the right panel has a peak at
  1~PeV~\cite{Kalashev:2013vba}.  {\bf Right:}~Predicted spectra of
  PeV neutrinos (red lines) compared with the flux measured by the
  IceCube experiment. The IceCube data points (red) are
  model-dependent 68\% confidence level flux estimates obtained by
  convolving the IceCube exposure with the predicted neutrino
  spectrum. The predicted spectra are shown for the sum of three
  flavors; each flavor contributes, roughly, 1/3. The solid, dashed,
  and dotted red lines correspond to different EBL
  models~\cite{Kneiske:2003tx,Stecker:2005qs,Stecker:2012ta,Inoue:2012bk}.
  The proton injection spectrum has a spectral index $\alpha = 2.6$
  and maximun energy $E_{p,{\rm max}} = 3 \times 10^{17}~{\rm
    eV}$. Also shown are the predicted gamma ray (lower curves below
  10~TeV) and cosmic ray (upper curve) fluxes. The cosmic ray data
  points above 10~PeV are based on KASCADE-Grande~\cite{Apel:2012rm}; the diffuse
  gamma ray background data points below 1~TeV are due to Fermi~\cite{Abdo:2010nz}.
  See text and Ref.~\cite{Kalashev:2013vba} for
  details.\label{fig:blazars}}
\end{figure}

Secondary gamma rays are generated in two types of interactions of
cosmic rays along the line of sight.  First, the proton interactions
with the CMB photons produce electron-positron pairs and give rise to
an electromagnetic cascades due to proton pair production (PPP) or
Bethe--Heitler process, $p \gamma_{_{\rm CMB}}\rightarrow
pe^+e^-$~\cite{Blumenthal:1970nn}.  Second, the proton interactions
with the extragalactic background light (EBL) can produce pions in the
reactions $p \gamma_{_{\rm EBL}} \rightarrow p\pi^0$ or $p
\gamma_{_{\rm EBL}} \rightarrow n\pi^+ $.  While the PPP process is
not associated with any neutrinos, the pion photoproduction generates
a neutrino flux related to the gamma ray flux.  The relative
importance of the two processes depends on the proton injection
spectrum.  Remarkably, the observed gamma ray spectrum is very robust
and does not depend on the spectrum of protons, as shown in
Fig.~\ref{fig:blazars} (left panel) and, in detail, in
Refs.~\cite{Essey:2009ju,Essey:2010er}.  This feature is particularly
appealing in application to the spectra of distant blazars, because
the shape of the spectrum is not model-dependent, and the data for
each distant blazar are explained with a one-parameter fit by varying
the total proton luminosity within the range allowed by AGN
energetics~\cite{Essey:2009zg,Essey:2009ju,Essey:2010er}.

While the spectra of gamma rays are practically independent of the
model parameters, as long as the maximal proton energy $E_{p, {\rm
    max}}> 10^{17}$~eV~\cite{Essey:2009ju,Essey:2010er}, the spectrum
of neutrinos does depend on $E_{p, {\rm max}}$, as well as on the EBL
model. The lowest-energy neutrinos are those produced just above the
threshold for pion production on EBL.  If $E_{p, {\rm max}} \sim
10^{17}$~eV, the proton spectrum cuts off just above the pion
production threshold, and the resulting neutrino spectrum has a peak
at 1~PeV, as shown in Fig.~\ref{fig:blazars} (left panel) by the solid
red line.  For higher values of $E_{p, {\rm max}}$, the peak moves to
higher energies.

However, according to detailed numerical
simulations~\cite{Sironi:2010rb,Sironi:2013ri}, acceleration beyond
$E_{p, {\rm max}}\sim 10^{17}$~eV in AGN jets would require some very
special conditions, such as magnetic field alignment in shocks.
Therefore, it is likely that the distribution of AGNs with respect to
the maximal proton energies is a decreasing function of $E_{\rm
  p,max}$, with the values $E_{\rm p,max} \gtrsim 10^{18}$~eV still
allowed, but uncommon.  The interactions of cosmic rays with EBL 
produce neutrinos via the reaction $p\gamma_{\rm EBL}\rightarrow p
\pi^+$, which has a sharp threshold around $E_{\rm th}\sim 10^{17}$~eV
(broadened by the energy distribution of the EBL photons).  As long as
the distribution of AGN with $E_{\rm p,max}$ decreases fast enough to
make the contribution of CMB photons unimportant, most neutrinos are
produced in interactions of protons near the threshold, $E_{\rm p}
\sim 10^{17}$~eV.  The neutrino spectrum is, therefore, limited by the
fraction $\sim (0.01 - 0.1)$ of the threshold energy from below and by
$\sim (0.01 - 0.1)\times E_{\rm p,max}$ from above, so that the
neutrino spectrum has a peak at $E_\nu \sim (0.01 - 0.1) \times
10^{17}\, {\rm eV}\sim 1\,{\rm PeV}$.

Taking into account a likely evolution of AGN with redshift, one obtains a spectrum for diffuse neutrino background produced by blazars as shown in 
Fig.~\ref{fig:blazars} (right panel)~\cite{Kalashev:2013vba}.  The consistency with IceCube results depends on the model of redshift evolution and on the model of extragalactic background light~\cite{Kalashev:2013vba}.

\subsection{Starburst Galaxies}

Collisions of cosmic rays with the radiation in galaxies undergoing
periods of rapid star formation, referred to as SBGs, are predicted to
yield significant fluxes of $\sim$~TeV-PeV
neutrinos~\cite{Loeb:2006tw}.  Among the supernovae that occur in such
galaxies, a small fraction produce ejecta with velocities that can be
as fast as 0.1$c$ and release kinetic energy at the level of $10^{52}$
erg \cite{Soderberg:2006vh, Liu:2013wia}. These extreme supernovae,
called hypernovae, and are able to accelerate protons up to $10^{17}$
eV. If these protons travel through a galactic medium of dense
infrared and optical radiation background, they will produce neutrinos
through photo-meson interactions with energies up to the
PeV-scale~\cite{Cholis:2012kq, Murase:2013rfa}.  For reasonable
assumptions, and calibrating to the observed fluxes of
gamma rays~\cite{Lacki:2010vs} and radio emission from such objects,
contributions to  the diffuse neutrino flux are expected to be on the order of $1 \times 10^{-8}$~GeV cm$^{-2}$ s$^{-1}$ sr$^{-1}$~\cite{Lacki:2010vs, Loeb:2006tw, Stecker:2006vz,Thompson:2006np} (see also~\cite{Torres:2004ui,DomingoSantamaria:2005qk,delPozo:2009wz, Rephaeli:2009ku,
  delPozo:2009mh, Persic:2008uj,Lacki:2009mj,Torres:2012xk} for detailed models on
multi-frequency emission from starburst galaxies
and~\cite{Acero:2009nb,Acciari:2009wq,Abdo:2009aa} for $\gamma$ ray observations).  The authors of Ref.~\cite{He:2013cqa} have
considered the case of ultra-luminous infrared galaxies, which are the
most intense and luminous galaxies among the SBGs with high gas
densities. The diffuse neutrino flux from hypernova remnants in these galaxies
can be at the level of $\sim 2 \times 10^{-9}$~GeV cm$^{-2}$ s$^{-1}$
sr$^{-1}$ with an estimated cut-off at $\sim 5$ PeV for an assumed CR
proton injected spectrum of $dN_{p}/dE_{p} \propto
E_{p}^{-2}$~\cite{He:2013cqa}.

Allowing for more optimistic acceleration conditions for protons in
hypernovae ($E_{p}$ up to $10^{18}$ eV and $5\times 10^{51}$ ergs in
CR protons), and for larger confinement timescales inside the SBGs
than \cite{He:2013cqa}\footnote{For CR protons the two important
  timescales that define the confinement time of CRs in a galaxy are
  the diffusion timescale $\tau_{\rm diff} \simeq h^{2}/4D$ and the
  advection timescale $\tau_{\rm adv} \simeq h/v_{\rm wind}$, where
  $h$ is the hight of the galaxy's gas disk, $D = (1/3)\cdot\lambda c$
  is the diffusion coefficient ($\lambda$ is the diffusion length),
  and $v_{\rm wind}$ is the velocity of the galactic wind. There are
  uncertainties in each of the $h$, $D$ and $v_{\rm wind}$ at the
  level of a factor of few, allowing for a wide range of values on the
  confinement time of the very high energy CR protons (from $\simeq$
  few $\times 10^{4}$ to few $\times 10^{5}$ years for 60 PeV
  protons). Also the timescale of energy losses for the CR protons can
  be of relevance under assumptions for very high gas galactic
  densities (see for more discussions of \cite{He:2013cqa} and
  \cite{Liu:2013wia}).}, the authors of Ref.~\cite{Liu:2013wia} have
suggested that hypernovae in SBGs could be responsible for the $\sim
10^{-8}$ GeV cm$^{-2}$ s$^{-1}$ sr$^{-1}$ PeV neutrino flux. Their
suggested neutrino spectrum has a power-law spectral slope of
$\Phi_\nu (E_\nu) \propto E_{\nu}^{-2}$, without any sharp cutoff
above the PeV energies, and with only a smooth softening of the
neutrino spectrum at $\sim 10$ PeV.

\subsection{Newborn Pulsars} 

Another proposed source-type is newborn pulsars.  Like GRBs,
 newborn pulsars are transients.  However, unlike GRBs,  newborn
pulsars accelerate the iron-rich surface elements on young neutron
stars, to populate the highest-energy cosmic rays with mainly heavy
elements like $^{56}$Fe.  Nearby pulsars show direct evidence of accelerated
electrons and positrons.  Their ability to accelerate hadrons is
speculative.

Heavy nuclei carry two advantages compared to light elements for a
given energy.  First of all, due to their lower energy per baryon,
heavy nuclei can travel hundreds of megaparsecs before losing their
energy by photo-disintegration processes on the cosmic
backgrounds~\cite{Puget:1976nz,Anchordoqui:1997rn,Stecker:1998ib,Epele:1998ia,Ahlers:2010ty}. Secondly,
nuclei of charge $Z$ can be accelerated to an 
energy typically $Z$ times larger than protons in a given electromagnetic configuration~\cite{Anchordoqui:1999cu,Aloisio:2009sj}. 
Results from the Pierre Auger Observatory indicate an increasing
average primary mass with energy above a few EeV (though this result
is not observed by HiRes or Telescope Array)~\cite{Anchordoqui:2013eqa}.

Pulsars have been suggested as possible accelerators of cosmic rays
since their discovery, due to their important rotational and magnetic
energy reservoirs~\cite{Gunn:1969ej}.  The fastest spinning young
neutron stars exhibit pulsar magnetic fields typically in the range
$10^{12}-10^{13}$~G.  Neutron stars with much larger surface magnetic
fields, {\it i.e.}, magnetars, have also been proposed as sources of
ultrahigh energy protons~\cite{Arons:2002yj}. Galactic pulsars have
been suggested as the sources of cosmic rays around the knee region up
to the
ankle~\cite{Karakula:1974rz,Bednarek:1997cn,Bednarek:2001av,Giller:2002tw,Bednarek:2004wp}. That iron nuclei accelerated in the fastest spinning young neutron
stars could explain the observed cosmic rays above the ankle in a
Galactic source scenario was proposed in~\cite{Blasi:2000xm}.

The acceleration mechanism in a young pulsar is unipolar induction:
In the out-flowing relativistic plasma, the combination of the fast star rotation and its strong magnetic field can induce, in principle, 
potential differences of order $\phi = \Omega^2 \mu/c^2$, 
where $\mu = BR^3_*/2$, $B$ is the surface dipole field strength and $R_*$ is the pulsar radius.
Provided that particles of charge $Z$ can experience a fraction $\eta$ of that potential, 
they will be accelerated to the energy 
\begin{equation}
E(\Omega)=Ze\phi\eta = 
   \frac{Z}{26} \frac{\eta}{0.03} \left(\frac{\Omega}{10^4{\rm s^{-1}}}\right)^2 \frac{\mu}{10^{30.5}{\rm cgs}} \times 10^{20}\,{\rm eV}\,.
\end{equation}

The potential success of a UHECR source scenario lies in its ability to reproduce these four observations: 
{\it (i)} the energy spectrum; {\it (ii)} the composition; 
{\it (iii)} the anisotropy, and 
{\it (iv)} a rate of sources consistent with the population studies inferred from other astronomical observations.
Newly-born pulsars are natural candidates to reproduce points {\it
  (ii)} and {\it (iii)}, due to their iron-peaked surface (if the
composition at the highest energies proves to be heavy as suggested by
Auger) and their transient nature.  Point {\it (iv)} is not daunting,
as newborn pulsars are copiously produced in supernovae.  Point {\it
  (i)} is challenged by the fact that the toy model of unipolar
induction generates a hard spectrum that does not fit the observed
UHECR spectrum.  However, the slope could be naturally softened during
the escape from the supernova envelopes of the 0.01\% of the ``normal''
(as opposed to binary millisecond) pulsar birth rate required to
achieve the requisite flux.  Results obtained in~\cite{Fang:2012rx} 
suggest that all four points could be reasonably achieved in the
extragalactic rotation-powered pulsar scenario.  The highest energy
protons and light elements can traverse only very dilute pulsar
envelopes.  However, iron nuclei appear able to escape from the
supernova envelope with energies above $10^{20}$~eV.  The escaped
spectrum displays a transition from light to heavy composition at a
few EeV, and, due to the production of secondary nucleons, a softer
slope than the initially injected one, two results which enable a good
fit to Auger observations.  The flux up to the ankle is mainly fit by
Galactic newborn pulsars, while the flux above the ankle mainly
derives from extragalactic newborns~\cite{Fang:2013cba}.

The transient nature of the source makes direct source identification
very difficult.  The deflection in the extragalactic magnetic fields
should indeed induce important time delays ($\sim 10^4$~yr for one
degree deflection over 100 Mpc) between charged particles and the
photons propagating in geodesics, so that the sources should already
be extinguished when cosmic rays are detected on Earth.  However,
neutrinos from a single close-by source born within $\sim 5$~Mpc may
be detectable at IceCube~\cite{Fang:2012rx}. Moreover, while the
overall background neutrino flux from newborn pulsars would be at
least an order of magnitude smaller for iron than for proton
injection, the resulting level of the diffuse neutrino flux might
still be detectable with the IceCube experiment. Future IceCube data
will provide a decisive test for the young pulsar origin of
UHECRs~\cite{Fang:2013vla}.

\subsection{Cosmogenic Neutrinos from Ultra-High Energy Cosmic Rays}

It is natural to ask whether interactions of cosmic rays of the highest observed energies, above $10^{18}$~eV, can generate 
a spectrum of neutrinos consistent with the data~\cite{Barger:2012mz}. This, however, is not the case~\cite{Roulet:2012rv,Laha:2013lka,Aartsen:2013bza}. 
There is a significant uncertainty in predicted spectra of cosmogenic neutrinos that must accompany the observed spectra of UHECR.  However, normalizing the expected neutrino flux to that observed by IceCube at 1~PeV leads to an excessive prediction for EeV neutrinos, which are not observed.  The spectral shape of requisite cosmic rays, implied by the observation of PeV neutrinos and by non-observation of EeV neutrinos, is not consistent with 
the observed spectrum of UHECR above 1 EeV~\cite{Roulet:2012rv,Laha:2013lka,Aartsen:2013bza}.  Cosmic rays at lower enegies, such as those discussed in Section~\ref{section:blazars}, evade the constraint because the spectrum of injected protons does not extend beyond 1~EeV.

\section{Cosmic Probes of  Fundamental Physics}
\label{section-5}

In this section we explore new physics processes which could give rise
to the observed neutrino flux. In particular we discuss: {\it (i)} the
potential of superheavy dark matter to produce a monochromatic
neutrino signal, interestingly not inconsistent with current IceCube
observations of two isolated events at about the same
energy~\cite{Feldstein:2013kka,Esmaili:2013gha,Bai:2013nga}; {\it
  (ii)} a model in which a leptoquark (of mass $\approx 0.6~{\rm
  TeV}$), coupling the tau-flavor to light quarks, enhances the CC
interaction for shower production at 1~PeV, and the NC interactions at
lower energies. This model is currently consistent with the possible
gap in the spectrum, and the paucity of muon
tracks~\cite{Barger:2013pla}; {\it (iii)} some exotic neutrino
property (such as neutrino decay or pseudo-Dirac neutrino states)
which could reduce the muon neutrino flux at high energies from
distant sources~\cite{Pakvasa:2012db}. However, it is important to
stress that with present statistics the observed neutrino flavor
ratios are consistent with the Standard Model, and there are not yet
signs of new physics in the data~\cite{Chen:2013dza}. To date,
neutrino observations have shown an absence of certain
Lorentz-Violating operators to extraordinary
energies~\cite{Stecker:2013jfa,Diaz:2013wia}, and no sign of Planck scale dissipative
phenomena~\cite{Liberati:2013usa}.

\subsection{Superheavy Dark Matter Decay} 

Some features of the IceCube results point to a tantalizing possibility that PeV neutrinos may come from decays of dark matter particles with masses of a few PeV~\cite{Feldstein:2013kka}.  These features may lack statistical significance at present, but they will be tested very soon with upcoming new data.    The lack of events above a PeV and, in particular, in the vicinity of the Glashow resonance, suggests that the spectrum should decrease significantly at the energy of a few PeV.  Spectra from dark matter decays (and annihilations) always exhibit a sharp cutoff determined by the particle mass.  Furthermore, the two PeV events appear to have identical energies, up to experimental uncertainties.  A line in the neutrino spectrum would be a “smoking gun” signature for dark matter.   A monochromatic neutrino line should be accompanied by a continuous spectrum of lower-energy 
neutrinos~\cite{Feldstein:2013kka}, which can explain both the PeV events and some of the sub-PeV events, a conclusion that 
emerges from particle physics considerations~\cite{Feldstein:2013kka} and appears to be in agreement with the data~\cite{Esmaili:2013gha,Esmaili:2012us}.  Finally, angular distribution of arrival directions is consistent with dark matter decay~\cite{Bai:2013nga}.

The existence of dark matter is confirmed by a number of independent observations, and many particle physics candidates have been proposed~\cite{Feng:2010gw}.  In general, both annihilations and decays of dark matter particles could be considered for explaining the IceCube results.  However, since the $m_{\rm DM}\sim$~PeV mass scale is determined by the PeV neutrino energies, annihilations can be ruled out~\cite{Feldstein:2013kka}.  
For dark matter with an annihilation cross section into neutrinos saturating the unitarity limit, $\sigma_{\rm Ann} \leq 4 \pi/(m_{\rm DM}^2 v^2)$, the event rate expected at a neutrino telescope of fiducial volume $V$ and nucleon number density $n_{\rm N}$ is 
\begin{eqnarray}
\Gamma_{\rm events}
\sim
V \, L_{\rm MW} \, n_{\rm N} \, \sigma_{\rm N} \,
\left(\frac{\rho_{\rm DM}}{m_{\rm DM}}\right)^2
\langle \sigma_{\rm Ann} v \rangle
\lesssim 10^{-3} \,
\left(\frac{1 \, {\rm PeV}}{m_{\rm DM}} \right)^{3.6} \,
(1~{\rm year})^{-1},
\end{eqnarray}
where we have used the fiducial volume of the IceCube experiment, and the energy-dependent neutrino-nucleon scattering cross section from Ref.~\cite{Gandhi:1998ri} for the energy $E\sim m_{\rm DM}$.  Obviously, dark matter annihilation cannot produce a sufficient number of events. 

However, decays of dark matter particles can produce the required flux for cosmologically acceptable decay times, much longer than 
the present age of the universe~\cite{Feldstein:2013kka,Esmaili:2013gha}. A systematic study of effective operators capable of producing the requisite decay signal leads to a number of very interesting dark matter candidates~\cite{Feldstein:2013kka}.  The list includes a gravitino with R-Parity 
violation, hidden sector gauge bosons, and singlet fermions and bosons
in extra dimensions.  

Even much heavier relic particles, with masses well above a PeV, can
generate the required neutrino spectrum from their decays if their
lifetime is much shorter than the present age of the
universe~\cite{Ema:2013nda}.  The spectrum of neutrinos is modified by
a combination of redshift and interactions with the background
neutrinos, and the observed spectrum can have a cutoff just above
1~PeV for a broad range of the relic particle masses, from $\sim
1$~PeV to $\sim 10$~EeV~\cite{Ema:2013nda}. Each of these
possibilities represents a new window on physics beyond the Standard
Model with profound implications for one's understanding of the
universe.

\subsection{Enhancement of Neutrino-Nucleon Cross Section}

A possible explanation of the PeV IceCube events is a resonant
enhancement of the neutrino cross-section. The Glashow resonance in
electron-antineutrinos scattering on electrons is one such
example~\cite{Glashow:1960zz}.  Shower events would result from the
hadronic decays of the produced
$W$-boson~\cite{Barger:2012mz,Bhattacharya:2012fh}. However, the
6.3~PeV energy of the Glashow resonance is too high to explain
the observed 1 PeV shower energies. 

Another resonance candidate is an $s$-channel leptoquark (LQ) in
neutrino scattering on light
quarks~\cite{Anchordoqui:2006wc,Berezinsky:1985yw}.  A leptoquark of
mass $\sim 0.6~{\rm TeV}$ that couples to $\tau$-lepton and down-quark
flavors provides a plausible explanation of the IceCube
data~\cite{Barger:2013pla}. The LQ resonance enhanced processes are
$\nu_\tau + q \to \hbox{\rm LQ } \to \tau + q’$ and $\nu_\tau + q \to
\hbox{\rm LQ } \to \nu_\tau + q$.

At PeV energies, upward-going neutrinos that pass through the Earth should
mainly be $\tau$-flavor, because the Earth is almost opaque to
electron-neutrinos and muon-neutrinos while $\tau$-neutrinos can be
regenerated via $\tau$-decays.  The $\tau$ decays to hadrons and
electrons, with a combined branching fraction of 82\%, produce shower
events, whereas only 18\% of $\tau$'s-decays give a muon-track.

How does this general expectation compare with the IceCube
observations?  The contribution from cascades and track topologies has
been studied in detail in~\cite{Winter:2013cla}. Above 20~TeV
deposited EM energy, there are fewer upward than downward events, as
expected from absorption of electron-neutrinos and muon-neutrinos by
the Earth.  Furthermore, all but one of the muon-track events are
upward or horizontal.  Above 150 TeV, there is only one muon-track
event, which is upward as compared to 6 shower events, 2 upward and 4
downward. The present statistics are low, but the IceCube data suggest
that mainly $\nu_\tau$ events are being seen above 150~TeV, in accord
with this leptoquark scenario.

The PeV IceCube events could be due to CC reactions with showers from
the hadronic $\tau$ decays and the hadron jet from the produced quark.
The observed shower energy would be a little less than the mass of the
leptoquark because the secondary neutrino from the $\tau$ decay is
undetected.  When the produced $\tau$ decays to a muon, giving a
track, or the $\tau$ decays to an electron, the shower energy is lower
than for events associated with the hadronic $\tau$-decays.  In the NC
reaction, the shower energy of the event will be approximately half
that of the CC reaction.  The shower energy gap between the PeV events
and the onset of lower energy events seems indicative of what is
expected from the LQ processes.

\begin{figure}[tbp]
\begin{minipage}[t]{0.49\textwidth}
\postscript{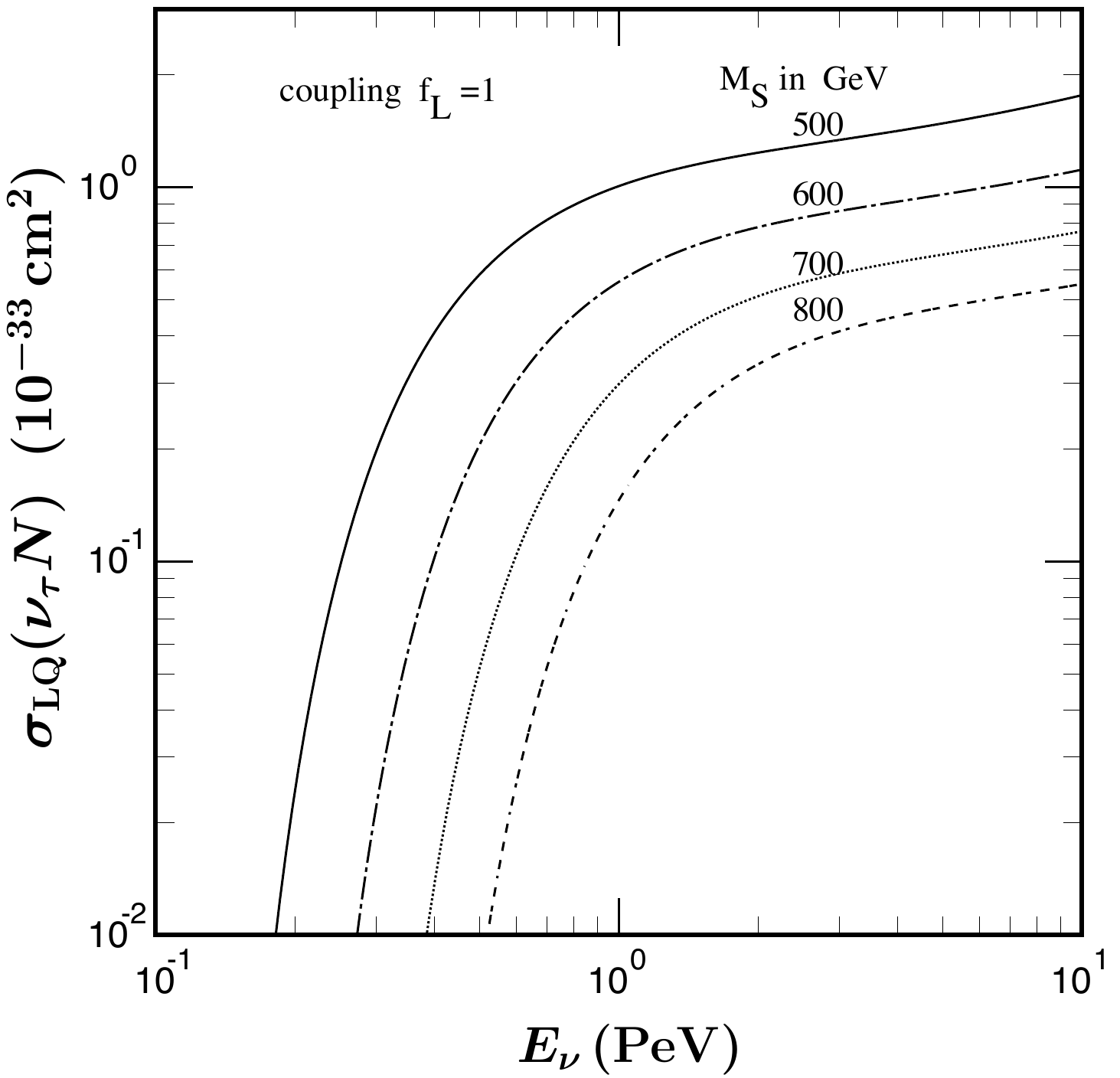}{0.99}
\end{minipage}
\hfill
\begin{minipage}[t]{0.49\textwidth}
\postscript{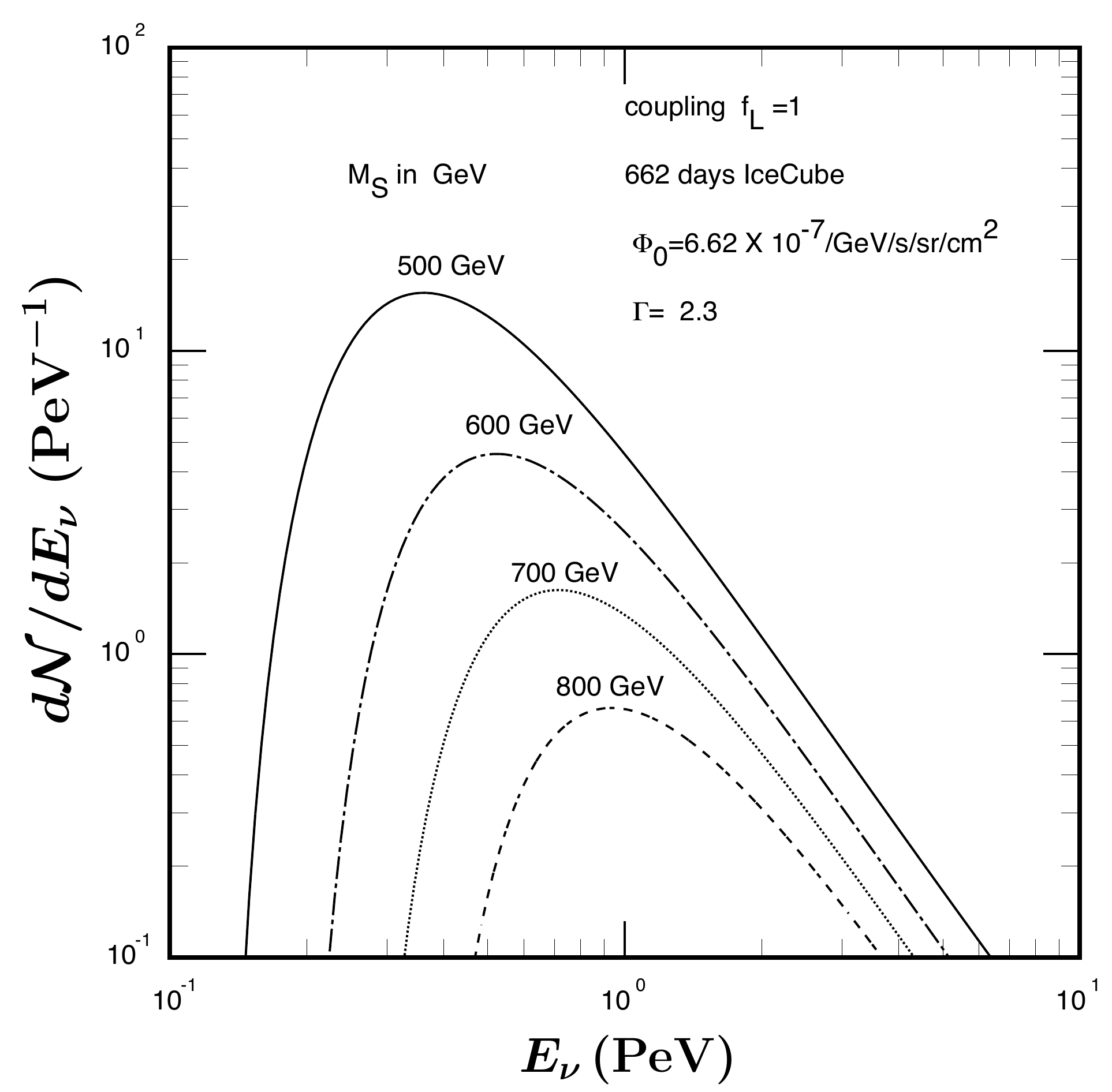}{0.99}
\end{minipage}
\caption{{\bf Left:} LQ production cross-section in $\nu_\tau N$
  scattering. {\bf Right:} Event rate distribution $d{\cal N}/dE_\nu$,
  from the LQ cross section convoluted with the flux of
  Eq.~(\ref{eqn:flux}). The CTEQ6.10 parton distributions at NLO are
  used in this calculation~\cite{Lai:2010vv}.}
\label{fig:LQ} 
\end{figure}

A general list of leptoquark models and the corresponding experimental
limits are given in~\cite{Rolli:2008zz}. For a
scalar leptoquark $S$ of charge $-\frac{1}{3}$, the Lagrangian interaction is
given by
\begin{equation} {\mathscr L}_{\rm LQ}= f_L S^\dagger (u,d)_L \
  \varepsilon\ \left(\begin{array}{c} \nu_\tau\\
      \tau\end{array}\right)_L + f_R S^\dagger u_R \tau_R + \hbox{
    h.c. } \, , \end{equation} where the Levi-Civita symbol
$\varepsilon$ antisymmetrizes the two $SU(2)$ doublets to match the
singlet $S$. The couplings $f_L,$ $f_R$ are the leptoquark couplings
to the left and right chiral quarks. In the narrow LQ width approximation, 
the neutrino cross-section is given by~\cite{Anchordoqui:2006wc,Alikhanov:2013fda,Doncheski:1997it} 
\begin{equation} 
\sigma_{\rm LQ}(\nu N)=\frac{\pi \, f_L^2}{2 \, M_S^2} \ x \, d_N (x, \mu^2) \ , 
\end{equation}
where $x = M_S^2/s$ is the parton fractional momentum, with $s = 2 m_N
E_\nu$. The down-quark parton distribution function $d_N(x,\mu^2)$ in
the target nucleon $N$ is evaluated at the scale $\mu^2 = M_S^2$ in
the leading order calculation. To obtain the corresponding rates for
each channel, we multiply the LQ production cross section (shown in
the left panel of Fig.~\ref{fig:LQ}) by the associated branching
fractions
\begin{eqnarray}
{\cal B}( S\to \nu_\tau d) = {\cal B}( S\to \tau_L u) &=&f_L^2/(2f_L^2+f_R^2)  \ , \nonumber \\
{\cal B}( S\to \tau_R u) &=&f_R^2/(2f_L^2+f_R^2)  \ .   \end{eqnarray}
The LQ width is found to be
\begin{equation}
\Gamma_{\rm LQ} = \frac{1}{16 \pi} \, M_S \, (2 f_L^2 + f_R^2) \,,
\end{equation}
which is a small fraction of its mass even for a unit coupling $f$, so
the narrow width approximation is justifiable.

As a benchmark, we consider the neutrino flux given in (\ref{eqn:flux}), with the power index $\Gamma=2.3$. For $t = 662$ days, the predicted number of events is given by
\begin{equation}
{\cal N}=  n_N t \Omega \int dE_{\nu} \, \sigma_{\rm LQ} 
\cdot {\cal B} (S \to ij) \cdot
\Phi_{\nu}(E_\nu) \ , 
\end{equation}
where $n_N = 6 \times 10^{38}$ is the effective target nucleon number
in IceCube and we take the solid angle of the full $4\pi$ coverage
($\Omega = 4 \pi$).  The event distribution $d{\cal N}/dE_\nu$, for
the coupling choice $f_L =1$, is shown in the right panel of
Fig.~\ref{fig:LQ}.  At a neutrino energy of $\sim 1~{\rm PeV}$ a few
cascade events are predicted for LQ mass of $\sim 0.6~{\rm TeV}$.

A leptoquark of 0.6 TeV mass can be probed at the
LHC~\cite{CiezaMontalvo:1998sk,Belyaev:2005ew,Blumlein:1996qp}. Based
on its pair production, the CMS/LHC search at 7
TeV~\cite{Chatrchyan:2012sv} for a scalar $\tau$-type LQ placed a
constraint $M_S \stackrel{>}{\sim} 525~{\rm GeV}$. Single LQ
production at the LHC occurs through the subprocesses $g u \to
\bar\tau S$ and $gd \to \bar\nu_\tau S$.  The down-type LQ, $S$,
subsequently decays into $ \tau u$ or $\nu_\tau d$, leading to the
final states $\bar\tau\tau u$, or $\bar\tau\nu_\tau d$, or
$\bar\nu_\tau\nu_\tau d$, etc. These subprocesses lead to distinctive
events of $\bar\tau\tau$ pair plus a jet or a monojet and missing
energy with or without a $\tau$.  Searches at LHC14 for these LQ
signals can confirm or reject the LQ interpretation of the PeV events
at IceCube.

\subsection{Neutrino Flavor Physics}

A natural question is whether the observation of 7 track ($T$) events
and 21 shower ($S$) events is consistent with the expected $1:1:1$
neutrino flavor signal. Denoting the fractional neutrino flux of
flavor $i$ by $\eta_{\nu_i}=\Phi_{\nu_i}/(\Phi_{\nu_e} +\Phi_{\nu_\mu}
+\Phi_{\nu_\tau})$, with $\Phi_{\nu_i}$ the neutrino flux of flavor
$i$, the ratio,
\begin{equation}
\frac{T}{T+S} \simeq \frac{(e_T/e_S)(\eta_{\nu_\mu} +0.2\eta_{\nu_\tau})}{(e_T/e_S)(\eta_{\nu_\mu} +0.2\eta_{\nu_\tau})+ \eta_{\nu_e} +0.8\eta_{\nu_\tau} +\sigma_{NC}/\sigma_{CC}}\,,
\end{equation}
where $e_T$ and $e_S$ are the efficiencies for detecting tracks and
showers, respectively.  Taking the efficiencies to be equal
(which is approximately true, since all 28 events have a vertex within the
instrumented volume), $\sigma_{NC}/\sigma_{CC}= 0.4$ for PeV
neutrinos, and $\eta_{\nu_e}=\eta_{\nu_\mu} =\eta_{\nu_\tau} =1/3$, we
get
\begin{equation}
\frac{T}{T+S}\simeq 0.286\,,
\end{equation}
which given the limited statistics is consistent with the observed
value of 0.25. Neverthless, with the promise of larger datasets, it is
worth considering deviations from the $1:1:1$ flavor mix, and there
are quite a few ways of achieving this. In the following, we draw from
Ref.~\cite{Pakvasa:2008nx}.

The simplest possibility is that initial flavor mix is not $1:2:0$. In
the damped muon case in which the initial flavor mix is $0:1:0$, the
final result after the oscillations are averaged out is $0.57:1:1$ on
arrival (for details, see~\ref{AII}). The ``beta'' beam which starts
out as $1:0:0$ becomes $2.5:1:1$ on arrival.  A``prompt'' beam from
heavy flavor decays which starts out as $1:1:0$ arrives as $1.27:1:1$.
These are sufficiently different from the universal mix that the
nature of the source can be easily distinguished. The two kinds of
production processes that lead to the initial flavor mix of $1: 2: 0$,
namely the $pp$ and $\gamma p$ interactions can also be distinguished
from each other, at least in principle~\cite{Anchordoqui:2004eb}.  In
the former case the flux of $\bar{\nu}_e$ relative to the total
neutrino flux is 1/6, whereas in the latter case it is 2/27; the
$\bar{\nu}_e$ flux can be measured at an incident energy of 6.3~PeV as
showers due to the Glashow resonance.

Neutrino decay is another way for the flavor mix to deviate
significantly from the democratic
mix~\cite{Beacom:2002vi,Beacom:2003zg} (see also~\cite{Anchordoqui:2007dp,Baerwald:2012kc}).  If the neutrino mass
hierarchy is normal and the source distances are large, the $\nu_2$
and $\nu_3$ mass eigenstates will have decayed away completely. For a
quasi-hierarchical mass spectrum, {\it i.e.}, $m_2, m_3\gg m_1$, the
daughter neutrino energy is much lower than that of the parent and the
final $\nu_1$ does not contribute to the flux at that energy; see {\it
  e.g.}, Ref.~\cite{Beacom:2002cb}. This may explain the absence of
$\nu_\mu$ events above 1~PeV~\cite{Pakvasa:2012db}.

In this picture, neutrinos originating from GRBs arrive at the earth
as pure $\nu_1$ whose flavor content is
$\Phi_{\nu_e}:\Phi_{\nu_\mu}:\Phi_{\nu_\tau} =|U_{e1}|^2:|U_{\mu
  1}|^2:|U_{\tau 1}|^2$~\cite{Pakvasa:1981ci}.  The current best fit
values for the neutrino mixing
parameters~\cite{Schwetz,GonzalezGarcia:2012sz,Fogli:2012ua} and the
unknown Dirac CP phase yield $|U_{\mu 1}|^2$ between 0.1 and 0.3 with
a central value of about 0.16.  This is a suppression beyond the
factor of two due to standard flavor oscillations, so that a
suppression of the muon neutrino flux by an order of magnitude is
possible.  Since the value of $|U_{e1}|^2$ lies between 0.65 and 0.72,
the $\nu_e$ flux is only slightly affected by the decays of $\nu_2$
and $\nu_3$. Note that $\Phi_{\nu_e}/\Phi_{\nu_\mu}$ ranges from 2.5 to 8 with a
central value of about 4, depending on the value of the phase
$\delta$.

Another possibility for deviations from the universal flavor mix arises in scenarios of pseudo-Dirac neutrinos in which each of the three neutrino mass eigenstates is a doublet
with mass differences smaller than $10^{-6}$~eV, thereby
evading detection~\cite{Wolfenstein:1981kw,Petcov:1982ya,Bilenky:1983wt}. In fact, the
only way to detect mass differences in the range $10^{-18}\, {\rm{eV}}^2 <\Delta
m^2 <10^{-12}$~eV$^2$ is by measuring flavor mixes of the high energy
neutrinos from cosmic sources. 

For large $L/E$, flavor ratios deviate from the
canonical value of 1/3 by
\begin{equation}
\delta P_\beta = \frac{1}{3}
\left [ \mid U_{\beta 1} \mid^2 \chi_1 \ + \mid U_{\beta 2} \mid^2
\chi_2 \ + \mid U_{\beta 3} \mid^2 \chi_3 \right ]\,,
\end{equation}
where $\chi_i = \sin^2 (\Delta m_i^2 L /4E)$, with the pseudo-Dirac mass-squared differences, 
\begin{equation}
\Delta m_i^2= (m_i^+)^2 - (m_i^-)^2 \,,
\end{equation} 
 of the nearly degenerate pair $\nu_i^+$ and
 $\nu_i^-$~\cite{Beacom:2003eu} (see also~\cite{Esmaili:2009fk,Esmaili:2012ac}).
The flavor ratios deviate from $1:1:1$ when one or two of the pseudo-Dirac oscillation modes is 
accessible.  In the limit where $L/E$ is so large that all three oscillating
factors have averaged to $1/2$, the flavor ratios return to $1:1:1$,
with only a net suppression of the measurable flux, by a factor of $1/2$. 

If neutrinos traverse regions with large magnetic fields and their
magnetic moments are large enough, the flavor mix can be
affected~\cite{Enqvist:1998un}. The main effect of the passage through a
magnetic field is the conversion of a given helicity into an equal
mixture of both helicity states.  This is also true in passage through
random magnetic fields~\cite{Domokos:1997cq}. It has been shown that a
magnetic field of 10 or more Gauss at the source can cause the
neutrinos to decohere as they traverse cosmic distances~\cite{Farzan:2008eg}.

If neutrinos are Dirac particles with comparable magnetic moments, 
then the effect of the spin-flip is to simply reduce the overall flux of all 
flavors by half, the other half becoming the sterile Dirac partners.
On the other hand, if neutrinos are Majorana particles, 
the flavor mix remains $1 : 1 : 1$ with the absolute flux unchanged.

What happens when large magnetic fields are present in or near the
neutrino production region?  In the case of Dirac neutrinos, the
situation is as above, and the outgoing flavor ratio remains $1 : 1 :
1$ with the absolute fluxes reduced by half.  In the case of Majorana
neutrinos, the initial flavor mix $1: 2: 0$ is unmodified at the
source because of the antisymmetry of the magnetic moment matrix, but
the final flavor mix after oscillations is still $1 : 1: 1$.

If neutrinos have 
flavor violating couplings to gravity then resonance
effects may allow one way transitions 
{\it e.g.}, $\nu_\mu \rightarrow \nu_\tau$ but not vice
versa~\cite{Minakata:1996nd,Barger:2000iv}. 
This can give rise to an anisotropic deviation of the 
$\Phi_{\nu_\mu}/\Phi_{\nu_\tau}$ from 1, becoming less than 1 for events coming from 
the direction of the Great
Attractor, while remaining 1 in other 
directions~\cite{Minakata:1996nd}. If such striking effects are not seen,
current bounds on such violations can be improved by
six to seven orders of magnitude.
 
Another possibility that can give rise to deviations of the flavor mix
from the canonical $1 : 1 : 1$ is the idea of mass-varying
neutrinos~\cite{Fardon:2003eh}, that was proposed to solve the cosmic
coincidence problem. Neutrino masses vary in such a way that the dark
energy density and neutrino energy density are related over cosmic
time scales.  The dark energy density is made neutrino-mass-dependent
by coupling a sterile neutrino and a light scalar field.  If the
sterile neutrino mixes with a flavor neutrino, the mass difference
varies along the path of propagation, with the potential for resonance
enhancement of the transition probability into the sterile neutrino,
and a resulting change in the flavor mix~\cite{Hung:2003jb}. For example, if
only one resonance is crossed {\it en route}, the heaviest (mostly) flavor
state may transform into the (mostly) sterile state, thus changing the
flavor mix to $1-|U_{e1}| ^2 : 1-|U_{\mu 1}|^2 : 1-|U_{\tau 1}|^2
\approx 0.4: 1:1$ for the inverted hierarchy and to $1-|U_{e3}| ^2:
1-|U_{\mu 3}| ^2 : 1-|U_{\tau3}|^2 \approx 2 : 1 : 1$ for the normal
hierarchy.

Complete quantum decoherence gives rise to a flavor mix of $1:1:1$, 
which is identical to the case of averaged oscillations. The
distinction is that complete decoherence always leads to this result,
whereas averaged oscillations give this result only for an initial
flavor mix of $1:2:0$.  Therefore, to find evidence for decoherence
requires a source which has a different flavor mix . An example is the
``beta'' beam source with an initial flavor mix of $1:0:0$. In this
case decoherence gives the universal $1:1:1$ mix whereas averaged
oscillations give $2.5:1:1$~\cite{Anchordoqui:2005gj}. The two cases can be
easily distinguished from each other.
 
Violations of Lorentz invariance and/or $CPT$ invariance
can change the final flavor mix from the
universal mix significantly. With a specific
choice of a modified dispersion relation due to Lorentz
Invariance Violation, the effects can be dramatic. For example,
the final flavor mix at sufficiently high energies can become
$7: 2: 0$~\cite{Hooper:2004xr}.

To summarize, a measurement of a flavor ratio different from $1:1:1$
would suggest new physics.  If measurements of the flavor mix at Earth
of high energy astrophysical neutrinos find it to be
\begin{equation}
\Phi_{\nu_e} : \Phi_{\nu_\mu} :\Phi_{\nu_\tau} = 
\alpha : 1 : 1\,,
\end{equation}
then: {\it (i)}~$\alpha \approx 1$  
confirms our knowledge of the neutrino
mixing matrix and our prejudice about the 
production mechanism; {\it (ii)}~$\alpha \approx 1/2$ indicates a pure $\nu_\mu$ source 
and conventional mixing; {\it (iii)}~$\alpha > 1$ indicates that neutrinos with a normal hierarchy are 
decaying; and {\it (iv)}~a value of $\alpha$ between 2.5 and 10, and a
deviation of the $\nu_\mu/\nu_\tau$ ratio from 1 (between 0.2 to 4) 
can yield valuable 
information about the $CP$ phase $\delta$, whereas 
a value of $\alpha$
between 0.7 and 1.5 can probe 
pseudo-Dirac 
$\Delta m^2$ smaller than $10^{-12}$~eV$^2$. These results have no dependence on the initial flavor mix, and are
consequently independent of the production model. One either learns
about the production mechanism and the initial flavor mix, or about
neutrino properties.

\section{Looking Ahead}
\label{section-6}

In summary,  a decades-long development program, culminating in the
fruits of the IceCube Collaboration, have introduced not only the
dawn of the age of neutrino astronomy and astrophysics but also
provides a beacon for future directions in particle physics. We
have reviewed the possible origins of the soon-to-be famous 28
IceCube neutrino events~\cite{Aartsen:2013bka,Aartsen:2013pza}. Thus far the IceCube excess is consistent with both
Galactic and extragalactic origin(s).

In particular, we showed that the IceCube neutrino excess is
consistent with optically thin Galactic sources producing an unbroken
power-law with a spectral index of $\Gamma =
2.3$~\cite{Anchordoqui:2013qsi}. A shallower spectrum would
overproduce events in the null region above $\sim 2~{\rm PeV}$, and
hence require a cutoff. We employed this hypothesis to argue that
cosmic neutrinos are more likely to arise from $pp$ interactions than
$p\gamma$ interactions. We also explored the validity of this
hypothesis considering other factors including the spectral shape and
anisotropy of baryonic CRs around and above 3~PeV predicted by various
Galactic magnetic field models as well as the compatibility of nearby
(Galactic) sources with bounds on the photon fraction measured by the
CASA-MIA Collaboration.  We conclude from the CR anisotropy searches
that a Galactic magnetic field model favoring a CR injection index of
$\alpha = 2.3$ is more likely to be correct than the usual $\alpha
\simeq 2$ of a Fermi engine. Furthermore, we find that existing photon
bounds do not rule out a Galactic origin at optically thin sources as
a viable model for the IceCube excess.  In all, given present
statistics, the Galactic hypothesis is plausible with an unbroken
power without cutoff. Note that these conclusions apply only if the
neutrino sources are Galactic, since, for instance, we do not know the
extragalactic CR power at PeV energies.

It is inspiring to realize that we are at the cusp of a new era of
multimessenger astronomy, now with 
cosmic neutrinos thrown into the
mix. Indeed a recent analysis~\cite{Murase:2013rfa} shows that a
bound on the steepness of an extragalactic $pp$-generated neutrino
spectrum can be estimated based on \textit{Fermi}-LAT observations of the
isotropic $\gamma$ ray background, showing $\Gamma \lesssim 2.1-2.2$
is required for consistency with extragalactic neutrino origin.  It is
notable then that IceCube spectrum alone will ultimately reveal
something about the source of cosmic neutrinos, even in the absence of
an observation of an excess in specific regions of the sky.

\begin{figure}[!t]
\centering
\postscript{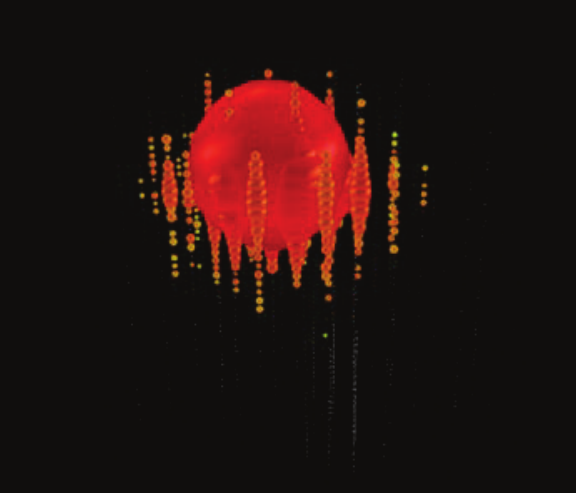}{0.8}
\caption{Event display showing Big Bird, with 378 optical modules hit. Each sphere shows a hit optical module.  The size of the spheres shows the number of photoelectrons observed by the DOM, while the color indicates the time, with red being earliest, and blue latest.  Figure courtesy of the IceCube Collaboration.
\label{fig:BigBird}}
\end{figure}

We have seen that various classes of extragalactic sources are
consistent with emission spectrum $\propto E_\nu^{-2}$ in the energy
range of interest.  These include GRBs~\cite{Cholis:2012kq},
AGNs~\cite{Stecker:2013fxa,Kalashev:2013vba,Kistler:2013my}, and
hypernova remnants in star forming galaxies~\cite{Liu:2013wia}. The
distribution of arrival directions would provide a way of
distinguishing among these models.  In the case of GRBs, coincidence
in time will be the key.  In the case of possible correlations with
AGN, we can learn about the intergalactic magnetic field via -
possible correlations with associated $\gamma$ rays, as a result of
photon showering during propagation to Earth.

In a few years of data taking IceCube will collect sufficient
statistics to ascertain whether there are structures in the
spectrum. A gap could be associated with NC and CC processes of a new
physics resonance~\cite{Barger:2013pla}. Alternatively, a neutrino
spectral line could then point to the decay or annihilation of the
elusive dark matter particles~\cite{Feldstein:2013kka}.

Examination of the data collected in 2012 has begun.  One very high
energy event, called Big Bird, appeared in the 10\% of the data that
was used to tune the selection cuts.  This event is shown in
Fig.~\ref{fig:BigBird}.  A total of 378 DOMs were hit, making it the
brightest neutrino event thus far observed. This is suggestive that
the energy spectrum will continue, in one form or another, beyond the
1~PeV limit found thus far, perhaps into into the sweet spot for
$\nu_\tau$-detection. The potential to access the region of high
sensitivity to $\nu_\tau$, together with growing statistics for
$\nu_\mu$ and $\nu_e$ events will wedge open a portal for flavor
physics exploration, making the coming era watershed years for
electroweak physics accessible via high energy cosmic
neutrinos~\cite{Pakvasa:2012db}.

The discovery of cosmic rays just over 100 years ago was not only
remarkable in its own right, but it also provided a cornerstone for
the field of particle physics.  Many of the most important early
breakthroughs in particle physics were achieved through observation of
cosmic rays, including the watershed discoveries of antimatter, the
pion, the muon, the kaon, and several other particles. In this
article, we have both reviewed the nascent field of cosmic neutrino
astronomy and considered some of the potential ways CR science will
once again point the way in the quest to understand Nature at its most
fundamental.

\section*{Acknowledgments}
We would like to thank Francis Halzen, Stephen King, Soeb Razzaque,
and the IceCube Collaboration for allowing us to use various figures
from their papers in this review.  LAA is supported by U.S. National
Science Foundation (NSF) CAREER Award PHY1053663 and by the National
Aeronautics and Space Administration (NASA) Grant No. NNX13AH52G. VB is supported by the U. S. Department of Energy (DoE)
Grant No. DE-FG-02- 95ER40896.  IC and DH are supported by DoE. HG is
supported by NSF Grant No. PHY-0757959.  AK is supported by DoE Grant
No. DE-SC0009937 and by the World Premier International Research
Center Initiative (WPI Initiative), MEXT, Japan.  JGL and SP are
supported by DoE Grant No. DE-FG02-04ER41291.  DM is supported by DoE
Grant No. DE–FG02–13ER42024.  TCP is supported by NSF Grant
No. PHY-1205854 and NASA Grant No. NNX13AH52G. TJW is supported by
DoE Grant No. DE-FG05-85ER40226.  Any opinions, findings, and
conclusions or recommendations expressed in this material are those of
the authors and do not necessarily reflect the views of the NSF, DoE,
or NASA.

\appendix

\section{Harmonic Analysis for Anisotropy Searches}
\label{AI}

Cosmic ray detectors which experience stable operation over a period
of a year or more attain a uniform exposure in right ascension,
$\alpha$. In such a case, the right ascension distribution of the flux
arriving at a detector can be characterized by the amplitudes and
phases of its Fourier expansion,
\begin{equation}
J (\alpha) = J_0 [1 + r \cos (\alpha - \phi)  + r' \cos (2 (\alpha - \phi')) + \dots \,] \, .
\end{equation}
For $N$ measurements $\alpha_i$, the first harmonic amplitude $r$ and
its phase $\phi$ can be determined by applying the classical
Rayleigh formalism~\cite{Linsley:1975kp},
\begin{equation}
r = \sqrt{x^2 + y^2}\,, \quad \quad \quad  \quad \quad \quad \quad \phi = {\rm arctan} \frac{y}{x} \, ,
\end{equation}
where
\begin{equation}
x = \frac{2}{{\cal N}} \sum_{i=1}^{N}  \, w_i \, \cos \alpha_i \,,
\quad \quad \quad  \quad \quad \quad \quad
y =
\frac{2}{{\cal N}} \sum_{i=1}^{N}  \, w_i\,\,
\, \sin \alpha_i\,,
\label{eqn:fh}
\end{equation}
${\cal N} = \sum_{i=1}^{N} w_i$ is the normalization factor, and the
weights, $w_i = \omega^{-1}(\delta_i)$, are the reciprocal of the
relative exposure, $\omega$, as a function of the declination,
$\delta_i$~\cite{Sommers:2000us}. As deviations from an uniform right
ascension exposure are small, the probability $P(>r)$ that an
amplitude equal or larger than $r$ arises from an isotropic 
distribution can be approximated by the cumulative distribution
function of the Rayleigh distribution $P(>r) = \exp (-k_0),$ where $k_0
= {\cal N} \, r^2/4$. 

The first harmonic amplitude of the  right ascension distribution
can be directly related to the amplitude $|\bm{\delta}|$ of a dipolar
distribution of the form 
\begin{equation}
J(\alpha,\delta) = (1+ |\bm{\delta}| \ \bm{\hat d}
\bm{\cdot} \bm{\hat u}) \,  J_0 \,, 
\end{equation}
where $\bm{\hat u}$ denotes the unit vector in the direction ($\alpha, \delta$)
of the sky  and $\bm{\hat d}$ denotes the unit
vector in the direction of the dipole. We can rewrite $x$, $y$, and
$\mathcal{N}$ as
\begin{eqnarray}
x&=& \frac{2}{\mathcal{N}} \int_{\delta_{\rm min}}^{\delta_{\rm max}} d\delta \int_0^{2\pi}
d\alpha \cos \delta \ J(\alpha,\delta) \ \omega(\delta) \cos \alpha, \nonumber \\  
y&=& \frac{2}{\mathcal{N}} \int_{\delta_{\rm min}}^{\delta_{\rm max}} d\delta \int_0^{2\pi}
d\alpha \cos \delta \ J(\alpha,\delta) \ \omega(\delta) \sin \alpha,  \label{corolo} \\ 
\mathcal{N}&=& \int_{\delta_{\rm min}}^{\delta_{\rm max}} d\delta \int_0^{2\pi}
d\alpha \cos \delta \ J(\alpha,\delta) \ \omega(\delta) \, .\nonumber
\end{eqnarray}
In (\ref{corolo}) we have neglected  the small dependence on right 
ascension in the exposure. Next, we write the angular dependence in 
$J(\alpha,\delta)$ as 
\begin{equation}
\bm{ \hat d} \bm{ \cdot} \bm{ \hat u_i} = \cos \delta_i \cos \delta_0 
\cos (\alpha_i-\alpha_0) + \sin \delta_i \sin \delta_0 \,,
\end{equation}
where $\alpha_0$ and $\delta_0$ are the right ascension and
declination of the direction where the flux is maximum, and $\alpha_i$
and $\delta_i$ are the right ascension and declination of the $i$th
event. Performing the $\alpha$ integration in (\ref{corolo}) it
follows that
\begin{equation} 
\label{eqn:amplitudes}
r=\left| \frac{A \, \delta_\perp}{1+B \delta_\parallel} \right| \,,
\end{equation}
where $\delta_\parallel= \bm{\delta} \sin{\delta_0}$ is the component of the dipole 
along the Earth rotation axis, and \mbox{$\delta_\perp= \bm{\delta}
\cos{\delta_0}$} is the component in the equatorial
plane~\cite{Aublin:2005nv}. The coefficients $A$ and $B$ can be
estimated from the data as the mean values of the cosine and the sine
of the event declinations,
\begin{equation}
A =\frac{\int d\delta\,\omega(\delta) \cos^2 \delta}
{\int d\delta\,\omega(\delta) \cos \delta} \quad \quad{\rm and} \quad \quad B
=\frac{\int d\delta\,\omega(\delta) \cos \delta \sin \delta} {\int
  d\delta\,\omega(\delta) \cos \delta} \, .
\end{equation}
For a dipole amplitude $|\bm{\delta}|$, the measured amplitude of the
first harmonic in right ascension $r$ thus depends on the region of
the sky observed, which is essentially a function of the latitude of
the observatory and the range of zenith angles
considered. In the case of a small $B \delta_\parallel$ factor, the
dipole component in the equatorial plane is obtained as
$\delta_\perp\simeq r/A$.  The phase $\phi$ corresponds to the right
ascension of the dipole direction $\alpha_0$.

\section{Cosmic Neutrino Flavor Ratio}
\label{AII}

The discovery of neutrino oscillations provoked quite a revolution in
elementary particle physics, demonstrating the need for physics beyond
the Standard Model.  The flavor oscillation patterns can be convincingly
interpreted as a non-trivial mixing among neutrino mass
eigenstates, with a small ``solar'' mass splitting $\Delta
m^2_\odot\simeq7.65\times10^{-5}~{\rm eV}^2$ and a large ``atmospheric''
splitting $\Delta m^2_{\rm atm}\simeq 2.40\times10^{-3}~{\rm eV}^2$~\cite{GonzalezGarcia:2012sz}.

\begin{figure}[tbp]
\postscript{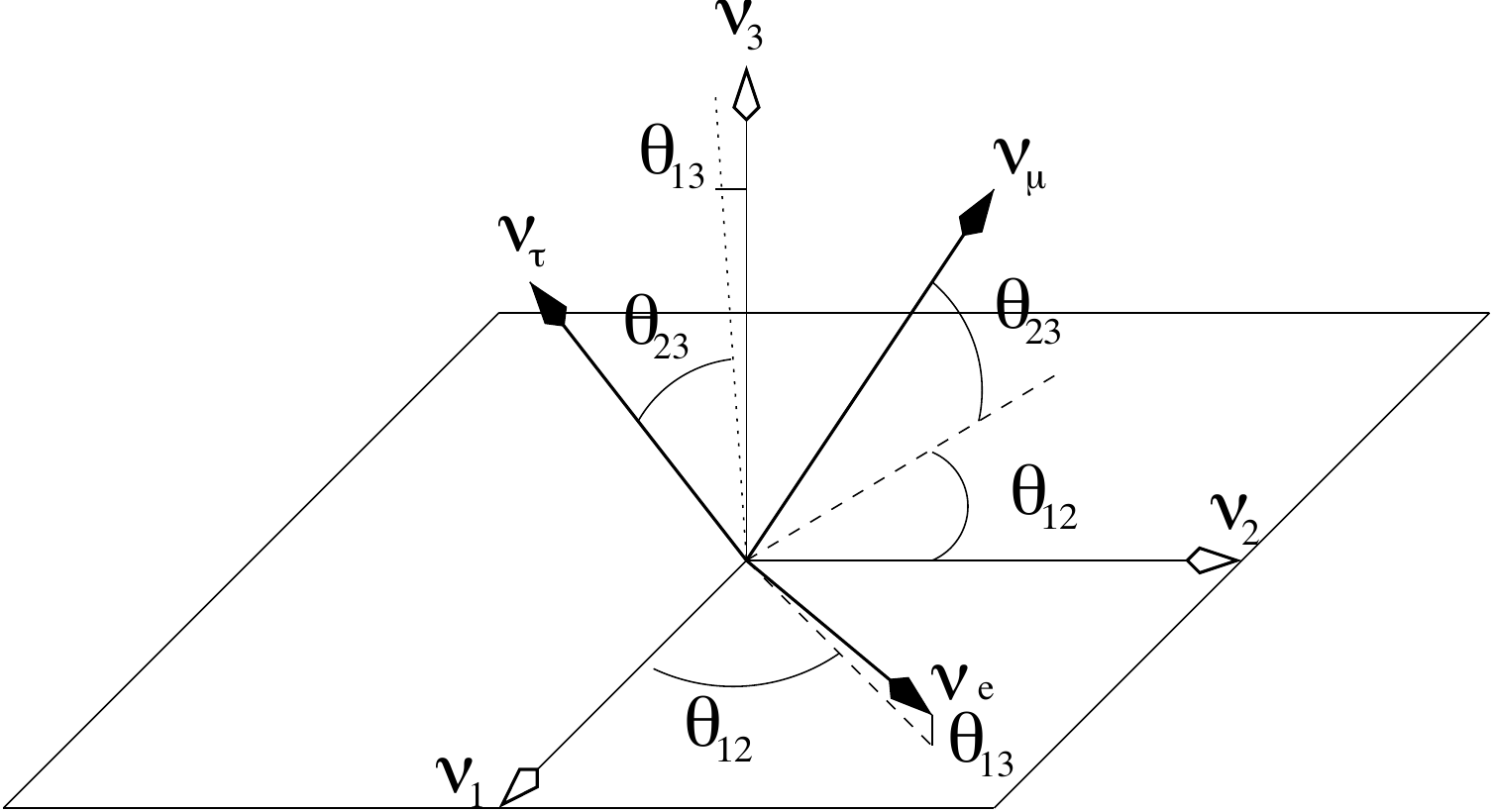}{0,7}
\caption{Display of the three mixing angles that characterize the orientation of 
the flavor axes relative to mass axes. From Ref.~\cite{King:2013eh}.}
\label{fig:SKing}
\end{figure}

The superposition of neutrino mass eigenstates $\nu_j$ ($j = 1, 2, 3,
\dots$) produced in association with the charged lepton of flavor
$\alpha,$
\begin{equation}\label{Uneutrino}
|\nu_\alpha\rangle =
\sum_j U_{\alpha j}^* |\nu_j\rangle,
\end{equation}
is the state we refer to as the neutrino of flavor $\alpha$, where
$U_{\alpha j}$'s are elements of the unitary neutrino mass-to-flavor
mixing matrix fundamental to particle physics, the so-called 
Pontecorvo-Maki-Nagakawa-Sakata (PMNS)
matrix~\cite{Pontecorvo:1957cp,Pontecorvo:1967fh,Maki:1962mu}. The unitary PMNS mixing matrix has 9
degrees of freedom, which are reduced to 6 after absorbing three global
phases into re-definitions of the three charged lepton states, $e, \mu, \tau$.
(For Majorana neutrinos, no further phases may be absorbed, while for Dirac neutrinos, 
two further relative phases among the three neutrinos may be absorbed by neutrino-field redefinitions.)
With six undetermined parameters, the neutrino mixing matrix $\mathbb U_{\rm PMNS}$ is conveniently
parametrized  by three Euler rotations
$\theta_{12}$, $\theta_{23}$, and $\theta_{13}$, and three
$CP$-violating phases $\delta$, $\alpha_1$ and $\alpha_2$,
\begin{equation}\label{stparam}
\mathbb{U}_{\rm PMNS} = R_{23}(\theta_{23})\, \left(\begin{array}{ccc}
    c_{13} & 0 & s_{13}e^{-i\delta} \\ 0 & 1 & 0 \\-s_{13}e^{i\delta}&0&c_{13} \end{array}
  \right)\,R_{12}(\theta_{12})\\\times {\rm diag}(e^{i\alpha_1/2},e^{i\alpha_2/2},1)\,,
\end{equation}
where we used the abbreviations $\sin\theta_{ij}=s_{ij}$ and
$\cos\theta_{ij}=c_{ij}$. $R_{ij}$ denotes a rotation in the
$\nu_i\nu_j$-plane, see Fig.~\ref{fig:SKing}. The ``Marjorana" phases $\alpha_1$ and $\alpha_2$ are 
unique to Majorana neutrinos, {\it i.e.} neutrinos which are their own anti-particles.
Note, that the phase $\delta$ (``Dirac phase'') appears only in combination with a
non-vanishing mixing angle $\theta_{13}$. Additional details are eloquently 
discussed in~\cite{Barger:2012pxa}. For the curious
public, Ref.~\cite{Weiler:2013rta} provides a very readable account.

The density matrix of a flavor state, $\rho_\alpha=|\nu_\alpha\rangle\langle\nu_\alpha| $, 
can be expressed in terms of mass eigenstates by
$\rho_\alpha=\sum_{i,j}U^*_{\alpha i} U_{\alpha j}|\nu_i\rangle\langle\nu_j|$. 
This is a pure quantum system, therefore the density matrix satisfies 
${\rm Tr}~\rho^2 \, = \,{\rm Tr}~\rho = 1.$
The time evolution of the density matrix,
\begin{equation}
\frac{\partial \rho}{\partial t} = - {\rm i}\, [H,\, \rho] \,\,,
\label{liouville}
\end{equation}
is governed by the Hamiltonian of the system, 
\begin{equation}\label{freeH}
H \simeq  \sum_i\frac{m_i^2}{2E_\nu}\Pi_i\,,
\end{equation}
where we have introduced the projection operator $\Pi_i \equiv |\nu_i\rangle\langle\nu_i|$.\footnote{Dissipative effects due to charged current
  interactions in matter can be simply included by an extra term
  $-\sum_\alpha \frac{1}{2\lambda_\alpha}\lbrace\Pi_\alpha,\rho\rbrace$
  to the r.h.s.~of Eq.~(\ref{liouville}), where $\Pi_\alpha =
  |\nu_\alpha\rangle\langle\nu_\alpha|$ and $\lambda_\alpha$ is the
  dissipation length.} Substituting the Hamiltonian~(\ref{freeH}) into (\ref{liouville})  we obtain
\begin{equation}\label{freeoscillation}
\frac{\partial \rho_{ij}}{\partial t} = \frac{{\rm i}\Delta m_{ij}^2}{2E_\nu}\rho_{ij}
\,,
\end{equation} 
where $\Delta m_{ij}^2\equiv m_i^2-m_j^2$.
 For the initial
condition $\rho_\alpha(0) = \Pi_\alpha$, the density matrix at a
distance $L$ is given by
\begin{equation}\label{freesolution}
\rho_\alpha(L)
 =  \sum_{i,j}U^*_{\alpha i}U_{\alpha j}\exp\left(\frac{{\rm i}\Delta m^2_{ij}L}{2E_\nu}\right) \, |\nu_i\rangle\langle\nu_j| \, .
\end{equation}
Therefore, after traveling a
distance $L$ an initial state $\nu_\alpha$ becomes a superposition of all
flavors, with probability of transition to flavor $\beta$ given by
$P_{\nu_\alpha \to \nu_\beta} = {\rm Tr} [\rho_\alpha(L) \Pi_\beta]$, or
equivalently~\cite{GonzalezGarcia:2007ib}
\begin{equation}
 P_{\nu_\alpha \to \nu_\beta} 
 =  \delta_{\alpha \beta} - 4 \sum_{i>j} \Re \, 
(U_{\alpha i}^*\, U_{\beta i}\, 
U_{\alpha j} \, U_{\beta j}^*) \, \sin^2 \Delta_{ij}   +  2
\sum_{i>j} \Im \, (U_{\alpha i}^*\, U_{\beta i}\, 
U_{\alpha j} \, U_{\beta j}^*) \, \sin 2 \Delta_{ij} \,\,.
\label{Palphabeta}
\end{equation}
The oscillation phase $\Delta_{ij}$ is conveniently parameterized as
\begin{equation}
\nonumber\Delta_{ij}  =  \frac{\Delta m_{ij}^2 L}{4E_\nu} 
  \simeq  1.27\,\left(\frac{\Delta m_{ij}^2}{\rm eV^2}\right)\left(\frac{L}{\rm km}\right)\left(\frac{E_\nu}{\rm GeV}\right)^{-1}\,.
\end{equation}
Note, that the third term in Eq.~(\ref{Palphabeta}) comprises
$CP$-violating effects, {\it i.e.} this term changes sign
for the antineutrino process $P_{\overline\nu_\alpha \to \overline\nu_\beta}$,
corresponding to the replacement $U\to U^*$. For the standard
parameterization (\ref{stparam}), the single $CP$-violating
contribution is attributable to the Dirac phase $\delta$;
oscillation experiments are {\it not} sensitive to Majorana phases.

For many years, the sparse data on the angle $\theta_{13}$ allowed
consistency with zero. However, in Spring of 2012, the angle was
definitively measured to be nonzero (but still small on the scale of
$\theta_{23}\sim 45^\circ$ and $\theta_{12}\sim 35^\circ$),
$\theta_{13} \approx 9^\circ$~\cite{An:2012eh,Ahn:2012nd,An:2012bu}.
At present, the low statistics statistics of IceCube limits its
capacity to disentangle neutrino flavors with sufficient precision to
be sensitive to small $\theta_{13}$. To simplify the following
discussion, we will adopt maximal mixing for atmospheric $\nu_\mu
\leftrightharpoons \nu_\tau$ neutrinos ({\it i.e.} $\theta_{23} \sim
45^\circ$) along with a negligible $|U_{e3}|^2 = \sin^2(\theta_{13})$.
The latter approximation allows us to ignore $CP$ violation and assume
real matrix elements.  (The small effects of nonzero $\theta_{13}$
have been investigated in~\cite{Fu:2012zr}. Other small corrections
that we need not consider here arise from the fact that flavor ratios
at injection deviate from whole numbers due to subtle particle physics
effects~\cite{Lipari:2007su,Pakvasa:2007dc}; see also~\cite{Esmaili:2009dz}.)

With our simplifying assumptions in mind, one can define a mass basis as follows,
\begin{equation}
|\nu_1 \rangle = \sin \theta_\odot |\nu^\star\rangle +  \cos \theta_\odot |\nu_e\rangle \,\,,
\end{equation}
\begin{equation}
|\nu_2 \rangle =  \cos \theta_\odot |\nu^\star\rangle  -\sin \theta_\odot |\nu_e\rangle \,\,,
\end{equation}
and
\begin{equation}
|\nu_3 \rangle = \frac{1}{\sqrt{2}} (|\nu_\mu \rangle + |\nu_\tau \rangle) \,\,,
\label{3rd}
\end{equation}
where $\theta_\odot \equiv \theta_{12} \approx 34^\circ$ is the solar mixing angle~\cite{Ahmed:2003kj}, and 
\begin{equation}
|\nu^\star\rangle = \frac{1}{\sqrt{2}} (|\nu_\mu\rangle - |\nu_\tau \rangle)
\label{orthogonal}
\end{equation}
is the eigenstate orthogonal to $|\nu_3 \rangle.$ Inversion of the neutrino mass-to-flavor 
mixing matrix leads leads to
\begin{equation}
|\nu_e \rangle = \cos \theta_\odot |\nu_1\rangle - \sin \theta_\odot |\nu_2 \rangle
\end{equation}
and
\begin{equation}
|\nu^\star \rangle = \sin \theta_\odot |\nu_1\rangle + \cos \theta_\odot |\nu_2 \rangle \,\,.
\end{equation}
Finally, by adding Eqs.~(\ref{3rd}) and (\ref{orthogonal}) one obtains the $\nu_\mu$ flavor eigenstate,
\begin{equation}
|\nu_\mu \rangle = \frac{1}{\sqrt{2}} \left[ |\nu_3 \rangle + \sin \theta_\odot |\nu_1 \rangle + 
\cos \theta_\odot |\nu_2\rangle \right] \,\,,
\end{equation}
and by subtracting these same equations the $\nu_\tau$ eigenstate.

For real PMNS matrix elements (\ref{Palphabeta}) becomes
\begin{equation}
P(\nu_\alpha \to \nu_\beta) = \delta_{\alpha \beta} - 4 \sum_{i>j} U_{\alpha i}\, U_{\beta i}\, 
U_{\alpha j} \, U_{\beta j} \, \sin^2 \Delta_{ij}\,\, .
\end{equation}
In addition, for $\Delta_{ij} \gg 1$, the phases will be erased by uncertainties in $L$ and $E$. Consequently,
averaging over $\sin^2 \Delta_{ij}$ one finds the decohered flavor-changing probability 
\begin{equation}
P(\nu_\alpha \to \nu_\beta) = \delta_{\alpha \beta} - 2 \sum_{i>j} U_{\alpha i}\, U_{\beta i}\, 
U_{\alpha j} \, U_{\beta j} \,.
\label{paco}
\end{equation}
Now, using $2 \sum_{1>j} = \sum_{i,j} - \sum_{i=j},$ Eq.~(\ref{paco}) can be re-written as
\begin{eqnarray}
P(\nu_\alpha \to \nu_\beta) & = & \delta_{\alpha \beta} -  \sum_{i,j} U_{\alpha i}\, U_{\beta i}\, 
U_{\alpha j} \, U_{\beta j} \, +  \sum_{i} U_{\alpha i}\, U_{\beta i}\, 
U_{\alpha i} \, U_{\beta i} \nonumber \\
 & = & \delta_{\alpha \beta} - \left( \sum_{i} U_{\alpha i}  U_{\beta i} \right)^2 + \sum_{i}   
U_{\alpha i}^2  U_{\beta i}^2\,.
\label{PP}
\end{eqnarray}
Since $\delta_{\alpha \beta}$ = $\delta_{\alpha \beta}^2,$ the first and second terms in (\ref{PP}) 
cancel each other, yielding
\begin{equation}
P(\nu_\alpha \to \nu_\beta) = \sum_{i} U_{\alpha i}^2 \,\,U_{\beta i}^2 \,\,. 
\end{equation}
In matrix notation, we have 
\begin{equation}
P(\nu_\alpha \to \nu_\beta) = \mathbb{ P} \ \mathbb{P}^{\rm T}\,,
\label{propgn1}
\end{equation}
where the decohered neutrino propagation matrix is
\begin{equation}
\mathbb{P} \equiv 
\left(
\begin{array}{ccc}
|U_{e1}|^2 & |U_{e2}|^2 & |U_{e3}|^2  \\
|U_{\mu 1}|^2 & |U_{\mu 2}|^2 & |U_{\mu 3}|^2  \\
|U_{\tau 1}|^2 & |U_{\tau 2}|^2 & |U_{\tau 3}|^2  \\
\end{array}
\right) \,.
\label{propgn2}
\end{equation}
(It is seen that decoherence returns the quantum mechanical realm to that of classical overlap probabilities.) 

The probabilities for flavor oscillation are then easily calculated to be
\begin{equation}
P(\nu_\mu \to \nu_\mu) = P(\nu_\tau \to \nu_\tau) = P(\nu_\mu \leftrightarrow \nu_\tau)= \frac{1}{8}\, [4-\sin^2(2\theta_\odot) ] \,\,,
\label{p1}
\end{equation}
\begin{equation}
P(\nu_\mu \leftrightarrow \nu_e) 
= P(\nu_e \leftrightarrow \nu_\tau) = \frac{1}{4} \sin^2(2\theta_\odot) \,\,,
\label{p2}
\end{equation} 
and
\begin{equation}
P(\nu_e \to \nu_e) = 1-\frac{1}{2} \sin^2(2\theta_\odot) \,\,,
\label{p3}
\end{equation}
with $\sin^2(2\theta_\odot)\sim 8/9$.

Neutrinos from astrophysical sources are expected to arise dominantly
from the decays of pions and their muon daughters, which results in
initial flavor ratios $N_{\nu_e} : N_{\nu_\mu}: N_{\nu_\tau}$ of
nearly $1:2:0$.  Using (\ref{p1}), (\ref{p2}), and (\ref{p3}),  
it is straightforward to verify that the neutrinos will arrive 
at Earth with equipartition on the three flavors, $1:1:1$. The prediction for a pure $\bar \nu_e$ source,
originating via neutron $\beta$-decay, has different implications for
the flavor ratios; namely, a source flavor ratio $1:0:0$ yields Earthly ratios $\approx 5 : 2 : 2$~\cite{Anchordoqui:2003vc}.
And finally, the ``damped muon'' source, wherein muon energy-losses at the source effectively terminate 
the pion decay chain at $\pi^\pm\to \mu^\pm + \stackrel{(-)}{\nu_\mu}$, evolves the initial $0:1:0$ flavor ratios to 
$4:7:7$.

\end{document}